\documentclass[12pt,draftclsnofoot,onecolumn]{IEEEtran}

\ifx\pdfoutput\undefined
\usepackage[dvips]{graphicx}
\else
\usepackage[pdftex]{graphicx}
\fi
\usepackage{epstopdf}
\usepackage{amssymb,amsmath,color,graphicx, amsbsy}
\usepackage{graphicx}
\usepackage{subfigure}
\usepackage{cite}
\usepackage{algorithm}
\usepackage{algorithmic}
\usepackage{amsmath}
\usepackage{amsfonts}
\usepackage{amssymb}
\usepackage{graphicx}

\DeclareMathOperator*{\argmax}{arg\,max}
\newtheorem{theorem}{Theorem}
\newtheorem{assumption}{Assumption}

\newtheorem{definition}{Definition}

\newtheorem{problem}{Problem}
\newtheorem{game}{Game}
\newtheorem{proposition}{Proposition}

\IEEEaftertitletext{\vspace{-2ex}}%

\ifodd 0
\newcommand{\rev}[1]{{\color{red}#1}} 
\newcommand{\com}[1]{\textbf{\color{blue} (COMMENT: #1)}} 
\else
\newcommand{\rev}[1]{#1}
\newcommand{\com}[1]{}
\fi

\begin{document}



\title {\fontsize{22pt}{22pt}\selectfont{Inter-Session Network Coding with Strategic Users: A Game-Theoretic Analysis of Network
Coding}}




\author{Amir-Hamed Mohsenian-Rad, Jianwei Huang, Vincent W.S.~Wong, \\ Sidharth Jaggi, and Robert Schober \thanks{A.H.~Mohsenian-Rad, V.~W.S.~Wong, and
R.~Schober are with the ECE Department, University of British
Columbia, Vancouver, Canada, email: \textit{\{hamed, vincentw,
rschober\}@ece.ubc.ca}. J.\ Huang and S.~Jaggi are with the
Information Engineering Department, Chinese University of Hong Kong,
Hong Kong, email: \textit{\{jwhuang, jaggi\}@ie.cuhk.edu.hk}. Part
of this paper has been accepted for presentation at \emph{IEEE
International Conference on Communications} (\emph{ICC'09}),
Dresden, Germany, June 2009. } }

\maketitle

\vspace{-1cm}



\begin{abstract}
\thispagestyle{empty}

A common assumption in the existing network coding literature is
that the users are \textit{cooperative} and do \textit{not} pursue
their own interests. However, this assumption can be violated in
practice. In this paper, we analyze \textit{inter-session network
coding} in a wired network using game theory. We assume that the
users are \textit{selfish} and act as \textit{strategic} players to
maximize their own utility, which leads to a resource allocation
\emph{game} among users. In particular, we study network coding with
strategic users for the well-known \emph{butterfly} network topology
where a bottleneck link is shared by several network coding and
routing flows. We prove the existence of a Nash equilibrium for a
wide range of utility functions. We also show that the number of
Nash equilibria can be large (even \emph{infinite}) for certain
choices of system parameters. This is in sharp contrast to a similar
game setting with traditional packet forwarding where the Nash
equilibrium is always unique. We then characterize the worst-case
efficiency bounds, i.e., the \emph{Price-of-Anarchy} (PoA), compared
to an \textit{optimal} and \emph{cooperative} network design. We
show that by using a novel \textit{discriminatory pricing} scheme
which charges encoded and forwarded packets differently, we can
improve the PoA in comparison with the case where a \emph{single}
pricing scheme is being used. However, regardless of the
discriminatory pricing scheme being used, the PoA is still worse
than for the case when network coding is not applied. This implies
that, although inter-session network coding can improve performance
compared to ordinary routing, it is significantly \emph{more
sensitive} to users' strategic behavior. For example, in a butterfly
network where the side links have \emph{zero} cost, the efficiency
at certain Nash equilibria can be as low as 25\%. If the side links
have \emph{non-zero} cost, then the efficiency at some Nash
equilibria can further reduce to only 20\%. These results generalize
the well-known result of guaranteed 67\% worst-case efficiency for
traditional packet forwarding networks.

\vspace{0.4cm}


\vspace{1cm}

\end{abstract}

\vspace{-0.45cm}

\section{Introduction} \label{sec:introduction}

\setcounter{page}{1}


Since the seminal paper by Ahlswede \emph{et al.}
\cite{ahlswede2000nif}, a rich body of work has been reported on how
network coding can improve performance in both wired and wireless
networks \cite{Chen2007, traskov2006, eryilmaz2007, wang2007}.
Network coding can be performed by \emph{jointly} encoding multiple
packets either from the \emph{same} user or from \emph{different}
users. The former is called \emph{intra-session} network coding
\cite{ahlswede2000nif, Chen2007} while the latter is called
\emph{inter-session} network coding \cite{traskov2006, eryilmaz2007,
wang2007}. A common assumption in most network coding schemes in the
literature is that the users are \textit{cooperative} and do
\textit{not} pursue their own interests. However, this assumption
can be violated in practice. Therefore, assuming that the users are
\textit{selfish} and \textit{strategic}, in this paper we ask the
following key questions: (a) \emph{What is the impact of users'
strategic behavior on network performance?} (b) \emph{How does this
impact change with different pricing schemes that we may potentially
choose for each link?}




It is widely accepted that \textit{pricing} is an effective approach
in terms of improving the efficiency of network resource allocation,
especially in \textit{distributed} settings. In \cite{Ke98}, Kelly
\emph{et al.} showed that if users are \emph{price takers} (i.e.,
they treat network prices as \emph{fixed}), efficient resource
allocation can be achieved by properly setting \emph{congestion
prices} on each of the shared links. Recently, Johari \emph{et al.}
studied how the results on efficiency can change in both
capacity-constrained \cite{JoTs03} and capacity-unconstrained
\cite{johari_jsac2006} networks if users are \emph{price
anticipators} who realize that the price is directly impacted by
each individual user's behavior. In this case, users play a
\textit{game} with each other, and the efficiency of resource
allocation is characterized by the Nash equilibrium of the game. A
key performance metric is the \emph{Price-of-Anarchy (PoA)}, which
measures the \emph{worst-case efficiency loss} at a Nash equilibrium
due to users' price anticipating behavior. The PoA is equal to 1 if
there is \emph{no} efficiency loss. A smaller PoA denotes a higher
efficiency loss. Other recent work on resource allocation games and
the  PoA include \cite{chen2009, harks2009, sanghavi2004oad,
acemoglu2007cps, RoTa02, johari2005, yang2006}. To the best of our
knowledge, none of the previous works along this line study price
anticipation in \emph{network coding} systems.


The game theoretic analysis of network coding has received limited
attention in the literature, e.g., in \cite{bhadra2006mcs,
Li2008,liang2006mgm, li2007mcm, zhang2008dgt}.
All results in \cite{bhadra2006mcs,
Li2008,liang2006mgm,li2007mcm,zhang2008dgt} focus on the case of
\emph{intra-session} network coding, whereas we consider
\emph{inter-session} network coding in this paper. In
\cite{Price2008}, a game theoretic analysis for \emph{inter-session}
network coding of \emph{unicast} flows in a single bottleneck link
is considered. It is shown that in some classes of \textit{two-user}
networks, it is possible to use a rate allocation mechanism to
enforce cooperation among users.
%
 In this paper, we assume that there are $N \geq 2$ users, two of which use network coding while the rest only use routing.
 This helps us to better understand the interaction between network coding and routing
flows. In fact, we show that the performance degrades when both
network coding and routing sessions share the same link. Our results
are also different from those in \cite{Price2008} since we consider
the capacity-unconstrained case instead of the capacity-constrained
case as in \cite{Price2008}. In fact, due to the focus on the
capacity region of the network coding scheme, the work in
\cite{Price2008} did not consider the impact of the \textit{utility
functions} of the users, the cost of the side links, price
anticipation, price discrimination, and the PoA.

%
%


The key contributions of this paper are as follows:
\begin{itemize}
  \item \emph{New problem formulation}: We formulate the problem of maximizing the
  \emph{network aggregate surplus}, i.e., the total utility of all users minus the network
  cost, for inter-session network coding. As far as we know, such a problem has not been studied in the literature before.

  \item \emph{Innovative pricing schemes}:  We consider two pricing schemes: \emph{non-discriminatory pricing} and \emph{discriminatory pricing}.
The first one is the traditional approach in networks with
routing-only users. It charges all packets with the same price.
  The second pricing scheme is a novel generalization of the first one where the encoded and forwarded packets are charged with different prices.
  We show that due to the special properties of network coding, discriminatory pricing is more reasonable in terms of reflecting the \emph{actual load} generated by each user.

  \item \emph{Characterization of Nash equilibria}: We prove that the existence of a Nash equilibrium
for the formulated game is always guaranteed; however, there can be
\emph{many} (even \emph{infinite}) Nash equilibria in the resource
allocation game with inter-session network coding.

  \item \emph{Calculation of the PoA in a butterfly network with zero-cost side links}: We show that, among the two
  aforementioned pricing schemes, a properly chosen discriminatory pricing leads to a better PoA compared with
  the non-discriminatory approach. We also show that the PoA is always smaller (i.e., worse) compared
with the case without network coding. In particular, at certain Nash
equilibria, the PoA can be as low as 25\%, which is less than the
well-known result of guaranteed 67\% worst-case efficiency
in \cite{johari_jsac2006} for
packet forwarding networks.

  \item \emph{Calculation of the PoA in a butterfly network with non-zero-cost side
links}: We further show that if the side links in the studied
butterfly network topology have non-zero cost, then the PoA can
further reduce to only 20\%. This occurs due to the fact that in
this case \emph{none} of the users have an incentive to participate
in network coding. This implies that if the users have strategic
behavior, then it is important to design mechanisms to encourage
users to perform network coding; otherwise, we cannot benefit from
the advantages of network coding.

\end{itemize}

The key results of this paper together with a comparison with the
related state-of-the-art results \emph{without} considering network
coding in \cite{johari_jsac2006} are summarized in Table I.

The rest of the paper is organized as follows.
In Section \ref{sec:single_link:routing}, we review some recent
results on resource allocation games with routing. In Section
\ref{sec:single_link:coding}, we extend those results to the case
when two users can jointly perform inter-session network coding in a
butterfly network where the side links have zero cost. We further
extend our analysis to the case where the side links have non-zero
cost in Section \ref{sec:single_link:coding:nonzerosidelinks}.
Conclusions and future work are discussed in Section
\ref{conclusion}. The key notations we used in this paper are
summarized in Table \ref{tab:notations}.






\vspace{-0.1cm}

\section{Background: Resource Allocation Game with Routing Flows} \label{sec:single_link:routing}

\begin{figure}[t]
\vspace{1cm}
\begin{center}
{\scalebox{0.9}{\includegraphics*{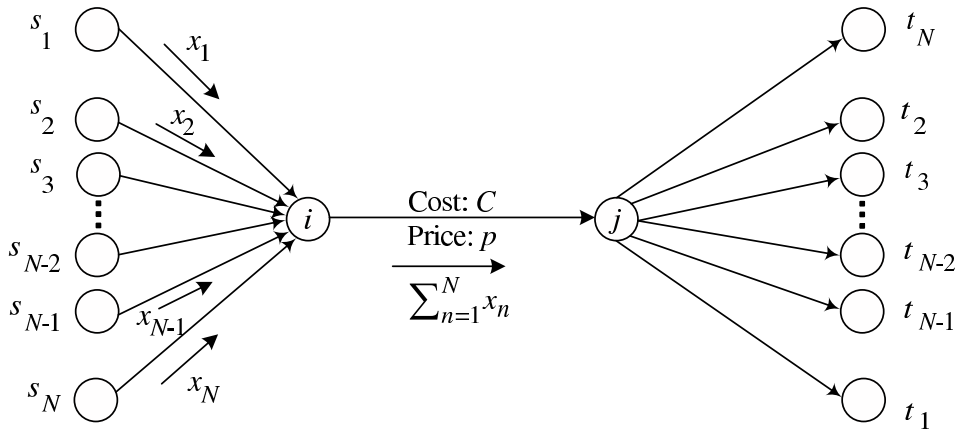}}}\vspace{-0.1cm}
\caption{A single wireline bottleneck link shared by $N$
\emph{routing} flows \cite{johari_jsac2006}. Transmission of packets
over bottleneck link at each unit of data rate incurs at cost of
$C(\sum_{n=1}^N x_n)$ and imposes price $p(\sum_{n=1}^N x_n)$.
Elastic data rates $x_1, \ldots, x_N$ are selected by users $1,
\ldots, N$, respectively.} \vspace{-0.7cm} \label{fig:f1}
\end{center}
\end{figure}

In this section, we consider a resource allocation game in which
multiple end-to-end users compete to send their packets through a
single shared link as in Fig.~\ref{fig:f1}. By construction, no
inter-session network coding is performed in this case. This is a
well-known problem which has been widely studied in
\cite{sanghavi2004oad, JoTs03, johari_jsac2006, Ke98, chen2009,
harks2009}. Here, we summarize the key results in
\cite{johari_jsac2006}, which present the proper terminology and
serve as a benchmark for our later discussions.

In Fig. \ref{fig:f1}, a set of users $\mathcal{N} = \{1, \ldots,
N\}$ \emph{shares} the bottleneck link $(i, j)$ between nodes $i$
and $j$. All packets that arrive at node $i$ are simply
\emph{forwarded} to node $j$ through link $(i, j)$.
%
For each user $n \in \mathcal{N}$, we denote the transmitter
and receiver nodes by $s_n$ and $t_n$, respectively. Let $x_n$
denote the transmission rate of user $n \in \mathcal{N}$.
We assume that each user $n \in \mathcal{N}$ has a \emph{utility
function} $U_n$, representing its degree of satisfaction based on
its achievable data rate $x_ n$. On the other hand, the shared link
has a \emph{cost function} $C$, which depends on the total rate
(i.e., $\sum_{n \in \mathcal{N}} x_n$). As in \cite{johari_jsac2006,
chen2009, harks2009}, we make the following common assumptions
throughout this paper:
%
%

\begin{assumption}[Users' Utility Functions] \label{assumption1}
For each user $n \! \in \! \mathcal{N}$, the utility function
$U_n(x_n)$ is \emph{concave}, \emph{non-negative},
\emph{increasing}, and \emph{differentiable}.
\end{assumption}

\begin{assumption}[Link Cost and Price Functions] \label{assumption2}
There exists a \emph{differentiable}, \emph{convex}, and
\emph{non-decreasing} function $p(q)$ over $q \geq 0$, with $p(0)
\geq 0$ and $p(q) \rightarrow \infty$ as $q \rightarrow \infty$,
such that for each $q \geq 0$, the cost is modeled as $C(q) =
\int_0^q p(z) \: d z$. Here, $C(q)$ is \emph{convex} and
\emph{non-decreasing}. In particular, we assume that there exists $a
> 0$ such that $p(q) = a q$ and $C(q) = \int_0^q a \: z \: d z =
\frac{a}{2} q^2$. That is, the cost function $C(q)$ is
\emph{quadratic} and the price function $p(q)$ is \emph{linear}.
Notice that linear price functions are the \emph{only} price
functions that satisfy the four well-known axioms of
\emph{rescaling}, \emph{consistency}, \emph{additivity}, and
\emph{positivity} for cost-sharing systems\footnote{The first axiom,
i.e., \emph{rescaling}, requires that the prices should be
\emph{independent} of the \emph{units} of measurement. The second
axiom, i.e., \emph{consistency}, requires that two users having the
same effect on the cost should face the same price. The third axiom,
i.e., \emph{additivity}, requires that if the cost function can be
decomposed, then so should the price function. Finally, the fourth
axiom, i.e., \emph{positivity}, requires that if the cost is
positive, then the price should be at least non-negative. Notice
that the last axiom reflects a notion of fairness toward the service
provider. \cite{samet}.} \cite{samet, chen2009}.
\end{assumption}

Assumption \ref{assumption1} is often used to model applications
with \emph{elastic} traffic, e.g., for remote file transfer using
the file transfer protocol (FTP) \cite{Ke98}.
Examples of utility functions which satisfy Assumption
\ref{assumption1} include the well-known class of $\alpha$-fair
utility functions with $\alpha \in (0, 1)$ \cite{mo_2000}.
Assumption \ref{assumption2} is also a common assumption in the
network resource management literature (cf. \cite{shetty2008,
johari2005, chen2009}). In practice, the cost function $C$ may
reflect the \emph{actual cost} (e.g., in dollars) of transmitting
units of data over link $(i, j)$ or simply the \emph{delay} that the
packets experience over link $(i, j)$. The more the aggregate data
on the link, the higher is the average queueing delay.

Let $\boldsymbol{x} = (x_1, \ldots, x_N)$. Given \emph{complete}
knowledge and \emph{centralized} control of the network in Fig.
\ref{fig:f1}, an efficient rate allocation can be characterized as a
solution of the following problem:

\vspace{0.1cm}

\begin{problem}[Surplus Maximization with Routing] \label{problem1}
\begin{equation}
\begin{aligned}
\nonumber
& \underset{\boldsymbol{x}}{\text{maximize}} \ & \textstyle  \sum_{n=1}^N U_n\left( x_n \right) - C\left(\sum_{n=1}^N x_n \right) \label{r_problem1_line1} \\
\nonumber & \text{subject to} \ \: & x_n \geq 0, \ \ \ \ n = 1,
\ldots, N. \ \ \ \ \ \ \;  \end{aligned}
\end{equation}
\end{problem}
\vspace{0.3cm}

\noindent The objective function in Problem \ref{r_problem1_line1}
is the \emph{network aggregate surplus} \cite{microeconomic_book}.
Network aggregate surplus maximization is a common network design
objective (cf. \cite{shetty2008, johari2005, chen2009, harks2009}).
Clearly, Problem \ref{problem1} is a \emph{convex} optimization
problem. In general, since the utility functions are local to the
users and are \emph{not} known to each link, efficient resource
allocation can be achieved via \emph{pricing}. Given the rate vector
$\boldsymbol{x}$ from the users, the shared link $(i, j)$ can set a
\emph{single} price
\begin{equation}
\abovedisplayskip 3pt \belowdisplayskip 3pt
\textstyle
\mu(\boldsymbol{x}) = p \left( \sum_{n = 1}^N x_n \right)
\label{r_pricing}
\end{equation}
for each unit of data rate it carries. Each user $n \in \mathcal{N}$
then pays $x_n \mu(\boldsymbol{x})$ for its data rate $x_n$.

Next, we analyze how the users determine their rates based on the
price set by link $(i, j)$. First, assume that the users are
\emph{price takers}, i.e., they do \emph{not} anticipate the effect
of a change of their rates on the resulting price. In that case,
each user $n \in \mathcal{N}$ selects its rate $x_n$ to maximize its
\emph{own} surplus, i.e., utility minus payment, by solving the
following \emph{local} problem \cite{Ke98}:
\begin{equation}
\abovedisplayskip 3pt \belowdisplayskip 3pt \max_{x_n \geq 0} \
\left( U_n(x_n) - x_n \mu \right) \ \ \ \ \ \Rightarrow \ \ \ \ \
x_n = {U^\prime_n}^{-1}(\mu), \label{data_rate_price_taking}
\end{equation}
where ${U^\prime_n}^{-1}$ denotes the inverse of the derivative of
utility function $U_n$ and price $\mu$ is as in (\ref{r_pricing}).
\com{I think we should also consider the lower boundary of 0?} From
the first fundamental theorem of welfare economics, if each user $n
\! \in \! \mathcal{N}$ selects its rate as in
(\ref{data_rate_price_taking}), then the network aggregate surplus
is maximized at equilibrium \cite[p. 326]{microeconomic_book}.

Next, we consider \emph{price anticipating} users: each user can
anticipate the effect of its selected data rate on the resulting
price. In this case, each user $n \! \in \! \mathcal{N}$ no longer
selects its rate as in (\ref{data_rate_price_taking}). Instead, it
\emph{strategically} selects $x_n$ to maximize its surplus given the
knowledge that the price $\mu(\boldsymbol{x})$ is set according to
(\ref{r_pricing}) and is \emph{not} fixed.
%
%
Clearly, the decision made by user $n$ also depends on the rates
selected by other users, leading to a \emph{resource allocation
game} among all users:

\vspace{0.1cm}

\begin{game}[Resource Allocation Game Among Routing Flows]
\label{game1}  $ \ \ $

\begin{itemize}
\item \emph{Players}: Users in set $\mathcal{N}$.

\item \emph{Strategies}: Transmission rates $\boldsymbol{x}$ for all users.

\item \emph{Payoffs}: $P_n(x_n ; \boldsymbol{x}_{-n})$ for
each user $n \! \in \! \mathcal{N}$, where
\begin{equation}
\abovedisplayskip 3pt \belowdisplayskip 3pt
\nonumber
\textstyle
P_n(x_n ; \boldsymbol{x}_{-n}) = U_n(x_n) - x_n \: p\left( \sum_{n=1}^N x_n \right), 
\end{equation}
and $\boldsymbol{x}_{-n}$ denotes the vector of selected data rates
for all users \emph{other than} user $n$.
\end{itemize}

\end{game}

In Game \ref{game1}, each user $n \in \mathcal{N}$ selects its rate
$x_n \geq 0$ to \emph{maximize} its payoff $P_n(x_n;
\boldsymbol{x}_{-n})$. A \emph{Nash equilibrium} of Game \ref{game1}
is defined as a non-negative rate vector $\boldsymbol{x}^\ast =
(x_1^\ast, \ldots, x_N^\ast)$ such that for each user $n \in
\mathcal{N}$, we have
%
$ P_n( x^\ast_n; \boldsymbol{x}^\ast_{-n} ) \geq P_n( \bar{x}_n;
\boldsymbol{x}^\ast_{-n} )$ for \emph{any} $\bar{x}_n \geq 0$.
%
In a Nash equilibrium $\boldsymbol{x}^\ast$, no user $n \in
\mathcal{N}$ can increase its payoff by \emph{unilaterally} changing
its strategy $x_n$.

\begin{definition} \label{definition1}
Let $\boldsymbol{x}^S \! = \! (x_1^S, \ldots, x_N^S)$ be an optimal
solution for Problem \ref{problem1} and $\boldsymbol{x}^\ast$ be a
Nash equilibrium for Game \ref{game1} for the \emph{same} choice of
system parameters. The \emph{efficiency} at Nash equilibrium
$\boldsymbol{x}^\ast$ is defined as the \emph{ratio} of the network
aggregate surplus at $\boldsymbol{x}^\ast$ to the network aggregate
surplus at $\boldsymbol{x}^S$:

\vspace{-0.4cm}

\begin{equation}
\frac{\sum_{n=1}^N U_n\left( x^\ast_n \right) - C \!
\left(\sum_{n=1}^N x^\ast_n \! \right)}{ \sum_{n=1}^N U_n\left(
x^S_n \right) \! - \! C \! \left(\sum_{n=1}^N x^S_n \! \right)}.
\label{r_bound}
\end{equation}
\end{definition}

\vspace{0.5cm}

\begin{definition} \label{definition2}
The \emph{price-of-anarchy} $\text{PoA}\left(\text{Game
\ref{game1}}, \text{Problem \ref{problem1}}\right)$ is defined as
the \emph{worst-case} efficiency of a Nash equilibrium of Game
\ref{game1} among \emph{all} possible selections of system
parameters (i.e., number of users, utility, cost, and price
functions) as long as Assumptions \ref{assumption1} and
\ref{assumption2} hold.
\end{definition}

%
%
%

\vspace{0.1cm}

\rev{The following key result is based on \cite[Theorem
3]{johari_jsac2006}}:

\vspace{0.1cm}

\begin{theorem}  \label{theorem1}
Suppose Assumptions \ref{assumption1} and \ref{assumption2} hold.
%
%
(a) Game \ref{game1} always has a \emph{unique} Nash equilibrium.
(b) We have
%
\begin{equation} \label{PoA_game_1_problem_1}
\text{PoA}\left(\text{Game \ref{game1}}, \text{Problem
\ref{problem1}}\right) = \frac{2}{3}.
\end{equation}
%
%


\end{theorem}


\vspace{0.15cm}

From Theorem \ref{theorem1}, for \emph{any} choice of parameters,
the network aggregate surplus at a Nash equilibrium of Game
\ref{game1} is guaranteed to be \emph{at least} $\frac{2}{3} \approx
67\%$ of the \emph{optimal} network aggregate surplus. Notice that
the PoA indicates how bad the network performance can become due to
strategic behavior of the users. In the rest of this paper, we
generalize Theorem \ref{theorem1} to the case where some of the
users can perform inter-session network coding. We show that such a
generalization is non-trivial and the results are drastically
different from those in Theorem \ref{theorem1} in several aspects.

\vspace{0.1cm}

\section{Resource Allocation Game with Inter-Session Network Coding
and Routing Flows: The Case with Zero Costs for Side Links}
\label{sec:single_link:coding}

In this section, we reformulate Problem \ref{problem1} and Game
\ref{game1} in a network scenario where a bottleneck link is shared
not only by routing flows, but also by some inter-session network
coding flows. We then extend the results in Theorem \ref{theorem1}
according to a \emph{new} network resource allocation game. We show
that the new game setting may have \emph{multiple} Nash equilibria.
In addition, the 67\% efficiency bound in Theorem \ref{theorem1} is
no longer guaranteed. In fact, although the efficiency loss is still
\emph{bounded}, the performance at some Nash equilibria is only 25\%
of the optimal performance.


\subsection{Problem Formulation} \label{sec:single_link:coding:problem_formulation}

\begin{figure}[t]
\begin{center}
{\scalebox{0.9}{\includegraphics*{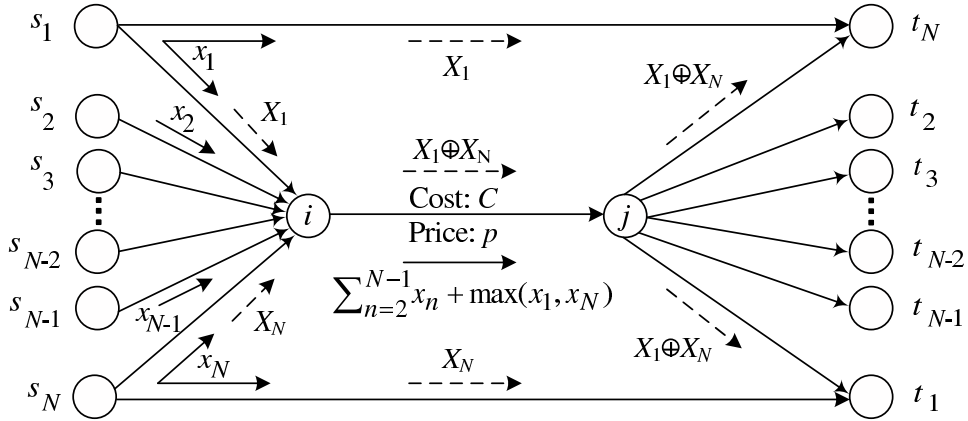}}} \caption{A
butterfly network (cf. \cite{khreishah2008, wang2007}) with a single
bottleneck link shared by $N$ flows from $N$ users and two side
links. Transmission of packets over bottleneck link at each unit of
data rate incurs at cost of $C\left(\sum_{n=2}^{N-1} x_n + \max(x_1,
x_N)\right)$ and imposes a price $p\left(\sum_{n=2}^{N-1} x_n +
\max(x_1, x_N)\right)$. Since the source node $s_1$ of user 1 is
located closer to the destination node $t_N$ of user $N$ (and vice
versa), users 1 and $N$ can jointly perform \emph{inter-session
network coding} and reduce the traffic load on the shared bottleneck
link $(i, j)$ (and reduce cost $C$). The side links $(s_1, t_N)$ and
$(s_N, t_1)$ are assumed to be free of charge in this setting. Here,
$x_1$ and $x_N$ denote the data rates of source nodes $s_1$ and
$s_N$, respectively. On the other hand, $X_1$ and $X_N$ denote the
actual packets/symbols sent from nodes $s_1$ and $s_N$,
respectively. The notation $X_1 \oplus X_N$ indicates a network
coded packet/symbol obtained by jointly encoding packets $X_1$ and
$X_N$.} \vspace{-0.5cm} \label{fig:f2}
\end{center}
\end{figure}

Consider the modified network model in Fig.~\ref{fig:f2}. The
network topology in this figure is called a \emph{butterfly} network
in the network coding literature  \cite{khreishah2008, wang2007}.
The network in Fig. \ref{fig:f2} is similar to the one in Fig.
\ref{fig:f1}, except that it includes two direct \emph{side} links
$(s_1, t_N)$ and $(s_N,t_1)$. In this scenario, the source node of
user 1 is located closer to the destination node of user $N$ than to
its own destination node (and vice versa). Thus, users 1 and $N$ can
perform \emph{inter-session} network coding. In this setting, we can
distinguish two different types of users in the system:

\begin{itemize}
\item \emph{Network Coding Users}: Users 1 and $N$, who \emph{can} perform
inter-session network coding.

\item \emph{Routing Users}: Users $2, \ldots, N-1$, who \emph{cannot} perform
inter-session network coding.
\end{itemize}

Let $X_1$ and $X_N$ denote packets sent from source nodes $s_1$ and
$s_N$, respectively. Node $i$ can encode packets $X_1$ and $X_N$
jointly, and then send out the resulting encoded packet, denoted by
$X_1 \oplus X_N$, towards node $j$ (and from there towards $t_1$ and
$t_N$). Given the \emph{remedy} data $X_1$ from the side link $(s_1,
t_N$) and the \emph{remedy} data $X_N$ from the side link $(s_N,
t_1)$, nodes $t_N$ and $t_1$ can \emph{decode} the encoded packets
that they receive. In fact, in this setting, nodes $t_1$ and $t_N$
can decode both $X_1$ and $X_N$. Clearly, the benefit of network
coding is to reduce the traffic load on link $(i, j)$ (thus reducing
the cost) while achieving the \emph{same} data rates compared to the
case that no network coding is performed. It is worth mentioning
that although the network coding scenario in Fig.~\ref{fig:f2} is
simple, it is the building block for more general network coding
scenarios. For example, in \cite{traskov2006, indexcoding2008,
Chen2007} the network is modeled as a \emph{superposition} of
several butterfly networks. Thus, understanding this model is the
key to understand more general networks. Further to Assumptions
\ref{assumption1} and \ref{assumption2}, in this section, we also
assume that:

\begin{assumption}[Zero Cost for Side Links] \label{assumption3}
The two side links $(s_{1},t_{N})$ and $(s_{N},t_{1})$ in
Fig.~\ref{fig:f2} always have \emph{zero} cost and impose
\emph{zero} prices.
\end{assumption}

For example, if the link cost is used to model the link \emph{delay}
and the side links $(s_1, t_N)$ and $(s_N, t_1)$ have a higher
capacity than the shared link $(i, j)$, then the costs of the side
links can be neglected. The case where the side links have
\emph{non-zero} cost is studied in Section
\ref{sec:single_link:coding:nonzerosidelinks}.

For the network in Fig. \ref{fig:f2}, the network aggregate surplus
maximization problem becomes:

\vspace{0.1cm}
\begin{problem} [Surplus Maximization with Network Coding and Zero-Cost Side Links]\label{problem2}
\begin{equation}
\nonumber
\begin{aligned}
& \underset{\boldsymbol{x}}{\text{maximize}} & \! \sum_{n=1}^N U_n\left( x_n \right) \! - C \! \left(\sum_{n=2}^{N-1} x_n \! + \! \max( x_1, x_N) \! \right) \label{nc_problem1_line1} \\
& \text{subject to} & \! x_n \geq 0, \ \ n = 1, \ldots, N. \ \ \ \ \
\ \ \ \ \ \ \ \ \ \ \ \ \ \ \ \ \
\end{aligned}
\end{equation}
\end{problem}
\vspace{0.2cm}
Comparing Problems \ref{problem1} and \ref{problem2}, we can see
that the cost term $C(\sum_{n=1}^N x_n)$ in Problem \ref{problem1}
 is now replaced by a new cost term $C(\sum_{n=2}^{N-1} x_n
 + \max( x_1, x_N) )$. The intuition behind the objective
function in Problem \ref{problem2} is as follows. Since $x_1$ and
$x_N$ are selected \emph{independently} by users 1 and $N$, in
general, we may have $x_1 \neq x_N$. Thus, regardless of the choice
of an \emph{efficient} network coding scheme, the intermediate node
$i$ can perform network coding \emph{only} at rate $\min(x_1, x_N)$.
Those packets which are \emph{not} encoded (e.g., at rate $x_1 -
\min(x_1, x_N)$ if $x_1 \geq x_N$, and at rate $x_N - \min(x_1,
x_N)$ if $x_1 \leq x_N$) are simply \emph{forwarded}, leading to an
aggregate rate of $\sum_{n=2}^{N-1} x_n + \max(x_1, x_N)$ on link
$(i, j)$. Note that if $x_1 = x_N$, then \emph{all} packets from
users 1 and $N$ are jointly encoded. In fact, this is the case at
optimality as the following result suggests:

\vspace{0.2cm}

\begin{theorem} \label{theorem2}
Let $\boldsymbol{x}^S \! = \! ( x^S_1, \ldots, x^S_N )$ be an
\emph{optimal} solution for Problem \ref{problem2}. We have
%
$x^S_1 = x^S_N$.
%
\end{theorem}

\vspace{0.2cm}

The proof of Theorem \ref{theorem2} is given in Appendix
\ref{app:proof_theorem2}. From Theorem \ref{theorem2}, users $1$ and
$N$ should have \emph{equal} rates at optimality. Notice that since
Problem \ref{problem2} is a convex optimization problem, it can be
solved in a centralized fashion using convex programming techniques
\cite{convex_book}. Distributed resource allocation can also be done
via pricing as explained next.

Following the same pricing scheme as in Section
\ref{sec:single_link:routing}, the shared link may apply a
\emph{single} price for \emph{all} (i.e., coded and routed) packets:
%
\begin{equation}
\textstyle \mu(\boldsymbol{x}) = p \left( \sum_{n=2}^{N-1} x_n +
\max( x_1, x_N ) \right). \label{price_NC}
\end{equation}
Each user $n$ pays $x_n \mu(\boldsymbol{x})$. However, this leads to
\emph{double charging} for \emph{encoded} packets. Note that each
encoded packet includes the data from \emph{both} users 1 and $N$.
Thus, the \emph{single} pricing model in (\ref{price_NC}) leads to
more payment from the users than the actual link cost. This can be
avoided by \emph{price discrimination}, i.e., charging the routed
and network-coded packets with \emph{different} prices.

Let $\mu(\boldsymbol{x})$ in (\ref{price_NC}) denote the price to be
charged for routed packets. \rev{Under the discriminatory pricing
scheme,} we define another price $\delta(\boldsymbol{x})$ for
network coded packets. In general, we have
%
\begin{equation}
\abovedisplayskip 3pt \belowdisplayskip 3pt \delta(\boldsymbol{x}) =
\beta \: \mu(\boldsymbol{x}), \label{delta_mu_relationship}
\end{equation}
where $0 \! < \! \beta \! \leq \! 1$ is a pricing parameter. If
$\beta \! = \! 1$, then there is only a single price. If $\beta \! <
\! 1$, then the encoded packets are charged less than the routed
packets as they carry more information compared to routing packets
of the same size. In this paper, we focus on the case of $\beta \! =
\! \frac{1}{2}$. This is the
%
%
only choice of $\beta$ that avoids \emph{over}- or
\emph{under}-charging with two network coding flows.
%
%
%
%
Based on the this pricing scheme, user 1 pays
%
$\min(x_1, x_N) \delta(\boldsymbol{x}) + \left( x_1 - \min(x_1, x_N)
\right) \mu(\boldsymbol{x})$. 
%
That is, it pays for transmission of its encoded packets at a price
of $\delta(\boldsymbol{x})$ and for transmission of its forwarded
(not coded) packets at a price of $\mu(\boldsymbol{x})$.
From (\ref{delta_mu_relationship}), the \emph{total} payment by user
1 is
\begin{equation} \label{payoff_user_1_discrimination}
\abovedisplayskip 3pt \belowdisplayskip 3pt \left( x_1 - (1 - \beta)
\min(x_1, x_N) \right) \mu(\boldsymbol{x}).
\end{equation}
A similar payment model applies to user $N$. Notice that each user
$n = 2, \ldots, N-1$ pays $x_n \mu(\boldsymbol{x})$.

We are now ready to define a resource allocation game for the
network setting in Fig. \ref{fig:f2}, when users can anticipate
prices $\mu$ and $\delta$ according to (\ref{price_NC}) and
(\ref{delta_mu_relationship}), respectively:

\vspace{0.1cm}

\begin{game} \label{game2}
\emph{(Resource Allocation Game with Inter-session Network Coding
and Routing Flows where the Side Links Have Zero Costs and Zero
Prices)}

\begin{itemize}
\item \emph{Players}: Users in set $\mathcal{N}$.

\item \emph{Strategies}: Transmission rates $\boldsymbol{x}$ for all users.

\item \emph{Payoffs}: $Q_n(x_n ; \boldsymbol{x}_{-n})$ for each user
$n \in \mathcal{N}$. The network coding users $1$ and $N$ have
\begin{equation}
Q_1(x_1; \boldsymbol{x}_{-1}) = U_1(x_1) - \left( x_1 - (1-\beta)
\min(x_1, x_N) \right) p \left( \textstyle \sum_{r=2}^{N-1} x_r +
\max( x_1, x_N ) \right), \! \label{Q_1_payoff};
\end{equation}
\begin{equation}
Q_N(x_N; \boldsymbol{x}_{-N}) = U_N(x_N) - \left( x_N \! -
 \! (1\!-\!\beta) \min(x_1, x_N) \right) p \left( \textstyle
\sum_{r=2}^{N-1} x_r + \max( x_1, x_N ) \right), \!
\end{equation}
and each routing user $n \in \mathcal{N} \backslash
\{1, N\}$ has
\begin{equation}
\abovedisplayskip 4pt \belowdisplayskip 4pt \nonumber Q_n(x_n ;
\boldsymbol{x}_{-n}) = U_n(x_n) - x_n p \left( \textstyle
\sum_{r=2}^{N-1} x_r + \max( x_1, x_N ) \right).
\end{equation}

\end{itemize}

\vspace{0.1cm}

\end{game}

Comparing Games \ref{game1} and \ref{game2}, we can see that Game
\ref{game2} introduces significantly more complex payoff functions.
In the rest of this section, we answer the following questions:
\begin{enumerate}
\item Does Game \ref{game2} always (i.e., for \emph{any} choice of system parameters) have a Nash equilibrium?

\item If a Nash equilibrium exists for Game \ref{game2}, is it always unique?

\item What is the \emph{worst-case} efficiency (i.e., the PoA) at a Nash equilibrium
of Game \ref{game2}?
\end{enumerate}
%


\vspace{-0.5cm}

\subsection{Existence and Non-uniqueness of Nash Equilibria}

A Nash equilibrium of Game \ref{game2} with both routing and
inter-session network coding flows can be defined as a data rate
selection vector $\boldsymbol{x}^\ast \succeq \boldsymbol{0}$, where
the inequality is coordinate-wise, such that for all users $n \in
\mathcal{N}$, we have
%
$ Q_n( x^\ast_n; \boldsymbol{x}^\ast_{-n} ) \geq Q_n( \bar{x}_n;
\boldsymbol{x}^\ast_{-n} )$ for any choice of $\bar{x}_n \geq 0$.

\vspace{0.1cm}

\begin{theorem} \label{theorem4}
There exists \emph{at least} one Nash equilibrium in Game
\ref{game2}.


\end{theorem}

\vspace{0.1cm}

The proof of Theorem \ref{theorem4} is given in Appendix
\ref{app:proof_theorem4}. The key to prove Theorem \ref{theorem4} is
to apply \emph{Rosen's existence theorem for concave N-person games}
\cite[Theorem 1]{rosen}. In this regard, we show that for \emph{all}
users $n \! \in \! \mathcal{N}$, the payoff function $Q_n(x_n;
\boldsymbol{x}_{-n})$ is a \emph{concave} function with respect to
$x_n$, even though $Q_1$ and $Q_N$ are \emph{not} differentiable due
to the $\max$ and $\min$ functions.

From Theorem \ref{theorem4}, the existence of Nash equilibria for
the resource allocation game is still guaranteed when network coding
is applied.
However, as we will see in Section \ref{sec:best_response}, there
can multiple Nash equilibria in this case. This can change the
results on efficiency loss and the PoA.

\vspace{-0.3cm}

\subsection{Users' Best Responses}\label{sec:best_response}

The strategic behavior of users can be modeled based on their
\emph{best responses}. In this regard, each user $n \in \mathcal{N}$
selects its data rate as $x^B_n$ to \emph{maximize} its own payoff
$Q_n$, given $\boldsymbol{x}_{-n}$:
%
%
\begin{equation}
\abovedisplayskip 3pt x^B_n(\boldsymbol{x}_{-n}) \ = \ \argmax_{x_n
\geq 0} \ Q_n(x_n; \boldsymbol{x}_{-n}), \ \ \ \ \forall \;\! n \in
\mathcal{N}. \label{best_response_definition}
\end{equation}
Since the problem in (\ref{best_response_definition}) is
\emph{convex}, we can readily show the following for \emph{routing}
users.

\vspace{0.1cm}

\begin{proposition} \label{proposition1}

For each routing user $n \! \in \! \mathcal{N} \backslash \{1, N\}$,
the best response $x_n^B(\boldsymbol{x}_{-n})$ is obtained as the
value of $x_n$ which satisfies the following equation (bounded below
by 0):
\begin{equation} \label{best_response_proposition_routing_equation}
U_n^{\:\prime}(x_n) - a \left( \sum_{r=2, r\neq n}^{N-1} x_r +
\max(x_1, x_N) \right) - 2 a x_n = 0.
\end{equation}
%
%
%

\vspace{0.1cm}

\end{proposition}


Recall that the \emph{linear} pricing parameter $a$ is defined in
Assumption \ref{assumption2}. The proof of Proposition
\ref{proposition1} is given in Appendix
\ref{app:proof_proposition1}. The key idea is to take the derivative
of the payoff $Q_n(x_n; \boldsymbol{x}_{-n})$ with respect to $x_n$
and solve the resulted Karush-Kuhn-Tucker (KKT) optimality condition
\cite{convex_book}.

%
%
%
%
%
%
%
Obtaining the best response functions for network coding users $1$
and $N$ is more complex, mostly due to the non-differentiability of
the payoff functions $Q_1(x_1; \boldsymbol{x}_{-1})$ and $Q_N(x_N;
\boldsymbol{x}_{-N})$. In fact, network coding user $1$ should
\emph{separately} examine two scenarios:

\noindent (a) Selecting its strategy (data rate) $x_1$ to be
\emph{greater} than or equal to $x_N$:
\begin{equation}
\tilde{x}^B_1(\boldsymbol{x}_{-1}) = \argmax_{x_1 \geq x_N} \
U_1(x_1) - \left( x_1 - (1-\beta) x_N \right) \: a \left( \sum_{n =
2}^{N-1} x_n + x_1 \right). \label{user_1_first_problem}
\end{equation}
%
(b) Selecting its strategy (data rate) $x_1$ to be \emph{less} than
or equal to $x_N$:
\begin{equation}
\hat{x}^B_1(\boldsymbol{x}_{-1}) = \argmax_{0 \leq x_1 \leq x_N} \ \
U_1(x_1) - \beta x_1 \ a \left( \sum_{n = 2}^{N-1} x_n + x_N
\right). \label{user_1_second_problem}
\end{equation}
%
%
The intuition behind the objective functions in
(\ref{user_1_first_problem}) and (\ref{user_1_second_problem}) is as
follows. In (\ref{user_1_first_problem}), since the strategy of user
1, i.e., its data rate $x_1$, is \emph{lower bounded} by $x_N$, we
have: $\min(x_1, x_N) = x_N$ and $\max(x_1, x_N) = x_1$. Thus, the
payoff function $Q_1(x_1; \boldsymbol{x}_{-1})$ in
(\ref{Q_1_payoff}) reduces to the objective function in
(\ref{user_1_first_problem}). On the other hand, in
(\ref{user_1_second_problem}), since the data rate $x_1$ is
\emph{upper bounded} by $x_N$, we have: $\min(x_1, x_N) = x_1$,
$\max(x_1, x_N) = x_N$, and $x_1 - (1 - \beta) \min(x_1, x_N) = x_1
- (1-\beta) x_1 = \beta x_1$. Thus, the payoff function $Q_1(x_1;
\boldsymbol{x}_{-1})$ in (\ref{Q_1_payoff}) reduces to the objective
function in (\ref{user_1_second_problem}).

Given $\tilde{x}^B_1(\boldsymbol{x}_{-1})$ and
$\hat{x}^B_1(\boldsymbol{x}_{-1})$, if
$Q_1(\tilde{x}^B_1(\boldsymbol{x}_{-1}); \boldsymbol{x}_{-1}) \geq
Q_1(\hat{x}^B_1(\boldsymbol{x}_{-1}); \boldsymbol{x}_{-1})$, then
user 1 selects its best response $x^B_1(\boldsymbol{x}_{-1}) \! = \!
\tilde{x}_1^B(\boldsymbol{x}_{-1})$; otherwise, it selects
$x^B_1(\boldsymbol{x}_{-1}) \! = \!
\hat{x}_1^B(\boldsymbol{x}_{-1})$. The best response for user $N$ is
obtained similarly. We can show the following for network coding
users.

\vspace{0.1cm}

\begin{proposition} \label{proposition2}
%

For network coding user $1$, the data rate
$\tilde{x}_1^B(\boldsymbol{x}_{-1})$ is obtained as the value of
$x_1$ that satisfies the following equation (bounded below by $x_N$)
\begin{equation} \label{best_response_proposition_nc_equation1}
U_1^\prime (x_1) - a \left( \sum_{n=2}^{N-1} x_n + x_1 \right) + a
(1 - \beta) x_N - a x_1 = 0.
\end{equation}
Furthermore, if the utility function $U_1(x_1)$ is
\emph{non-linear}, then $\hat{x}_1^B\: (\boldsymbol{x}_{-1})$ is
obtained as the value

\noindent of $x_1$ that satisfies the following equation (bounded
between 0 and $x_N$)
\begin{equation} \label{best_response_proposition_nc_equation2}
U_1^\prime (x_1) - \beta a \left( \sum_{n = 2}^{N-1} x_n + x_N
\right) = 0.
\end{equation}
When the utility function $U_1(x_1)$ is \emph{linear} (i.e.,
$U^\prime_1(x_1)$ is a \emph{constant} for all $x_1 \geq 0$), we
have $\hat{x}_1^B(\boldsymbol{x}_{-1}) \! = \! x_N$, if
$U^\prime_1(x_1)
> \beta a ( \sum_{n=2}^{N-1} x_n \! + \! x_N )$; and
$\hat{x}_1^B(\boldsymbol{x}_{-1}) \! = \! 0$, if $U^\prime_1(x_1) <
\beta a ( \sum_{n=2}^{N-1} x_n \! + \! x_N )$. In this case, if
$U^\prime_1(x_1) = \beta a ( \sum_{n=2}^{N-1} x_n + x_N )$, then
$\hat{x}_1^B(\boldsymbol{x}_{-1})$ is \emph{any} value between 0 and
$x_N$.

\end{proposition}

\vspace{0.1cm}

The proof of Proposition \ref{proposition2} is given in Appendix
\ref{app:proof_proposition2}.
The key idea is to solve the KKT optimality conditions
\cite{convex_book} for the optimization problems in
(\ref{user_1_first_problem}) and (\ref{user_1_second_problem}) for
user 1 (and $N$).
We can see that the best responses in Propositions
\ref{proposition1} and \ref{proposition2} only depend on the
\emph{first derivatives} of the utility functions. We will use this
key observation to characterize the Nash equilibrium in Section
\ref{Nash_PoA_section}.


\subsection{Nash Equilibrium and Price-of-Anarchy}
\label{Nash_PoA_section}

Given the best response functions in Section
\ref{sec:best_response},
we are now ready to characterize the Nash equilibria. 
Let $\mathcal{X}^\ast$ denote the set of \emph{all} Nash equilibria
of Game \ref{game2}. Recall that set $\mathcal{X}^\ast$ has \emph{at
least} one member as shown in Theorem \ref{theorem4}. By definition,
for any Nash equilibrium $\boldsymbol{x}^\ast \in \mathcal{X}^\ast$,
given $\boldsymbol{x}^\ast_{-n}$, the best response for user $n \in
\mathcal{N}$ is the same as its strategy at Nash equilibrium
\cite{game_theory_book}. That is, $x_n^B(\boldsymbol{x}^\ast_{-n}) =
x_n^\ast$ for all $n \in \mathcal{N}$. Thus, all Nash equilibria of
Game \ref{game2} can be obtained using the best response equations
in Propositions \ref{proposition1} and \ref{proposition2}. Recall
from Section \ref{sec:best_response} that the best responses in
(\ref{user_1_second_problem}),
(\ref{best_response_proposition_nc_equation2}),
(\ref{best_response_proposition_nc_equation2}) only depend on the
\emph{first derivatives} of the utility functions. Therefore, for
each Nash equilibrium $\boldsymbol{x}^\ast \! \in \!
\mathcal{X}^\ast$, if we define the following \emph{linear} utility
functions:
\begin{equation}
\abovedisplayskip 6pt \belowdisplayskip 6pt  \bar{U}_n(x_n) \; = \;
U^\prime_n(x^\ast_n) \; x_n, \ \ \ \ \forall \;\! n \in \mathcal{N},
\label{linear_utilities_class}
\end{equation}
then $\boldsymbol{x}^\ast$ continues to be a Nash equilibrium for a
\emph{new} game with \emph{new} utilities $\bar{U}_1(x_1), \ldots,
\! \bar{U}_N(x_N)$. In fact, $\boldsymbol{x}^\ast$ is a Nash
equilibrium for the \emph{family} of games with utility functions
$U_1(x_{\!1}), \ldots, \! U_N(x_N)$ having their first derivatives
equal to $U^\prime_1(x^\ast_1), \ldots, U^\prime_N(x^\ast_N\!)$ at
Nash
equilibrium, respectively. 

\vspace{0.1cm}

\begin{theorem} \label{theorem3}
For each Nash equilibrium $\boldsymbol{x}^\ast \in \mathcal{X}^\ast$
of Game \ref{game2} and any optimal solution $\boldsymbol{x}^S$ of
Problem \ref{problem2}, the following inequality holds:
\begin{equation} 
\abovedisplayskip 10pt \belowdisplayskip 8pt \frac{\sum_{n=1}^N
U_n(x^\ast_n) \! - \! C \! \left( \sum_{n=2}^{N-1} x_n^\ast \! + \!
\max(x^\ast_1,x^\ast_N) \right)}{\sum_{n=1}^N U_n(x^S_n) \! - \! C
\! \left( \sum_{n=2}^{N-1} x_n^S \! + \! \max(x^S_1,x^S_N) \right)}
\ \geq \frac{\sum_{n=1}^N \bar{U}_n( x^\ast_n ) \! - \! C \!\left(
\sum_{n=2}^{N-1} x_n^\ast \! + \! \max(x^\ast_1,x^\ast_N) \right) }{
\max_{\tilde{q} \geq 0} \left[ \: \sigma \: \tilde{q} - C
(\tilde{q}) \: \right] }, \label{efficiency_loss_nonlinear_linear}
\end{equation}
%
%
where
%
$ \sigma = \max\left\{ U^\prime_2(x^\ast_2), \ldots,
U^\prime_{N-1}(x^\ast_{N-1}), U^\prime_1(x^\ast_1) +
U^\prime_N(x^\ast_N) \right\}$.

\end{theorem}

\vspace{0.3cm}

%
%
%

The proof of Theorem \ref{theorem3} is given in Appendix
\ref{app:proof_theorem3}.
Notice that $\max_{\tilde{q} \geq 0} \left[ \: \sigma \: \tilde{q} -
C(\tilde{q}) \: \right]$ denotes the optimal objective value of
Problem \ref{problem2} for the special case of having \emph{linear}
utility functions (see Appendix \ref{app:proof_theorem3}).
Therefore, for the inequality in
(\ref{efficiency_loss_nonlinear_linear}), the right hand side
denotes the efficiency for \emph{linear} utility functions while the
left hand side denotes the efficiency for \emph{any} utility
functions, assuming that the rest of the system parameters (i.e.,
number of users, cost function, and price functions) are the same.
This leads us to the following helpful theorem.

\vspace{0.1cm}

\begin{theorem} \label{theorem5}
The \emph{worst-case} efficiency at a Nash equilibrium of Game
\ref{game2} with respect to the optimal solution of Problem
\ref{problem2} occurs when the utility functions are \emph{linear}
for all users. That is,
\begin{equation}
\abovedisplayskip 3pt \belowdisplayskip 3pt U_n(x_n) = \gamma_n x_n,
\ \ \ \ \ \forall \;\! n \in \mathcal{N}, \label{linear_utilities}
\end{equation}
where utility parameter $\gamma_n > 0$ for all users $n \in
\mathcal{N}$.

\end{theorem}

\vspace{0.1cm}

From Theorem \ref{theorem5}, to obtain the PoA for Game \ref{game2}
for \emph{arbitrary} choices of utility functions (as long as they
satisfy Assumption \ref{assumption1}), it is \emph{enough} to only
analyze the case when all utility functions are \emph{linear}. This
key observation can make our analysis significantly more tractable.
Notice that for the case of linear utilities, we have
$U^\prime_n(x_n) \! = \! \gamma_n$ for all $n \in \mathcal{N}$. As a
result, the best responses for all users can be obtained in
\emph{closed form} using Propositions \ref{proposition1} and
\ref{proposition2}.

Next, we obtain the exact value(s) of the Nash equilibrium(s) and
PoA for Game \ref{game2}.
%

\vspace{0.1cm}

\begin{theorem} \label{theorem6}
Suppose Assumptions \ref{assumption1}, \ref{assumption2}, and
\ref{assumption3} hold. Also assume that the utility functions are
\emph{linear}. Consider the case where $N \geq 2$ and let
$\boldsymbol{x}^\ast$ denote the Nash equilibrium for Game 2.
Without loss of generality, assume that $\gamma_1 \geq \gamma_N$.
For notational simplicity, we also define
\begin{equation}
q^\ast = \sum_{n = 2}^{N-1} x^\ast_n.
\end{equation}

\noindent (a) If $\gamma_N \leq \gamma_1 \leq \left( 1 +
\frac{1}{\beta}\right) \gamma_N - \beta a q^\ast$, then
\begin{equation}
\max\left\{ 0, \frac{\gamma_1 - a q^\ast}{a (1 + \beta)}\right\}
\leq x^\ast_1 = x^\ast_N \leq \max\left\{ 0, \frac{\gamma_N - \beta
a q^\ast}{\beta a} \right\}. \label{theorem6_part_a}
\end{equation}

\noindent (b) If $\left( 1 + \frac{1}{\beta} \right) \gamma_N -
\beta a q^\ast \leq \gamma_1 \leq \frac{2}{\beta} \gamma_N - a
q^\ast$, then
\begin{equation}
x_1^\ast = \frac{\gamma_N}{\beta a} - q^\ast, \ \ \ \ \ \ \ x^\ast_N
= \frac{\frac{2}{\beta} \gamma_N - \gamma_1}{a (1 - \beta)} -
\frac{q^\ast}{1 - \beta}. \label{theorem6_part_b}
\end{equation}

\noindent (c) If $\gamma_1 \geq \frac{2}{\beta} \gamma_N - a
q^\ast$, then
\begin{equation}
x_1^\ast = \max\left\{ 0, \frac{\gamma_1}{2 a} -
\frac{q^\ast}{2}\right\}, \ \ \ \ \ \ \ \ x_N^\ast = 0.
\label{theorem6_part_c}
\end{equation}

\noindent (d) For any choice of system parameters in (a)-(c), the
routing users have the following rates\footnote{Notice that for each
routing user $n \in \mathcal{N} \backslash \{1, N\}$, the strategy
at Nash equilibrium, i.e., data rate $x^\ast_n$, only depends on
$q^\ast$ and $x_1^\ast$, but \emph{not} $x_N^\ast$. In fact, since
we have assumed that $\gamma_N \leq \gamma_1$, we indeed have
$x_N^\ast \leq x_1^\ast$, as shown in
(\ref{theorem6_part_a})-(\ref{theorem6_part_c}). Therefore,
$\max(x_1^\ast, x_N^\ast) = x_1^\ast$ and neither the cost function
nor the price functions for the shared link $(i, j)$ depend on data
rate $x_N^\ast$. }
\begin{equation}
x_n^\ast = \left\{ \!\! \begin{array}{ll} 0, & \text{if } \gamma_n
\leq a (q^\ast + x^\ast_1), \\ \frac{\gamma_n}{a} - q^\ast -
x_1^\ast, & \text{otherwise}, \end{array} \right. \ \ \ \ \ \ \ \ \
\ \ \ n = 2, \ldots, N-1. \label{theorem6_part_d}
\end{equation}

\vspace{0.1cm}

\end{theorem}

\vspace{0.1cm}

The proof of Theorem \ref{theorem6} is given in Appendix
\ref{app:proof_theorem6}. From Theorem \ref{theorem6}(a), if the
slopes of the linear utility functions for users 1 and $N$ (i.e.,
$\gamma_1$ and $\gamma_N$) are identical or close, then at Nash
equilibrium, users 1 and $N$ choose to have the same data rates and
there are \emph{infinite} Nash equilibria.
Theorem \ref{theorem6}(b) and \ref{theorem6}(c) show that if
$\gamma_1$ and $\gamma_N$ are \emph{not} close, then users 1 and $N$
will choose \emph{different} rates at the Nash equilibrium.
Comparing this with the results in Theorem \ref{theorem2}, we shall
expect a drastic efficiency loss, especially if $\gamma_1 \geq
\frac{2}{\beta} \gamma_N \! - \! a q^\ast$ as it results in
$x^\ast_N = 0$. We also notice that the Nash equilibrium directly
depends on the value of the pricing parameter $\beta$.


To study the properties of Nash equilibria of Game 2, we consider
two different cases:

\subsubsection{Two Users Case} 
\begin{figure}[vtbp]
 \begin{center}
   \mbox{
     \subfigure{\scalebox{0.8}{\includegraphics*{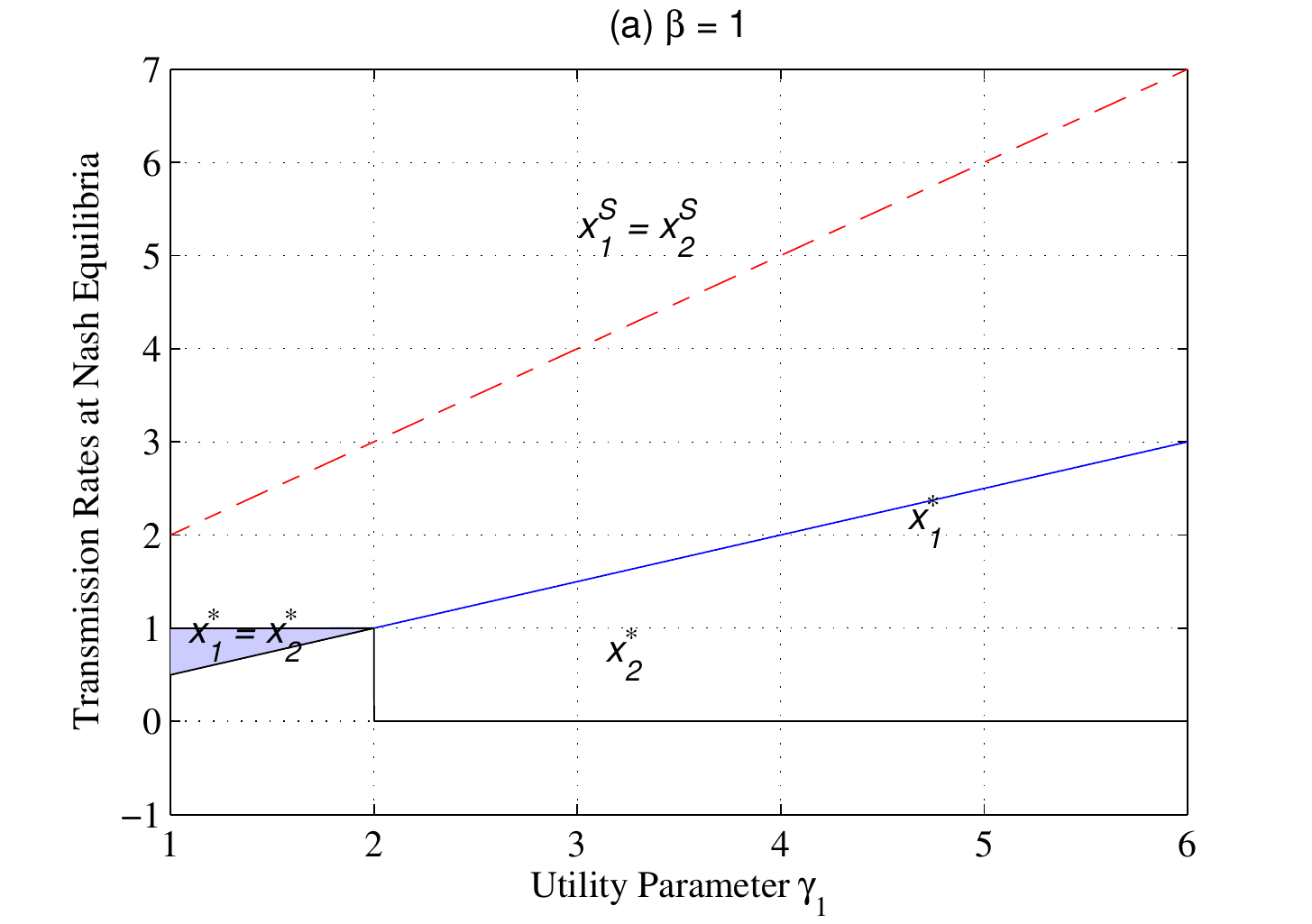}}} }
     \\
   \mbox{
     \subfigure{\scalebox{0.8}{\includegraphics*{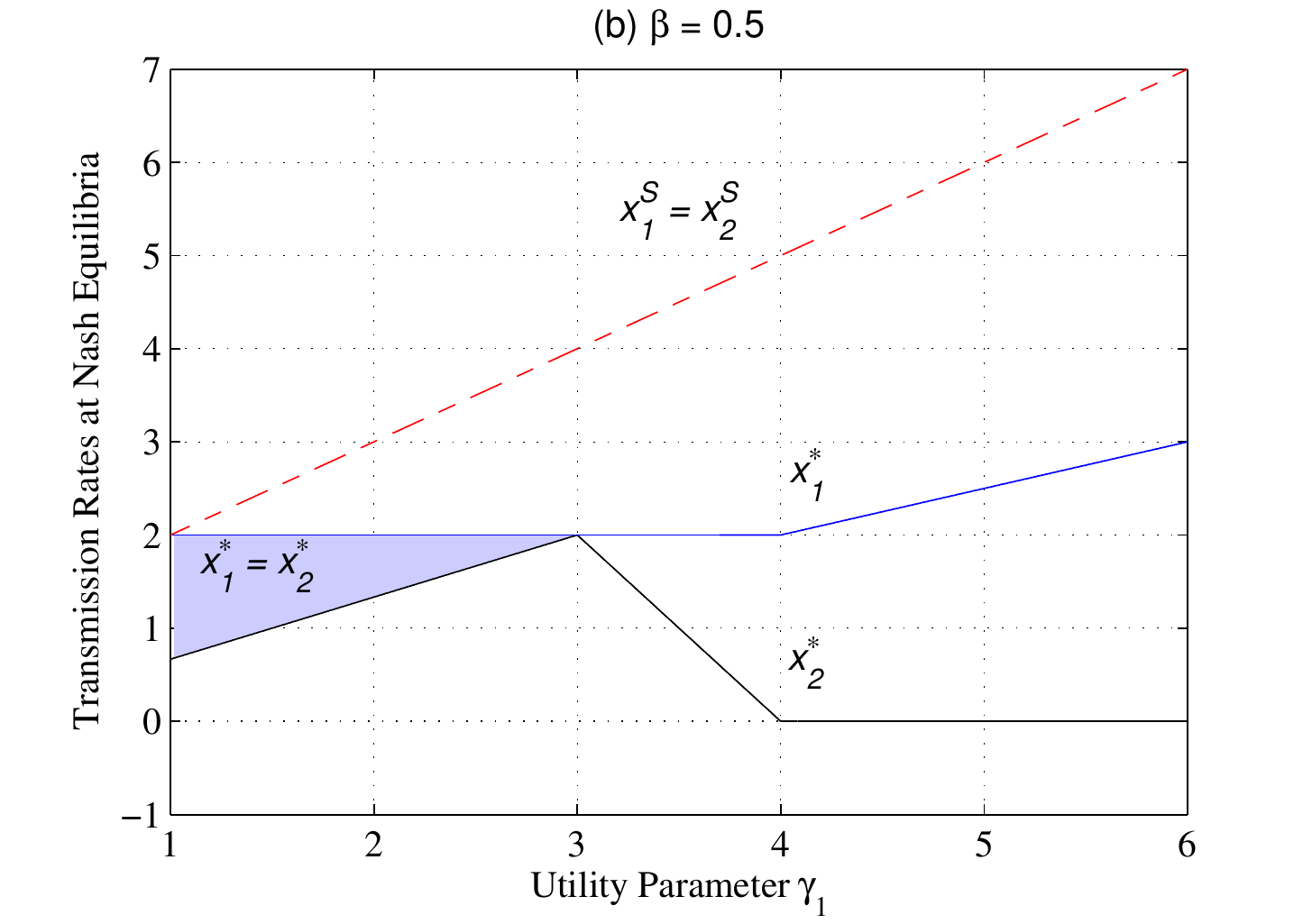}}}
     }
   \caption{Nash equilibria for the resource allocation game in Fig. 2, i.e., Game \ref{game2}, when there are $N=2$ (network coding) users: (a) Without price discrimination, i.e., $\beta = 1$, (b) With price discrimination when $\beta = \frac{1}{2}$. The rest of the parameters are as follows:
   $a = 1$, $\gamma_1 \geq \gamma_2$, and $\gamma_2 = 1$. If $\gamma_1$ and $\gamma_2$ are close (e.g., $\gamma_2 \leq \gamma_1 \leq 2 \gamma_2$ for $\beta = 1$ and $\gamma_2 \leq \gamma_1 \leq 3 \gamma_2$ for $\beta = \frac{1}{2}$), there are multiple Nash equilibria as indicated by the
   \emph{shaded} area, following the model in (\ref{theorem6_part_a}) with $q^\ast = 0$. For example, if $\beta = \frac{1}{2}$ and $\gamma_1 = \gamma_2 = 1$, then
   \emph{any} choice of data rates $x_1^\ast = x_2^\ast$ between $\frac{2}{3}$ and 2 is a Nash equilibrium. Therefore, in this case, there is an \emph{infinite} number of Nash equilibria. Recall from Theorem \ref{theorem1}(a) that in Game \ref{game1}, i.e.,
   the resource allocation with routing-only users, the Nash equilibrium is unique. This is one of the key differences between Game \ref{game1} and Game \ref{game2}. }
   \label{fig:f3}
 \end{center}
\end{figure}

Consider the butterfly network in Fig. \ref{fig:f2} and assume that $N = 2$. In this case,
the network includes two network coding users and \emph{no} routing
users. We can obtain the Nash equilibria using Theorem
\ref{theorem6} by setting $q^\ast = 0$. The Nash equilibria when
$\beta = 1$ and $\beta = \frac{1}{2}$ are shown in Fig.
\ref{fig:f3}(a) and (b), respectively. We can see that the data
rates at the Nash equilibria are always less than the optimal rates,
except when $\gamma_1 = \gamma_2$ and $\beta = \frac{1}{2}$. In
addition, in many cases (i.e., $\gamma_1 \geq 2 \gamma_2$ for $\beta
= 1$ and $\gamma_1 \geq 3 \gamma_2$ for $\beta = \frac{1}{2}$) we
have $x_2^\ast < x_1^\ast$. This leads to further deviation from the
optimal performance. Recall from Theorem \ref{theorem2} that at
optimality, the data rates of users 1 and $N$ should be
\emph{equal}.
We can show the following in the two-user case:


\vspace{0.1cm}

\begin{theorem} \label{theorem8}
In a network as in Fig. \ref{fig:f2} with $N = 2$, under the
\emph{single} pricing scheme ($\beta = 1$),
\begin{equation} \label{PoA_game_2_problem_2_N2_beta1}
\text{PoA}\left(\text{Game \ref{game2}}, \text{Problem
\ref{problem2}}\right) = \frac{1}{3},
\end{equation}
and under the \emph{discriminatory} pricing scheme with $\beta =
\frac{1}{2}$,
\begin{equation} \label{PoA_game_2_problem_2_N2_beta_0_5}
\text{PoA}\left(\text{Game \ref{game2}}, \text{Problem
\ref{problem2}}\right) = \frac{12}{25}.
\end{equation}


%
%
%

\end{theorem}

\vspace{0.2cm}

The proof of Theorem \ref{theorem8} is given in Appendix
\ref{app:proof_theorem8}.
Here, $\text{PoA}\left(\text{Game \ref{game2}}, \text{Problem
\ref{problem2}}\right)$ denotes the \emph{lowest} (i.e.,
\emph{worst-case}) ratio of the network aggregate surplus at a Nash
equilibrium of Game \ref{game2} to the network aggregate surplus at
the optimal solution of Problem \ref{problem2}.
 Theorem \ref{theorem8} extends the results on
efficiency bounds for routing flows in Theorem \ref{theorem1} to the
case where two inter-session network coding users share a link. We
can see that even for this simple scenario, the efficiency bound in
Theorem \ref{theorem1} \emph{cannot} be guaranteed anymore. From
 Theorem \ref{theorem8}, inter-session network coding with no price
discrimination can reduce the PoA from 0.67 down to $\frac{1}{3} \!
\approx \! 0.33$. On the other hand, even if we use price
discrimination by setting $\beta = \frac{1}{2}$, i.e., users $1$ and
$N$ split the price of encoded packets, the PoA improves only to
$\frac{12}{25} = 0.48$. This implies that inter-session network
coding is significantly \emph{more sensitive} to strategic users.
Thus, unlike the case of routing networks, a simple pricing scheme
(even with price discrimination) may not be sufficient to encourage
cooperation in inter-session network coding systems.

It is worth mentioning that the above results do \emph{not} imply
superiority of routing versus network coding. In fact, we can verify
that at \emph{any} Nash equilibrium of Game \ref{game2}, the network
surplus is higher than or equal to the network surplus at the Nash
equilibrium of Game \ref{game1} for the \emph{same} parameters. In
other words, non-cooperative network coding results in an
\emph{absolute} performance which is \emph{no worse} than the
\emph{absolute} performance of non-cooperative routing. However, the
\emph{relative} performance in non-cooperative network coding
compared to optimal cooperative network coding is worse than the
relative performance in the routing-only case.



\begin{figure}[t]
\begin{center}
{\scalebox{0.8}{\includegraphics*{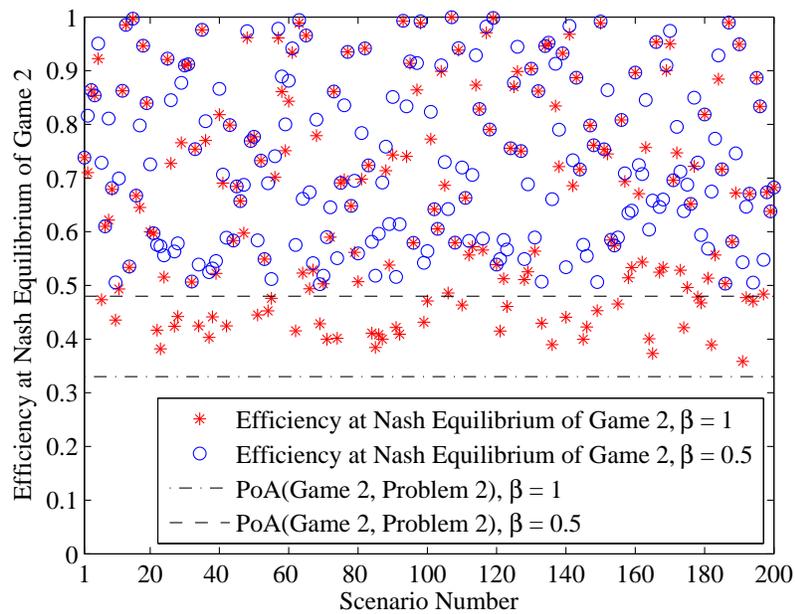}}} \vspace{-0.2cm}
\caption{Efficiency at a Nash equilibrium of 200 randomly generated
resource allocation game scenarios when the network topology is as
in Fig. \ref{fig:f2} and the number of users $N = 2$. Notice that
for each scenario, the efficiency is obtained as the ratio of the
network aggregate surplus at Nash equilibrium of Game \ref{game2}
and the network aggregate surplus at optimal solution of Problem
\ref{problem2}.
Here, we set either $\beta = 1$ or $\beta = \frac{1}{2}$, where
$\beta$ is the price discriminating parameter. For each scenario,
the pricing parameter $a \in (0, 10)$ is selected randomly. The
utility functions $U_1$ and $U_2$ are chosen to be $\alpha$-fair
(cf. \cite{mo_2000}) with randomly selected choices of utility
parameter $\alpha \in (0, 1)$. Note that the PoA is defined as the
\emph{worst-case} efficiency. We can see that if $\beta = 1$, the
PoA is equal to $\frac{1}{3} \approx 0.33$. On the other hand, if
$\beta = \frac{1}{2}$, then the PoA is equal to $\frac{12}{25} =
0.48$. These results confirm Theorem \ref{theorem8}.}
\vspace{-0.3cm} \label{fig:f5}
\end{center}
\end{figure}

Numerical results on efficiency of the Nash equilibrium of Game
\ref{game2} for 200 \emph{randomly} generated scenarios with
different choices of system parameters in the two-user case are
shown in Fig. \ref{fig:f5}. In particular, in each scenario, the
utility functions of the users are chosen to be $\alpha$-fair (cf.
\cite{mo_2000}) with a randomly selected utility parameter $\alpha
\in (0, 1)$. We can see that by using price discrimination with
parameter $\beta = \frac{1}{2}$, the guaranteed worst-case
efficiency bound (i.e., the PoA) improves from $0.33$ to $0.48$. For
the rest of this paper, we focus on the case with $\beta =
\frac{1}{2}$. That is, the network coding users \emph{split} the
charge of transmitting their jointly encoded packets.

\subsubsection{General Case} 
\begin{figure}[vtbp]
 \begin{center}
   \mbox{
     \subfigure{\scalebox{0.8}{\includegraphics*{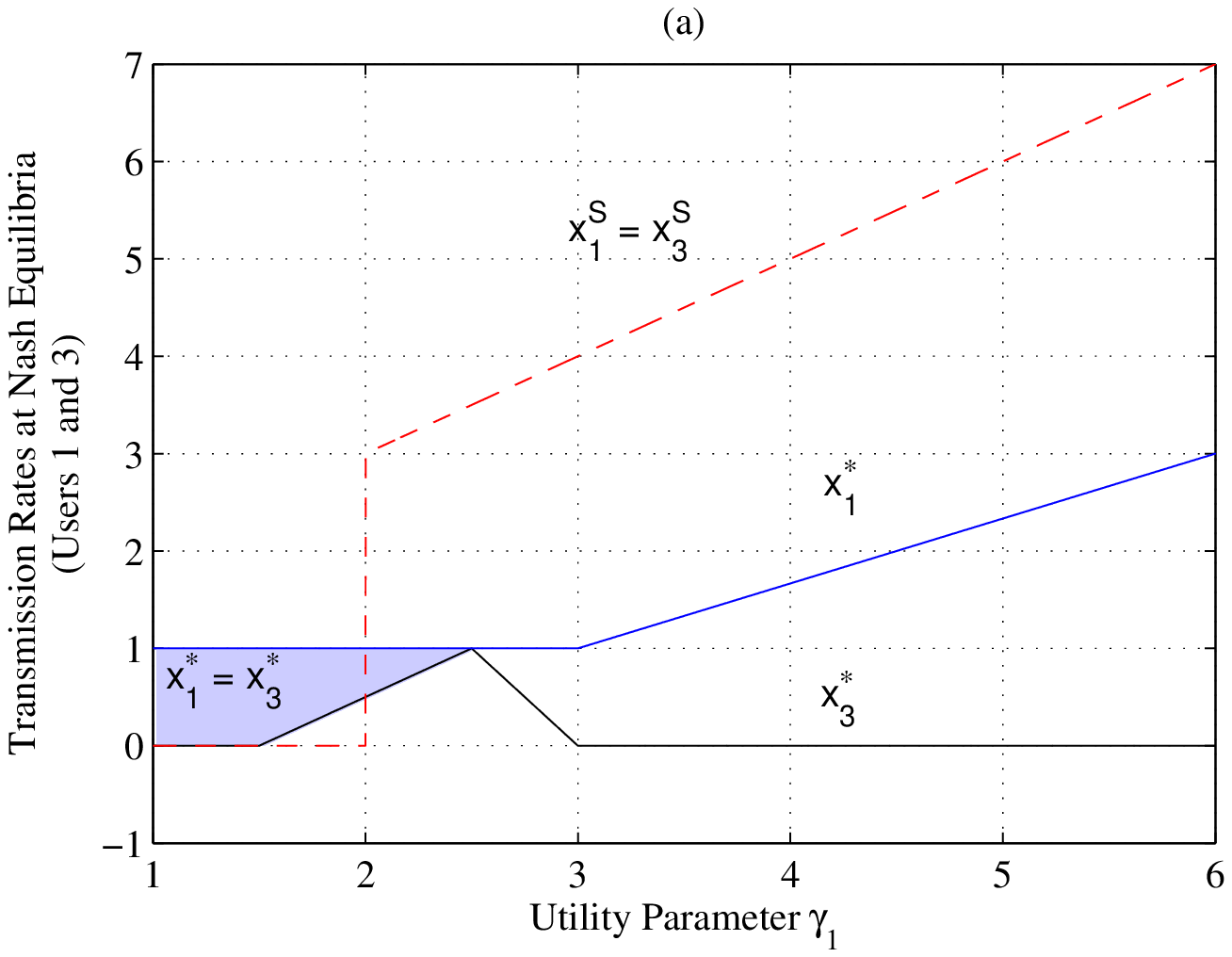}}} }
     \\
   \mbox{
     \subfigure{\scalebox{0.8}{\includegraphics*{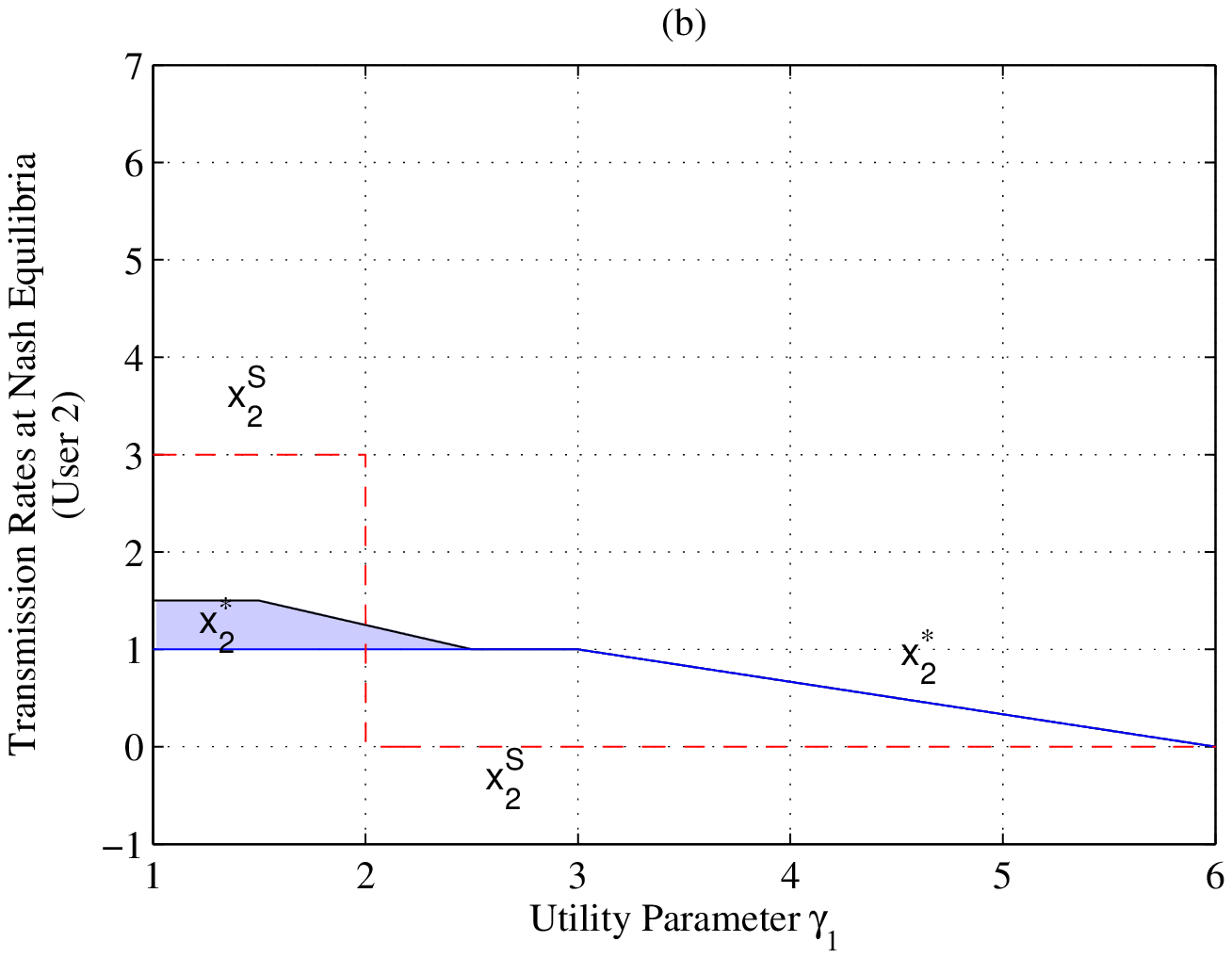}}}
     }
   \caption{Nash equilibria for the resource allocation game in Fig.
   \ref{fig:f2} when $N=3$, $a = 1$, $\beta = \frac{1}{2}$, $\gamma_1 \geq \gamma_3$, $\gamma_3
= 1$, and $\gamma_2 = 3$.
    Users 1 and 3 jointly perform
    inter-session network coding and user 2 is a routing user: (a)
    Transmission rates for users 1 and 3, (b) Transmission rates for
    user 2. If $\gamma_1 \leq 2$, then $\gamma_2 \geq \gamma_1 + \gamma_3$. In that case, at optimality of Problem \ref{problem2}, link $(i, j)$ should only
    carry the packets from routing user 2. If $\gamma_1 > 2$, then $\gamma_2 < \gamma_1 + \gamma_3$. In that case, at optimality, link $(i, j)$ should only
    carry the packets from network coding users 1 and 3.      }
   \label{fig:f6}
 \end{center}
\end{figure}

Next, consider the case where the topology is as in Fig.
\ref{fig:f2} and there are $N > 2$ users in the network. The
presence of both network coding and routing users makes the analysis
more complex. To see this, consider the network in Fig. \ref{fig:f2}
and assume that $N \! = \! 3$, $a \! = \! 1$, $\beta = \frac{1}{2}$,
$\gamma_1 \geq \gamma_3$, $\gamma_3 = 1$, and $\gamma_2 = 3$. In
this case, users 1 and 3 are the network coding users and user 2 is
a routing user. From Theorem \ref{theorem6}, the Nash equilibria are
obtained as shown in Fig. \ref{fig:f6}. Comparing the results in
Fig. \ref{fig:f6} with those in Fig. \ref{fig:f3}, we can see that
adding an extra routing user forces the network coding users 1 and 3
to \emph{reduce} their data rates at Nash equilibrium. However,
there still exist multiple (infinite) Nash equilibria when
$\gamma_1$ and $\gamma_3$ are close. It is easy to numerically
verify that in this scenario, the \emph{worst-case} efficiency at
Nash equilibrium of Game \ref{game2} is 46.5\%. Comparing this with
the results in Theorem \ref{theorem8}, we can expect that adding
routing users will further reduce the PoA. In fact, we can show the
next theorem in a \emph{general} case:

\vspace{0.3cm}


%

\begin{theorem} \label{theorem9}

Consider a network coding system as in Fig. \ref{fig:f2} and assume
that $N \geq 2$.

\noindent (a) If the price discrimination parameter $\beta =
\frac{1}{2}$, we have
\begin{equation} \label{PoA_game_2_problem_2_N_general}
\text{PoA}\left(\text{Game \ref{game2}}, \text{Problem
\ref{problem2}}\right) = \frac{1}{4}.
\end{equation}

\noindent (b) The worst-case efficiency occurs when $N \to \infty$.
%
%
%
%
\end{theorem}

\vspace{0.3cm}

The proof of Theorem \ref{theorem9} is given in Appendix
\ref{app:proof_theorem9}.
Comparing the results in Theorem \ref{theorem9} with those in
Theorems \ref{theorem1} and \ref{theorem8}, we can see that a
resource allocation game with \emph{both} network coding and routing
users has a \emph{worse} PoA compared to the \emph{routing only} and
\emph{network coding only} cases.
%

\section{Resource Allocation Game with Inter-Session Network Coding
and Routing Flows: The Case with Non-zero Costs for Side Links}
\label{sec:single_link:coding:nonzerosidelinks}

In Section \ref{sec:single_link:coding}, we considered a network
coding scenario in a butterfly network where the side links have
\emph{zero} cost as stated in Assumption \ref{assumption3}. In this
section, we study the case where the side links have \emph{non-zero}
cost. We show that the results will be noticeably different. In
particular, the network coding users are no longer interested in
participating in network coding. This can further reduce the
efficiency to as low as only 20\% of the optimal network coding
performance.

\subsection{Problem Formulation}

\begin{figure}[t]
\begin{center}
{\scalebox{0.9}{\includegraphics*{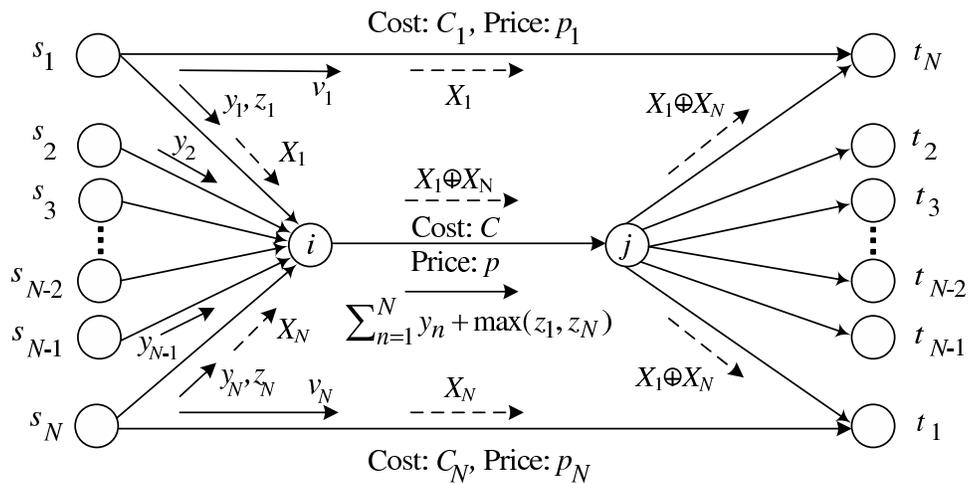}}} \caption{A
single link shared by $N$ flows. Users 1 and $N$ perform
inter-session network coding. The side links $(s_1, t_N)$ and $(s_N,
t_1)$ have non-zero cost. Thus, users 1 and $N$ need to pay for
their transmissions over links $(s_1, t_N)$ and $(s_N, t_1)$,
respectively. Here, $y_1$ and $z_1$ denote the data rate at which
source $s_1$ sends data to intermediate node $i$ \emph{marked} for
routing and network coding, respectively. Similarly, $y_N$ and $z_N$
denote the data rate at which source $s_N$ sends data to node $i$
\emph{marked} for routing and network coding, respectively. Node $i$
jointly encodes only those packets which are marked for network
coding. On the other hand, $v_1$ and $v_N$ denote the data rates at
which sources $s_1$ and $s_N$ send remedy packets over side links
$(s_1, t_N)$ and $(s_N, t_1)$, respectively. The routing users $2,
\ldots, N-1$ send packets to node $i$ all marked for routing with
rates $y_2, \ldots, y_{N-1}$, respectively. Notation $X_1 \oplus
X_N$ indicates a network coded packet/symbol obtained by jointly
encoding packets $X_1$ and $X_N$.} \vspace{-0.5cm} \label{fig:f7}
\end{center}
\end{figure}

Consider the network in Fig. \ref{fig:f7}. In this figure, the side
link $(s_1, t_N)$ has price $p_1$ and cost $C_1$ while the side link
$(s_N, t_1)$ has price $p_N$ and cost $C_N$. Suppose that Assumption
\ref{assumption2} also holds for the price and cost functions of
both side links. In addition, we make the following assumption.

\begin{assumption}[Non-Zero Cost for Side Links] \label{assumption4}
The side links $(s_{1},t_{N})$ and $(s_{N},t_{1})$ in
Fig.~\ref{fig:f7} always have \emph{non-zero} cost and impose
\emph{non-zero} prices. In particular, the side link $(s_{1},t_{N})$
has price $p_1(v_1) = a_1 v_1$ for $a_1 > 0$ and the side link
$(s_{N},t_{1})$ has price $p_N(v_N) = a_N v_N$ for $a_N > 0$.

\end{assumption}

Clearly, by sending remedy packets over side link $(s_1, t_N)$, user
1 is \emph{helping} user $N$ to decode the encoded packets it may
receive. However, due to non-zero cost at the side links, user 1
will be \emph{charged} for sending these remedy packets. A similar
statement is true for user $N$ when it sends remedy packets on the
side link $(s_N, t_1)$. Therefore, users 1 and $N$ may decide to
\emph{reduce} the rate at which they send the remedy packets on the
side links, compared to the rate at which they send their own data
packets to node $i$. In other words, they may decide to have
\emph{partial} or \emph{no} participation in network coding. Users 1
and $N$ can inform node $i$ about their decision using a simple
\emph{packet marking} scheme, e.g., by using a \emph{flag} in the
packet header. Let $y_1$ and $z_1$ denote the rate at which source
$s_1$ sends data to node $i$ \emph{marked} for routing and network
coding, respectively. Similarly, let $y_N$ and $z_N$ denote the data
rate at which source $s_N$ sends data to node $i$ \emph{marked} for
routing and network coding, respectively. Node $i$ may jointly
encode only those packets which are marked for network coding. If
the packet is marked for routing, then node $i$ simply forwards the
packet without modifying its content. Furthermore, let $v_1$ and
$v_N$ denote the data rates at which sources $s_1$ and $s_N$ send
remedy packets on side links $(s_1, t_N)$ and $(s_N, t_1)$,
respectively. The routing users $2, \ldots, N\!-\!1$ just send
routing packets at rates $y_2, \ldots, y_{N-1}$, respectively.

Given the above data rates, intermediate node $i$ encodes packets at
rate $\min(z_1, z_N)$ and \emph{forwards} the rest of packets at
rate $\sum_{n = 1}^N y_n + | z_1 - z_N|$. As a result, the total
rate on link $(i, j)$ becomes $\sum_{n = 1}^N y_n + \max(z_1, z_N)$.
Let $\boldsymbol{y} = (y_1, \ldots, y_N)$, $\boldsymbol{z} = (z_1,
z_N)$, and $\boldsymbol{v} = (v_1, v_N)$. For the butterfly network
in Fig. \ref{fig:f7}, the network aggregate surplus maximization
problem becomes

\begin{problem} [Surplus Maximization with Network Coding and Non-Zero-Cost Side Links]\label{problem3}
\begin{equation}
\nonumber
\begin{aligned}
& \underset{\boldsymbol{y}, \boldsymbol{z}, \boldsymbol{v}}{\text{maximize}} & \! \sum_{n=2}^{N-1} U_n\left( y_n \right)
+ U_1\left( y_1 + \min(z_1, v_N) \right) + U_N\left( y_N + \min(z_N,
v_1) \right) \\
& & - \: C \! \left(\sum_{n=1}^{N} y_n \! + \! \max( z_1,
z_N) \! \right) \! - C_1(v_1) - C_N(v_N) \ \ \ \ \ \ \ \ \ \ \ \ \ \ \ \ \label{nc_problem2_line1} \\
& \text{subject to} & \! y_n \geq 0, \ \ n = 1, \ldots, N, \ \ \ \ \
z_1, z_N, v_1, v_N \geq 0. \ \ \ \ \ \ \ \ \ \ \ \ \ \ \ \ \ \ \;
\end{aligned}
\end{equation}
\end{problem}

\vspace{0.5cm}

\noindent We can see that the objective function in Problem
\ref{problem3} is more complex than the one in Problem
\ref{problem2} and includes the cost of side links $(s_1, t_N)$ and
$(s_N, t_1)$.

Following a discriminatory pricing model as in Section
\ref{sec:single_link:coding:problem_formulation}, we can define a
resource allocation game for the network setting in Fig.
\ref{fig:f7}, when users are price anticipators:

\vspace{0.5cm}

\begin{game} \label{game3}
\emph{(Resource Allocation Game with Inter-session Network Coding
and Routing Flows and Non-zero Costs for Side Links)}

\begin{itemize}
\item \emph{Players}: Users in set $\mathcal{N}$.

\item \emph{Strategies}: Transmission rates $\boldsymbol{y}$, $\boldsymbol{z}$, and $\boldsymbol{v}$.

\item \emph{Payoffs}: $W_n(\cdot)$ for each user $n \in
\mathcal{N}$, where
%
\begin{equation}
\begin{split}
W_1(y_1, z_1, v_1 ; \boldsymbol{y}_{-1}, z_N, v_N) = & \:
U_1\left(y_1 + \min(z_1, v_N)\right) - v_1
p_1(v_1) \\
&
\!\!\!\!\!\!\!\!\!\!\!\!\!\!\!\!\!\!\!\!\!\!\!\!\!\!\!\!\!\!\!\!\!\!
\textstyle - \left( y_1 + z_1 - (1-\beta) \min(z_1, z_N) \right)
p\left(\sum_{r=1}^N y_r + \max(z_1, z_N) \right),
\end{split}
\end{equation}
\begin{equation}
\begin{split}
W_N(y_N, z_N, v_N ; \boldsymbol{y}_{-N}, z_1, v_1) = & \:
U_N\left(y_N + \min(z_N, v_1)\right) -
v_N p_N(v_N) \\
&
\!\!\!\!\!\!\!\!\!\!\!\!\!\!\!\!\!\!\!\!\!\!\!\!\!\!\!\!\!\!\!\!\!\!\!\!\!\!\!\!\!\!\!\!
\textstyle - \left( y_N + z_N - (1-\beta) \min(z_1, z_N) \right)
p\left(\sum_{r=1}^N y_r + \max(z_1, z_N) \right),
\end{split}
\end{equation}
%
%
\begin{equation}
W_n(y_n ; \boldsymbol{y}_{-n}) = U_n(y_n) - y_n p \left( \textstyle
\sum_{r=1}^{N} y_r + \max( z_1, z_N ) \right), \ \ \ \ \ \ n \in
\mathcal{N} \backslash \{1, N\}.
\end{equation}
\end{itemize}

\vspace{0.1cm}

\end{game}

Here, for each user $n \in \mathcal{N}$, we have
$\boldsymbol{y}_{-n} = (y_1, \ldots, y_{n-1}, y_{n+1}, \ldots,
y_N)$. Next, we study the efficiency and the worst-case efficiency
(i.e., the PoA) at Nash equilibria of Game \ref{game3}.


\vspace{-0.1cm}

\subsection{Users' Best Responses}

For network coding user 1, the best response is in form of $ \;
\left( \:\! y_1^B(\boldsymbol{y}_{-1}, z_N, v_N), \;
z_1^B(\boldsymbol{y}_{-1}, z_N, v_N), \right.$ $
\left.v_1^B(\boldsymbol{y}_{-1}, z_N, v_N) \right)$ which is
obtained as the solution of the following optimization problem
\begin{equation} \nonumber
\left( y_1^B(\boldsymbol{y}_{-1}, \! z_N, \! v_N),
z_1^B(\boldsymbol{y}_{-1}, \! z_N, \! v_N),
v_1^B(\boldsymbol{y}_{-1}, \! z_N, \! v_N) \right) = \!\!
\argmax_{y_1 \geq 0, \: z_1 \geq 0, \: v_1 \geq 0} W_1(y_1, z_1, v_1
; \boldsymbol{y}_{-1}, z_N, v_N).
\end{equation}

\noindent The best response for network coding user $N$, denoted by
$ \; \left( \:\! y_N^B(\boldsymbol{y}_{-N}, z_1, v_1), \:
z_N^B(\boldsymbol{y}_{-N}, z_1, v_1), \right. $ $
\left.v_N^B(\boldsymbol{y}_{-N}, z_1, v_1) \right)$ is obtained
similarly. We can show the following for users 1 and $N$.

\vspace{0.1cm}

\begin{proposition} \label{proposition3}

Users 1 and $N$ \emph{always} send zero remedy packets at the best
responses of Game \ref{game3}. That is,
we always have
%
$v_1^B(\boldsymbol{y}_{-1}, \! z_N, \! v_N) = 0$
%
and
%
$v_N^B(\boldsymbol{y}_{-N}, \! z_1, \! v_1) = 0$.

\end{proposition}

\vspace{0.1cm}

Proposition \ref{proposition3} can be proved by noticing that the
payoff $W_1(y_1, z_1, v_1 ; \boldsymbol{y}_{-1}, z_N, v_N)$ is
\emph{decreasing} in $v_1$ and $W_N(y_N, z_N, v_N ;
\boldsymbol{y}_{-N}, z_1, v_1)$ is \emph{decreasing} in $v_N$.
%
%
Clearly, if the network coding users do \emph{not} receive the
remedy data from the side links, they \emph{cannot} decode any
encoded packet they may receive through the shared link $(i,j)$. In
fact, we can further show the following.

\vspace{0.1cm}

\begin{proposition} \label{proposition4}

Users 1 and $N$ \emph{always} send zero network coding packets to
intermediate node $i$ as the best responses of Game \ref{game3}.
That is, 
%
%
$z_1^B(\boldsymbol{y}_{-1}, \! z_N, \! v_N) = 0$
and
%
%
$z_N^B(\boldsymbol{y}_{-N}, \! z_1, \! v_1) = 0$.

\end{proposition}

\vspace{0.1cm}

Notice that if $v_N = 0$, then $\min(z_1, v_N) = 0$ and the payoff
function for user $1$ reduces to $U_1\left(y_1\right) - v_1 p_1(v_1)
- \left( y_1 + z_1 - (1-\beta) \min(z_1, z_N) \right) p
(\sum_{r=1}^N y_r + \max(z_1, z_N) )$. In that case, the payoff
function is \emph{decreasing} in $z_1$. A similar statement is true
for network coding user $N$.

\vspace{-0.1cm}

\subsection{Nash Equilibrium and Price-of-Anarchy}

Given the results on users' best responses in Propositions
\ref{proposition3} and \ref{proposition4}, we can conclude that at
any Nash equilibrium of Game \ref{game3}, denoted by
$(\boldsymbol{y}^\ast, \boldsymbol{z}^\ast, \boldsymbol{v}^\ast)$,
we should indeed have
\begin{equation}
\abovedisplayskip 3pt \belowdisplayskip 3pt z_1^\ast = z_N^\ast =
v_1^\ast = v_N^\ast = 0. \label{NE_non_zero}
\end{equation}
In other words, at a Nash equilibrium of Game \ref{game3},
\emph{none} of the users perform network coding. In that case, the
Nash equilibria of Game \ref{game3} would be closely related to the
Nash equilibria of Game \ref{game1}. In fact, for any choice of
system parameters, if $\boldsymbol{x}^\ast$ is a Nash equilibrium of
Game \ref{game1}, then $\boldsymbol{y}^\ast = \boldsymbol{x}^\ast$,
$\boldsymbol{z}^\ast = \boldsymbol{0}$, and $\boldsymbol{v}^\ast =
\boldsymbol{0}$ would be a Nash equilibrium of Game \ref{game3} for
the same choice of system parameters. From this, together with the
results in Theorem \ref{theorem1}(a), we can conclude that Game
\ref{game3} always has a \emph{unique} Nash equilibrium. This leads
to the following theorem.

\vspace{0.2cm}

\begin{theorem} \label{theorem10}
The \emph{worst-case} efficiency of Game \ref{game3} occurs when the
utility functions are \emph{linear}.

\end{theorem}

\vspace{0.2cm}

The proof of Theorem \ref{theorem10} is similar to that of
\cite[Lemma 4]{johari_jsac2006}. From Theorem \ref{theorem10}, to
obtain the PoA for Game \ref{game3} for \emph{arbitrary} choices of
utility functions (as long as the utility functions satisfy
Assumption \ref{assumption1}), it is \emph{enough} to only analyze
the case where all utility functions are \emph{linear}. Furthermore,
we notice that if the side links have a very large cost compared to
the cost of the bottleneck link, the optimal performance is achieved
if network coding is \emph{not} applied. In that case, the
efficiency can be obtained by using Theorem \ref{theorem1}. Notice
that in this case, the \emph{optimal} network aggregate surplus for
Problem \ref{problem3} is the same as the \emph{optimal} network
aggregate surplus for Problem \ref{problem1}. In addition, the
network aggregate surplus is the same at the Nash equilibrium of
Game \ref{game3} and Game \ref{game1}. However, for general choices
of $a_1 > 0$ and $a_N
> 0$, obtaining the PoA requires further investigation of the
optimal solution of Problem \ref{problem3}.

\vspace{0.2cm}

\begin{theorem} \label{theorem11}
Let $\boldsymbol{y}^S = (y_1^S, \ldots, y_N^S)$, $\boldsymbol{z}^S =
(z_1^S, z_N^S)$, and $\boldsymbol{v}^S = (v_1^S, v_N^S)$ denote the
optimal solution for Problem \ref{problem3}. Assume that all utility
functions are linear with slope $\gamma_n$ for each user $n \in
\mathcal{N}$. Define $\gamma_{\max} = \max_{n \in \mathcal{N}}
\gamma_n$ and $\mathcal{M} = \{n: \gamma_n = \gamma_{\max}\}$ with
size $M = |\mathcal{M}|$.

\noindent (a) If $\gamma_1 + \gamma_N \geq \left( 1 + \frac{a_1 +
a_N}{a} \right) \gamma_{\max}$, then
\begin{equation}
z_1^S = z_N^S = v_1^S = v_N^S = \frac{\gamma_1 + \gamma_N}{a + a_1 +
a_N}, \ \ \ \ \ \ \ \ \ \ \ \ \ \ \ y_n^S = 0, \ \ \ \forall \;\! n
\in \mathcal{N}. \label{theorem11_a}
\end{equation}
\noindent (b) If $\gamma_{\max} \leq \gamma_1 + \gamma_N \leq \left(
1 + \frac{a_1 + a_N}{a} \right) \gamma_{\max}$, then
\begin{equation}
z_1^S \! = \! z_N^S \! = \! v_1^S \! = \! v_N^S \! = \!
\frac{\gamma_1 + \gamma_N - \gamma_{\max}}{a_1 + a_N}, \ \ \ \ y_n^S
\! = \! \left\{ \!\!
\begin{array}{ll} \frac{(a+a_1+a_N) \gamma_{\max} - a(\gamma_1 +
\gamma_N)}{a M (a_1+a_N)}, & \text{if } n \! \in \! \mathcal{M}, \\
\ 0, & \text{if } n \! \in \! \mathcal{N}\backslash \mathcal{M}.
\end{array} \right.
\label{theorem11_b}
\end{equation}
\noindent (c) If $\gamma_{\max} \geq \gamma_1 + \gamma_N$, then
\begin{equation}
z_1^S = z_N^S = v_1^S = v_N^S = 0, \ \ \ \ \ \ y_n^S = \left\{
\!\!\!
\begin{array}{ll} \frac{\gamma_{\max}}{a M}, & \text{if } n \! \in \! \mathcal{M}, \\
\ 0, & \text{if } n \! \in \! \mathcal{N} \backslash \mathcal{M}.
\end{array} \right.
\label{theorem11_c}
\end{equation}
\vspace{0.001cm}
\end{theorem}

Recall that the \emph{linear} pricing parameters $a$, $a_1$, and
$a_N$ are defined in Assumptions \ref{assumption2} and
\ref{assumption4}, respectively. The proof of Theorem
\ref{theorem11} is given in Appendix \ref{app:proof_theorem11}.
Combining (\ref{NE_non_zero})-(\ref{theorem11_c}) with the results
in \cite[Section II]{johari_jsac2006}, we can show the following
results on the efficiency loss in Game \ref{game3}.

\vspace{0.2cm}

\begin{theorem} \label{theorem12}

Consider a network coding system as in Fig. \ref{fig:f7} and assume
that $N \geq 2$.

\noindent (a) We have
\begin{equation} \label{PoA_game_3_problem_3}
\abovedisplayskip 6pt \belowdisplayskip 6pt
\text{PoA}\left(\text{Game \ref{game3}}, \text{Problem
\ref{problem3}}\right) = \frac{1}{5}.
\end{equation}

\noindent (b) The worst-case efficiency occurs when $N \to \infty$.

%
%
%
%
%
\end{theorem}
\vspace{0.3cm}

\begin{figure}[t]
\begin{center}
{\scalebox{0.8}{\includegraphics*{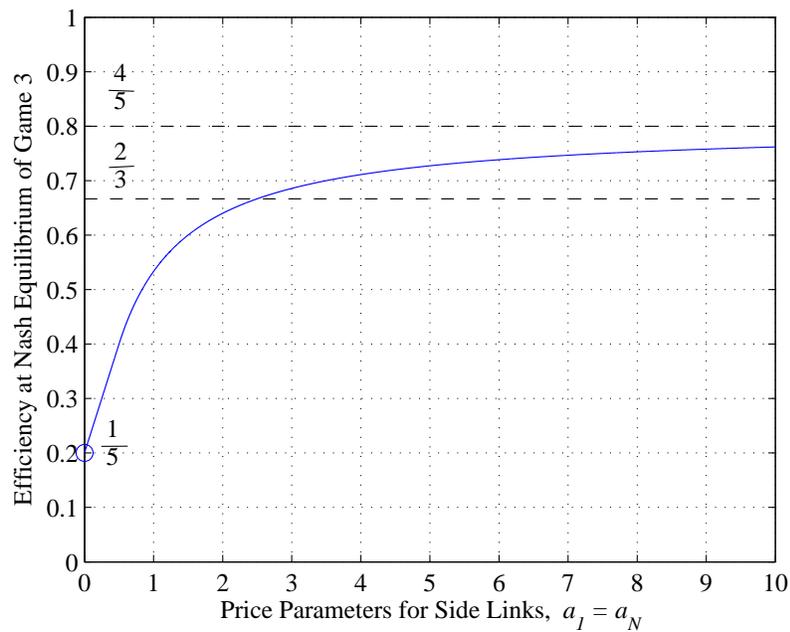}}} \vspace{-0.2cm}
\caption{Efficiency at Nash equilibrium of Game \ref{game3} when the
network topology is as in Fig. \ref{fig:f7} and we have: $N
\rightarrow \infty$, $\gamma_1 = \gamma_N = 1$, $\gamma_n =
\frac{4}{5}$ for all $n \in \mathcal{N} \backslash \{1, N\}$, and $a
= 1$. Side link price parameters $a_1 = a_N$ vary from 0
(\emph{non-inclusive}) to 10. Notice that for each choice of
parameters $a_1 = a_N$, the efficiency is obtained as the ratio of
the network surplus at Nash equilibrium of Game \ref{game3} and the
network surplus at optimal solution of Problem \ref{problem3}.
We can see that the worst-case efficiency in this case is equal to
$\frac{1}{5}$, as expected from Theorem \ref{theorem12}. As the
prices on the side links increase, network coding becomes less
beneficial and the efficiency increases to a value (in this case
$\frac{4}{5} = 0.8$) which is higher than $\frac{2}{3}$, as expected
from Theorem \ref{theorem1}.} \vspace{-0.3cm} \label{fig:f8}
\end{center}
\end{figure}

\begin{figure}[t]
\begin{center}
{\scalebox{0.8}{\includegraphics*{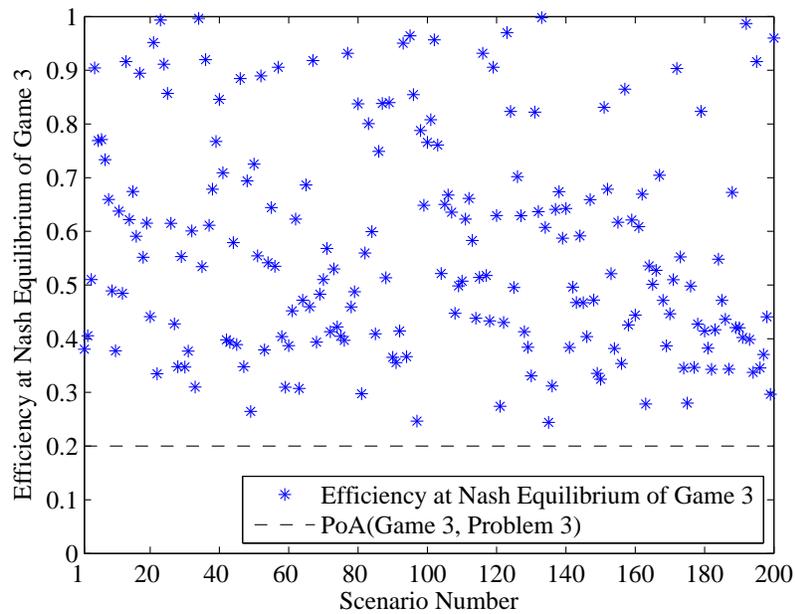}}} \vspace{-0.2cm}
\caption{Efficiency at Nash equilibrium of 200 randomly generated
resource allocation game scenarios when the network topology is as
in Fig. \ref{fig:f7} and where $N = 2$. For each scenario, the
pricing parameters $a \in (0, 10)$, $a_1 \in (0, 5)$, and $a_2 \in
(0, 5)$ are selected randomly. The utility functions $U_1$ and $U_2$
are chosen to be $\alpha$-fair (cf. \cite{mo_2000}) with randomly
selected choices of utility parameter $\alpha \in (0, 1)$. Note that
the PoA is defined as the \emph{worst-case} efficiency.
We can see that the efficiency of Game \ref{game3} is always lower
bounded by $\frac{1}{5} = 0.2$, i.e., the PoA is $\frac{1}{5}$, as
expected from Theorem \ref{theorem12}. } \vspace{-0.3cm}
\label{fig:f9}
\end{center}
\end{figure}

The proof of Theorem \ref{theorem12} is given in Appendix
\ref{app:proof_theorem12}.
%
Here, $\text{PoA}\left(\text{Game \ref{game3}}, \text{Problem
\ref{problem3}}\right)$ denotes the \emph{lowest} (i.e.,
\emph{worst-case}) ratio of the network surplus at a Nash
equilibrium of Game \ref{game3} to the network surplus at the
optimal solution of Problem \ref{problem3}. Comparing Theorem
\ref{theorem12} and Theorem \ref{theorem9}, we can see that a
non-zero cost at the side links can further reduce the PoA in a
network resource allocation game as the users do \emph{not} perform
network coding in this case. If the side link price parameters $a_1$
and $a_N$ are significantly greater than the bottleneck link price
parameter $a$, then network coding is not an optimal solution and
the efficiency loss follows from the results in Theorem
\ref{theorem1}. This is shown in Fig. \ref{fig:f8}. For the results
in this figure, the network topology is assumed to be as in Fig.
\ref{fig:f7}, where $N \rightarrow \infty$, $\gamma_1 = \gamma_N =
1$, $a = 1$, and $\gamma_n = \frac{4}{5}$ for all $n \in \mathcal{N}
\backslash \{1, N\}$. The side link price parameters $a_1 = a_N$
vary from 0 (\emph{non-inclusive}) to 10. If $a_1>0$ and $a_N>0$
tend to zero, the efficiency becomes as low as $\frac{1}{5} = 0.2$
as Theorem \ref{theorem12} suggests. As $a_1 = a_N$ increases and
tends to infinity, Problem \ref{problem3} becomes equivalent to
Problem \ref{problem1} (in terms of the optimal network aggregate
surplus)
and Game \ref{game3} becomes equivalent to Game \ref{game1} (in
terms of network aggregate surplus at Nash equilibrium) which leads
to an efficiency
higher than $\frac{2}{3} \approx 0.67$ as Theorem \ref{theorem1}
suggests (for the choice of parameters in Fig. \ref{fig:f8}, the
efficiency approaches $\frac{4}{5} = 0.8$). Numerical results on the
efficiency of the Nash equilibrium of Game \ref{game3} for 200
random scenarios with different choices of system parameters in the
\emph{two-user} case are also shown in Fig. \ref{fig:f9}. We can see
that the simulations confirm Theorem \ref{theorem12}.




\section{Conclusion} \label{conclusion}

This paper represents a first step towards understanding the impact
of strategic network coding users on the network resource allocation
efficiency. To gain insights, we focus on the case of the well-known
butterfly network topology where there is a single bottleneck link
in the network shared by several users. Two of the users have the
capability of performing inter-session network coding, and the rest
are routing only users. Even with this simple setup, the results are
dramatically different from the traditional routing-only case. In
particular, there can be many (even infinite) Nash equilibria in the
resulting resource allocation game. This is in sharp contrast to a
similar game setting with traditional packet forwarding where the
Nash equilibrium is always unique. Furthermore, we showed that the
efficiency loss can be significantly more severe than for the case
without network coding. The precise values of the PoA and the
efficiency loss depend on the pricing scheme used by the links.
Compared with the traditional single pricing approach, a novel
discriminatory pricing, which charges encoded and forwarded packets
differently can improve efficiency. However, regardless of the
discriminatory pricing scheme being used, the PoA is still worse
than for the case when network coding is not applied. This implies
that, although inter-session network coding can improve performance
compared to routing, it is significantly \emph{more sensitive} to
users' strategic behaviors. For example, in a butterfly network when
the side links have \emph{zero} cost, the efficiency at certain Nash
equilibria can be as low as 25\%. If the side links have
\emph{non-zero} cost, then the efficiency at some Nash equilibria
further reduces to only 20\%. These results generalize the
well-known result of guaranteed 67\% worst-case efficiency, shown by
Johari and Tsitsiklis in \cite{johari_jsac2006} for traditional
packet forwarding networks. This motivates our ongoing work of
mechanism design to encourage the strategic users to perform network
coding, e.g., by using a combination of reward and punishment in a
dynamic game setting.

 \appendix
 \subsection{Proof of Theorem \ref{theorem2}} \label{app:proof_theorem2}

 Let $\tilde{\boldsymbol{x}} = (\tilde{x}_1, \ldots, \tilde{x}_N)$
 denote \emph{any} arbitrary feasible solution for Problem
 \ref{problem2} such that $\tilde{x}_1 \neq \tilde{x}_N$. Without
 loss of generality, we assume that $\tilde{x}_1 > \tilde{x}_N$. We
 then define $\hat{\boldsymbol{x}} = (\hat{x}_1, \ldots, \hat{x}_N)$
 as \emph{another} feasible solution such that $\hat{x}_n = \tilde{x}_n$ for
 all $n \in \mathcal{N} \backslash \{1, N\}$ and $\hat{x}_1 =
 \hat{x}_N = \max(\tilde{x}_1, \tilde{x}_N) = \tilde{x}_1$. Since for
 all users, the utility functions are strictly increasing in
 transmission rates, we have
 \begin{equation}
\sum_{n = 1}^N U_n(\hat{x}_n) = \sum_{n = 1}^{N-1} U_n(\tilde{x}_n)
+ U_N(\hat{x}_N) > \sum_{n = 1}^{N-1} U_n(\tilde{x}_n) +
U_N(\tilde{x}_N) = \sum_{n = 1}^N U_n(\tilde{x}_n).
 \label{utility_compare}
 \end{equation}
%
 %
 On the other hand, since $\max(\hat{x}_1, \hat{x}_N) =
 \max(\tilde{x}_1, \tilde{x}_N)$,
 \begin{equation}
 C \left( \sum_{n=2}^{N-1} \hat{x}_n + \max( \hat{x}_1, \hat{x}_N)
 \right) = C \left( \sum_{n=2}^{N-1} \tilde{x}_n + \max( \tilde{x}_1, \tilde{x}_N)
 \right).
 \label{cost_compare}
 \end{equation}
 From (\ref{utility_compare}) and (\ref{cost_compare}), the feasible
 solution $\hat{\boldsymbol{x}}$ results in strictly higher
 network aggregate surplus compared to $\tilde{\boldsymbol{x}}$. Thus,
 vector $\tilde{\boldsymbol{x}}$ \emph{cannot} be an optimal solution for Problem
 \ref{problem2}. $\hfill \blacksquare$

\vspace{-0.2cm}

 \subsection{Proof of Theorem \ref{theorem4}} \label{app:proof_theorem4}

We prove the existence of Nash equilibrium for Game \ref{game2} by
using Rosen's existence theorem for $N$-person games \cite[Theorem
1]{rosen}. In this regard, we need to show that for each user $n \in
\mathcal{N}$, the payoff function $Q_n$ is \emph{continuous} and
\emph{concave} in data rate $x_n$. This is \emph{not} trivial,
particularly for network coding users due to the complexity of the
payoff functions.

For user $1$, given $\bar{\boldsymbol{x}}_{-1} = (\bar{x}_2, \ldots,
\bar{x}_N)$, we can
 rewrite the payoff function $Q_1$ as
 \begin{equation}
 Q_1(x_1; \bar{\boldsymbol{x}}_{-1}) = \left\{ \!\! \begin{array}{ll} G_1(x_1; \bar{\boldsymbol{x}}_{-1}), & \ \ \text{if } x_1 \leq \bar{x}_N, \\
 H_1(x_1; \bar{\boldsymbol{x}}_{-1}), & \ \ \text{if } x_1 > \bar{x}_N, \end{array} \right.
 \end{equation}
 where
 \begin{equation}
 \nonumber
 \textstyle G_1(x_1; \bar{\boldsymbol{x}}_{-1}) = U_1(x_1) - \beta x_1 \ p \left( \sum_{n = 2}^{N-1} \bar{x}_n + \bar{x}_N \right)\!,
 \end{equation}
 and
 \begin{equation}
 \nonumber
 \textstyle H_1(x_1; \bar{\boldsymbol{x}}_{-1}) \! = \! U_1(x_1) - \left(
 x_1 \! - \! (1 \! - \! \beta) \bar{x}_N \right) p \left( \sum_{n =
 2}^{N-1} \bar{x}_n \! + \! x_1 \right)\!.
 \end{equation}
 Clearly, $G_1(\bar{x}_N; \bar{\boldsymbol{x}}_{-1}) =
 H_1(\bar{x}_N;
 \bar{\boldsymbol{x}}_{-1})$. That is, $G_1(x_1;
 \bar{\boldsymbol{x}}_{-1})$ is continuous at $x_1 = \bar{x}_N$.
 Thus, since $G_1(\bar{x}_N, \bar{\boldsymbol{x}}_{-1})$ and
 $H_1(\bar{x}_N, \bar{\boldsymbol{x}}_{-1})$ are also continuous
 functions, function $Q_1(x_1, \bar{\boldsymbol{x}}_{-1})$ is
 continuous in $x_1$. However, $Q_1(x_1;
 \bar{\boldsymbol{x}}_{-1})$ may \emph{not} always be differentiable. Next,
 we show that $Q_1(x_1; \bar{\boldsymbol{x}}_{-1})$ is
 \emph{concave}. We need to show that for each $\tilde{x}_1 \geq 0$
 and $\hat{x}_1 \geq 0$ and for any $0 \leq \theta \leq 1$, we have
 \begin{equation}
  Q_1( \theta \hat{x}_1 + (1-\theta) \tilde{x}_1; \bar{\boldsymbol{x}}_{-1} ) \geq \theta Q_1( \hat{x}_1; \bar{\boldsymbol{x}}_{-1} ) + (1 - \theta) Q_1(\tilde{x}_1; \bar{\boldsymbol{x}}_{-1}).
 \label{concavity}
 \end{equation}
 Without loss of generality, we assume that $\tilde{x}_1 \! \leq \! \hat{x}_1$. We notice that if $\tilde{x}_1 \! \leq \! \hat{x}_1 \! \leq \! \bar{x}_N$ or $\bar{x}_N \leq \tilde{x}_1 \leq \hat{x}_1$,
then (\ref{concavity}) is directly obtained from the fact that $G_1$
and $H_1$ are
 concave. Thus, we only consider the case where $\tilde{x}_1 \leq
 \bar{x}_N \leq \hat{x}_1$. We consider two different scenarios for
 choice of $\theta$:
 \begin{equation}
 0 \leq \theta \leq \frac{\bar{x}_N - \tilde{x}_1}{\hat{x}_1 - \tilde{x}_1},
 \label{case_1_theta}
 \end{equation}
 and
 \begin{equation}
 \frac{\bar{x}_N - \tilde{x}_1}{\hat{x}_1 - \tilde{x}_1} \leq \theta \leq 1.
 \label{case_2_theta}
 \end{equation}
If (\ref{case_1_theta}) holds, then $\theta \hat{x}_1 + (1-\theta)
\tilde{x}_1 \leq \bar{x}_N$. In
 addition, since $\hat{x}_1 \geq \bar{x}_N$, we have
 \begin{equation}
 (1-\beta) \bar{x}_N \leq (1-\beta) \hat{x}_1 \ \ \ \ \ \ \Rightarrow \ \ \ \ \ \ \beta \hat{x}_1 \leq \hat{x}_1 - (1-\beta) \bar{x}_N,
 \label{line1}
 \end{equation}
 and
 \begin{equation}
p \left( \sum_{n = 2}^{N-1} \bar{x}_n + \hat{x}_1 \right) \ \geq \ p
\left( \sum_{n = 2}^{N-1} \bar{x}_n + \bar{x}_N
 \right).
 \label{line2}
 \end{equation}
 From (\ref{line1}) and (\ref{line2}), we have
 \begin{equation}
  G_1(\hat{x}_1) \geq H_1(\hat{x}_1).
 \label{g_h_compare_1}
 \end{equation}
Therefore,
 \begin{equation}
 \begin{split}
 Q_1( \theta \hat{x}_1 + (1-\theta) \tilde{x}_1; \bar{\boldsymbol{x}}_{-1} ) & = G_1( \theta \hat{x}_1 + (1-\theta) \tilde{x}_1; \bar{\boldsymbol{x}}_{-1} ) \\
 & \geq \theta G_1(\hat{x}_1; \bar{\boldsymbol{x}}_{-1}) + (1-\theta) G_1(\tilde{x}_1; \bar{\boldsymbol{x}}_{-1} ), \\
 & = \theta G_1(\hat{x}_1) + (1-\theta) Q_1(\tilde{x}_1; \bar{\boldsymbol{x}}_{-1} ), \\
 & \geq \theta H_1(\hat{x}_1; \bar{\boldsymbol{x}}_{-1}) + (1-\theta) Q_1(\tilde{x}_1; \bar{\boldsymbol{x}}_{-1} ), \\
 & = \theta Q_1(\hat{x}_1 ; \bar{\boldsymbol{x}}_{-1}) + (1-\theta) Q_1(\tilde{x}_1; \bar{\boldsymbol{x}}_{-1} ),
 \end{split}
 \label{concavity_user_1}
 \end{equation}

\vspace{0.2cm}

 \noindent where the second line results from concavity of $G_1$, the third
 line results from the fact that $Q_1(\tilde{x}_1; \bar{\boldsymbol{x}}_{-1} ) =
 G_1(\tilde{x}_1; \bar{\boldsymbol{x}}_{-1}
 )$, the fourth line results from (\ref{g_h_compare_1}), and the
 fifth line is due to $Q_1(\hat{x}_1; \bar{\boldsymbol{x}}_{-1}) = H_1(\hat{x}_1; \bar{\boldsymbol{x}}_{-1})$.
%
 Next assume that (\ref{case_2_theta}) holds. In that case, $\theta \hat{x}_1 +
 (1-\theta) \tilde{x}_1 \geq \bar{x}_N$. In addition, since
 $\tilde{x}_1 \leq \bar{x}_N$, we have
 \begin{equation}
 (1-\beta) \bar{x}_N \geq (1-\beta) \tilde{x}_1 \ \ \ \ \ \ \Rightarrow \ \ \ \ \ \ \beta \tilde{x}_1 \geq \tilde{x}_1 - (1-\beta) \bar{x}_N,
 \label{line3}
 \end{equation}
 and
 \begin{equation}
p \left( \sum_{n = 2}^{N-1} \bar{x}_n + \tilde{x}_1 \right) \ \leq \
p \left( \sum_{n = 2}^{N-1} \bar{x}_n + \bar{x}_N
 \right).
 \label{line4}
 \end{equation}
From (\ref{line3}) and (\ref{line4}), we have
 \begin{equation}
  G_1(\tilde{x}_1; \bar{\boldsymbol{x}}_{-1}) \leq H_1(\tilde{x}_1; \bar{\boldsymbol{x}}_{-1}),
 \label{g_h_compare_2}
 \end{equation}
Therefore, we can show that

 \begin{equation}
 \begin{split}
 Q_1( \theta \hat{x}_1 + (1-\theta) \tilde{x}_1; \bar{\boldsymbol{x}}_{-1} ) & = H_1( \theta \hat{x}_1 + (1-\theta) \tilde{x}_1; \bar{\boldsymbol{x}}_{-1} ) \\
 & \geq \theta H_1(\hat{x}_1; \bar{\boldsymbol{x}}_{-1}) + (1-\theta) H_1(\tilde{x}_1 ; \bar{\boldsymbol{x}}_{-1}), \\
 & = \theta Q_1(\hat{x}_1; \bar{\boldsymbol{x}}_{-1}) + (1-\theta) H_1(\tilde{x}_1 ; \bar{\boldsymbol{x}}_{-1}), \\
 & \geq \theta Q_1(\hat{x}_1; \bar{\boldsymbol{x}}_{-1}) + (1-\theta) G_1(\tilde{x}_1; \bar{\boldsymbol{x}}_{-1} ), \\
 & = \theta Q_1(\hat{x}_1; \bar{\boldsymbol{x}}_{-1}) + (1-\theta) Q_1(\tilde{x}_1; \bar{\boldsymbol{x}}_{-1} ).
 \end{split}
 \label{concavity_user_1_item2}
 \end{equation}

\noindent From (\ref{concavity_user_1}) and
(\ref{concavity_user_1_item2}), payoff $Q_1(x_1,
\bar{\boldsymbol{x}}_{-1})$ always satisfies (\ref{concavity}).
Thus, $Q_1(x_1,
 \bar{\boldsymbol{x}}_{-1})$ is concave. Similarly, we can show that $Q_N(x_N,
 \bar{\boldsymbol{x}}_{-N})$ is continuous and concave. The same
 statement is evidently true for all $n \! \in \! \mathcal{N}
 \backslash \{1, N\}$. Thus, Game \ref{game2} is a concave
 \emph{N-person} game and the existence of Nash equilibria follows
 from the Rosen's existence theorem \cite[Theorem 1]{rosen}. $\hfill
\blacksquare$


\vspace{-0.2cm}

 \subsection{Proof of Proposition \ref{proposition1}} \label{app:proof_proposition1}

 For each user $n \! \in \! \mathcal{N} \backslash \{1, N\}$, the KKT optimality conditions for problem
 (\ref{best_response_definition}) become
 \begin{equation}
 \textstyle U^\prime_n(x_n) - a ( \sum_{r=2, r \neq n}^{N-1} x_r \! + x_n \! + \! \max(x_1,
 x_N) ) - a x_n = - \lambda_n,
 \label{KKT1}
 \end{equation}
 and
 %
$ \lambda_n \: x_n \! = \! 0$.
 \label{KKT2}
 %
Here, $\lambda_n \! \geq \! 0$ denotes the \emph{Lagrange
multiplier}
 corresponding to inequality constraint $x_n \geq 0$. If $x_n > 0$,
 then $\lambda_n = 0$ and (\ref{KKT1}) reduces to
 (\ref{best_response_proposition_routing_equation}). Otherwise,
 (\ref{best_response_proposition_routing_equation}) cannot hold for
 any positive $x_n$ and we have $x_n = 0$. Recall that the best response
 is bounded below by zero.
  $\hfill
 \blacksquare$

%
%
%

\vspace{-0.2cm}

 \subsection{Proof of Proposition \ref{proposition2}} \label{app:proof_proposition2}

 For user $1$, we can write the KKT optimality
 conditions for optimization problem
 (\ref{user_1_first_problem}) as
 \begin{equation}
 \textstyle U^\prime_1(x_1) - a ( \sum_{n = 2}^{N-1} x_n + x_1
 ) - a ( x_1 - (1-\beta) x_N
 ) = - \lambda_1,
 \label{KKT1_user1}
 \end{equation}
 and
 %
$ \lambda_1 \: \left( x_1 - x_N \right) = 0$.
 %
 Here, $\lambda_1 \geq 0$ denotes the \emph{Lagrange
 multiplier} corresponding to constraint $x_1 \geq x_N$. If $x_1 > x_N$,
 then $\lambda_1 = 0$ and (\ref{KKT1_user1}) reduces to
 (\ref{best_response_proposition_nc_equation1}). Otherwise, $x_1$
 takes its lower bound value, i.e., $x_1 = x_N$. It is also easy to verify that
 (\ref{best_response_proposition_nc_equation2}) is simply the KKT condition for convex optimization problem
 (\ref{user_1_second_problem}), as long as the utility function $U_1(x_1)$ is \emph{non-linear} (i.e., concave, but \emph{not} linear).
 Therefore, it should hold for user
 1's best response data rate. If the utility function $U_1(x_1)$ is
 \emph{linear}, then the objective function in optimization problem
 (\ref{user_1_second_problem}) becomes
 \begin{equation}
\textstyle x_1 \left( U^\prime_1(x_1) - \beta \: a \left( \sum_{n =
2}^{N-1} x_n + x_N \right) \right),
\end{equation}
where $U^\prime_1(x_1) - \beta a ( \sum_{n = 2}^{N-1} x_n +  x_N )$
is a \emph{constant}. In this case, the best response depends on the
\emph{sign} of the multiplier term $U^\prime_1(x_1) - \beta a (
\sum_{n = 2}^{N-1} x_n + x_N )$ as explained in Proposition
\ref{proposition2}.
 $\hfill
 \blacksquare$

%
%

\vspace{-0.2cm}

 \subsection{Proof of Theorem \ref{theorem3}} \label{app:proof_theorem3}

At Nash equilibrium of Game \ref{game2}, we have
$x_n^B(\boldsymbol{x}^\ast_{-n}) = x_n^\ast$ for all users $n \in
\mathcal{N}$. From this, together with
(\ref{best_response_proposition_routing_equation}),
(\ref{best_response_proposition_nc_equation1}), and
(\ref{best_response_proposition_nc_equation2}) and also due to
$\beta \leq 1$, we can show that
%
%
%
\begin{equation}
\sum_{n=1}^N U_n^\prime( x^\ast_n ) x^\ast_n - C \left(
\sum_{n=2}^{N-1} x^\ast_n + \max( x^\ast_1, x^\ast_N ) \right) \geq
0. \label{theorem3_condition}
\end{equation}
That is, the network aggregate surplus is always \emph{non-negative}
at any Nash equilibrium of Game \ref{game2}. On the other hand,
using concavity of the utility functions, for each user $n \! \in \!
 \mathcal{N}$, we have
 \begin{equation}
 U_n(x^S_n) \leq U_n(x^\ast_n) + U^\prime_n(x^\ast_n) \left( x^S_n - x^\ast_n \right).
 \label{concavity_result}
 \end{equation}
 Again, by concavity, $U_n(0) \leq U_n(x^\ast_n) +
 U^\prime_n(x^\ast_n) (0 - x^\ast_n)$. Since according to Assumption \ref{assumption1}, $U_n(0) \geq 0$, we
 have $U^\prime_n(x^\ast_n) x^\ast_n \leq U_n(x^\ast_n)$.
 Applying this to all users $n \in \mathcal{N}$, we have
 \begin{equation}
 \textstyle \sum_{n = 1}^N \left( U_n(x^\ast_n) - U^\prime_n(x^\ast_n) x^\ast_n \right) \geq 0.
 \label{concavity_result_improved}
 \end{equation}
 For notational simplicity, we define
 \begin{equation}
\label{C_definition} C^\ast = \textstyle \ C \left( \sum_{n=2}^{N-1}
x^\ast_n + \max(x^\ast_1, x^\ast_N) \right) \ \ \text{ and } \ \ \
C^S = \textstyle \; C \left( \sum_{n=2}^{N-1} x^S_n + \max(x^S_1,
x^S_N) \right).
 \end{equation}
 We notice that
 \begin{equation}
 \begin{split}
 \textstyle \sum_{n = 1}^N U^\prime_n( x^\ast_n ) x^S_n - C^S & =
\textstyle \sum_{n = 2}^{N-1} U^\prime_n( x^\ast_n ) x^S_n +  \left(
U^\prime_1( x^\ast_1 ) + U^\prime_N( x^\ast_N ) \right) x_1^S- C^S \\
 & \leq \textstyle \sigma \left( \sum_{n = 2}^{N-1} x_n^S + \max( x_1^S, x_N^S ) \right) - C^S \\
 & \leq \max_{\tilde{q} \geq 0} \left[ \sigma \tilde{q} - C(\tilde{q})
\right],
 \label{last_line_proof}
 \end{split}
 \end{equation}
 where $\tilde{q}$ is an \emph{auxiliary} variable, the equality results from Theorem \ref{theorem2}, the first inequality is due to the definition of $\sigma$, and the second inequality results from (\ref{C_definition}).
 Clearly, $\max_{\tilde{q} \geq 0} \left[ \sigma \tilde{q} - C(\tilde{q})
 \right]$ in (\ref{last_line_proof}) provides an \emph{upper bound} for the \emph{optimal}
 network aggregate surplus when the utility functions are
 \emph{linear} as in (\ref{linear_utilities_class}). First assume
 that $\sigma = U^\prime_1( x^\ast_1 ) + U^\prime_N( x^\ast_N )$. In
 this case, if we select $x_n^S = 0$ for all $n \in
 \mathcal{N} \backslash \{1, N\}$ and select $x_1^S = x_N^S = \tilde{q}^S$, where $\tilde{q}^S
 = \argmax_{\tilde{q} \geq 0} \left[ \sigma \tilde{q} - C(\tilde{q})
 \right]$, then the network aggregate surplus reaches its upper bound $\max_{\tilde{q} \geq 0} \left[ \sigma \tilde{q} - C(\tilde{q})
 \right]$. Next, assume that $\sigma = U^\prime_m( x^\ast_m )$ for
 some $m \in \mathcal{N} \backslash \{1, N\}$. In that case, if we select $x_n^S = 0$ for all $n \in
 \mathcal{N} \backslash \{m\}$ and select $x_m^S = \tilde{q}^S$, then the network aggregate surplus reaches its upper bound $\max_{\tilde{q} \geq 0} \left[ \sigma \tilde{q} - C(\tilde{q})
 \right]$. Therefore, we can conclude that $\max_{\tilde{q} \geq 0} \left[ \sigma \tilde{q} - C(\tilde{q})
 \right]$ is indeed equal to the optimal objective value of
Problem \ref{problem2} for the special case of having \emph{linear}
utility functions. Finally, from
(\ref{concavity_result})-(\ref{last_line_proof}),
 \begin{equation}
 \begin{split}
 1 & \geq \frac{\sum_{n=1}^N U_n(x^\ast_n) - C^\ast}{\sum_{n=1}^N U_n(x^S_n) - C^S} \\
 & = \frac{ \sum_{n=1}^N \left( U_n(x^\ast_n) - U^\prime_n(x^\ast_n) x^\ast_n \right) + \sum_{n=1}^N U^\prime_n(x^\ast_n) x^\ast_n - C^\ast }{ \sum_{n = 1}^N U_n( x_n^S ) - C^S  } \\
 & \geq \frac{ \sum_{n=1}^N \left( U_n(x^\ast_n) - U^\prime_n(x^\ast_n) x^\ast_n \right) + \sum_{n=1}^N U^\prime_n(x^\ast_n) x^\ast_n - C^\ast }{ \sum_{n=1}^N \left( U_n(x^\ast_n) - U^\prime_n(x^\ast_n) x^\ast_n \right) + \sum_{n = 1}^N U^\prime_n( x^\ast_n ) x_n^S - C^S  } \\
 & \geq \frac{ \sum_{n=1}^N \left( U_n(x^\ast_n) - U^\prime_n(x^\ast_n) x^\ast_n \right) + \sum_{n=1}^N U^\prime_n(x^\ast_n) x^\ast_n - C^\ast }{ \sum_{n=1}^N \left( U_n(x^\ast_n) - U^\prime_n(x^\ast_n) x^\ast_n \right) + \max_{\tilde{q} \geq 0} \left[ \sigma \tilde{q} - C(\tilde{q}) \right]  } \\
 & \geq \frac{ \sum_{n=1}^N U^\prime_n(x^\ast_n) x^\ast_n - C^\ast }{ \max_{\tilde{q} \geq 0} \left[ \sigma \tilde{q} - C(\tilde{q}) \right]  } \geq 0,
 \end{split}
 \label{first_line_proof}
 \end{equation}
 where the third line results from (\ref{concavity_result}), the
 fourth line results from (\ref{last_line_proof}), and the last line results from (\ref{theorem3_condition}), (\ref{concavity_result_improved}), and (\ref{last_line_proof}) and also the fact that $\max_{\tilde{q} \geq 0} \left[ \sigma \tilde{q} - C(\tilde{q}) \right]$ is always non-negative. $\hfill \blacksquare$

\subsection{Proof of Theorem \ref{theorem6}} \label{app:proof_theorem6}

 We notice that since the utility functions are linear, we have: $U^\prime_n(\cdot) = \gamma_n$ for all $n \in \mathcal{N}$. Next, we prove that due to $\gamma_1 \geq \gamma_N$, we should always have $x_1^\ast \geq x_N^\ast$.
 To show thus, assume that $x_1^\ast <
 x_N^\ast$. Since $U^\prime_1(x_1) = \gamma_1$, from Proposition \ref{proposition2}, the inequality $x_1^\ast <
 x_N^\ast$ implies that
 \begin{equation}
\gamma_1 \leq \beta a \left( q^\ast + x_N^\ast \right).
\label{contradiction_line1}
 \end{equation}
Furthermore, since $U^\prime_N(x_N) = \gamma_N$, from
 (\ref{best_response_proposition_nc_equation1}) and after reordering the terms, we have\footnote{Notice that (\ref{best_response_proposition_nc_equation1}) in its current format is formulated for user $1$; however, it can be reformulated similarly for user $N$.}
\begin{equation}
  \gamma_N = a q^\ast + 2 a x_N^\ast - a (1 - \beta) x_1^\ast.
\label{contradiction_line2}
\end{equation}
From (\ref{contradiction_line1}), (\ref{contradiction_line2}), and
due to $\gamma_N \leq \gamma_1$, it is required that
\begin{equation}
\begin{split}
& a q^\ast + 2 a x^\ast_N - a (1 - \beta) x_1^\ast \leq \beta a
(q^\ast + x_N^\ast) \\ & \ \ \ \ \ \ \Rightarrow \ \ q^\ast(1 -
\beta) + x_N^\ast (2 - \beta) - (1 - \beta) x_1^\ast \leq 0.
\end{split} \label{greater_equal}
\end{equation}
If $\beta = 1$, then the inequality in (\ref{greater_equal}) reduces
to $x_N^\ast \leq 0$ which \emph{contradicts} the assumption that
$x_N^\ast
> x_1^\ast \geq 0$. On the other hand, if $0 < \beta < 1$, then we can further show that
\begin{equation}
q^\ast(1 - \beta) + x_N^\ast (2 - \beta) - (1 - \beta) x_1^\ast \;
\geq \; q^\ast(1 - \beta) + (x_N^\ast - x_1^\ast) (1 - \beta) \;
> \; 0, \label{less}
\end{equation}
where the last (strict) inequality results from the assumption that
$x_N^\ast > x_1^\ast$. It is clear that (\ref{less})
\emph{contradicts} (\ref{greater_equal}). Thus, for any $0 < \beta
\leq 1$, the data rate $x_N^\ast$ \emph{cannot} be greater than
$x_1^\ast$ and we \emph{always} have $x_N^\ast \leq x_1^\ast$. Next,
we prove each part of the theorem separately.


Part (a): We first prove that
%
$x_1^\ast = x_N^\ast$ by contradiction.
%
If $x_1^\ast \neq x_N^\ast$, then since $x_1^\ast \geq x_N^\ast$, we
should have $x_1^\ast > x_N^\ast$. In that case, from Proposition
\ref{proposition2},
\begin{equation}
x^\ast_1 =  \frac{\gamma_1 - a q^\ast + a (1 - \beta) x_N^\ast}{2 a}
> x^\ast_N \ \ \ \ \Rightarrow \ \ \ \gamma_1 > (1 + \beta) a
x^\ast_N + a q^\ast, \label{proof_theorem6_part_a_line1}
\end{equation}
and
\begin{equation}
\gamma_N \leq \beta a x^\ast_1 + \beta a q^\ast.
\label{proof_theorem6_part_a_line2}
\end{equation}
From (\ref{proof_theorem6_part_a_line1}) and
(\ref{proof_theorem6_part_a_line2}),
\begin{equation}
\begin{split}
\gamma_N & \leq \beta a q^\ast + \frac{\beta}{2} \left( \gamma_1 - a q^\ast + a (1-\beta) x_N^\ast \right) \\
& < \beta a q^\ast + \frac{\beta}{2} \left( \gamma_1 - a q^\ast + (1
- \beta) \frac{\gamma_1 - a q^\ast}{1 + \beta} \right)
%
%
 = \frac{\beta}{1 + \beta} \left( \gamma_1 + \beta a q^\ast
\right).
\end{split}
\end{equation}
This implies that
\begin{equation}
\left( 1 + \frac{1}{\beta} \right) \gamma_N - \beta a q^\ast <
\gamma_1.
\end{equation}
However, this contradicts the assumption in this scenario that
$\gamma_1 \leq \left( 1 + 1 / \beta \right) \gamma_N - \beta a
q^\ast$. Thus, we indeed have
%
$ x_1^\ast = x_N^\ast$.
%
From this, together with Proposition \ref{proposition2}, we have
\begin{equation}
 \gamma_1 \leq \beta a x_1^\ast + a q^\ast + a x_1^\ast = a q^\ast + (1 + \beta) a x_1^\ast \ \ \ \ \Rightarrow \ \ \ \ \frac{\gamma_1 - a q^\ast}{a ( 1 + \beta)} \leq x_1^\ast,
\end{equation}
and
\begin{equation}
 \gamma_N \geq \beta a q^\ast + \beta a x_1^\ast \ \ \ \ \Rightarrow \ \ \ \ x_1^\ast \leq \frac{\gamma_N - \beta a q^\ast}{\beta a}.
\end{equation}
Notice that the transmission rates are \emph{non-negative}. Thus,
the best responses are as in (\ref{theorem6_part_a}).

Part (b): We first notice that the condition in this scenario holds
if and only if
\begin{equation}
\left( 1 + \frac{1}{\beta} \right) \gamma_N - \beta a q^\ast \leq
\frac{2}{\beta} \gamma_N - a q^\ast \ \ \ \ \Rightarrow \ \ \ \
\gamma_N \geq \beta a q^\ast.
\end{equation}
On the other hand, since $\gamma_1 \leq \frac{2}{\beta} \gamma_N - a
q^\ast$, we have $\frac{2}{\beta} \gamma_N - \gamma_1 - a q^\ast
\geq 0$ and $x_N^\ast$ in (\ref{theorem6_part_b}) is \emph{non-
negative}. Since $x_1^\ast \geq x_N^\ast$, this also implies
non-negativity of $x_1^\ast$. Next, we consider two cases:

Case I) Assume that $x^\ast_N > 0$. Following similar steps as in
the proof of Part (a), we can show that in this case,
%
$x_1^\ast > x_N^\ast$. 
%
From this, together with Proposition \ref{proposition2},
\begin{equation}
x_1^\ast = \frac{\gamma_1 - a q^\ast + a (1 - \beta) x_N^\ast}{2 a},
\label{proof_theorem6_part_b_line1}
\end{equation}
and
\begin{equation}
 \gamma_N = \beta a q^\ast + \beta a x_1^\ast \ \ \ \ \ \Rightarrow \ \ \ \ \ x_1^\ast = \frac{\gamma_N - \beta a q^\ast}{\beta a} = \frac{\gamma_N}{\beta a} - q^\ast.
\label{proof_theorem6_part_b_line2}
\end{equation}
Replacing (\ref{proof_theorem6_part_b_line2}) in
(\ref{proof_theorem6_part_b_line1}), we have
\begin{equation}
 \frac{\gamma_N}{\beta a} - q^\ast = \frac{\gamma_1 - a q^\ast + a (1 - \beta) x_N^\ast}{2 a} \ \ \ \ \ \Rightarrow \ \ \ \ \ x^\ast_N = \frac{\frac{2}{\beta} \gamma_N - \gamma_1 - a q^\ast}{a (1- \beta)}.
\end{equation}

Case II) Assume that $x_N^\ast = 0$. In that case, from Proposition
2, we have
\begin{equation}
 x_1^\ast = \frac{\gamma_1 - a q^\ast}{2 a},
\label{proof_theorem6_part_b_line3}
\end{equation}
and
\begin{equation}
\gamma_N \leq \beta a q^\ast + \beta a x_1^\ast.
\label{proof_theorem6_part_b_line4}
\end{equation}
Replacing (\ref{proof_theorem6_part_b_line3}) in
(\ref{proof_theorem6_part_b_line4}), we have
\begin{equation}
\gamma_N \leq \frac{\beta}{2} + \frac{\beta a q^\ast}{2} \ \ \ \
\Rightarrow \ \ \ \ \gamma_1 \geq \frac{2}{\beta} \gamma_N - a
q^\ast. \label{proof_theorem6_part_b_line4_another}
\end{equation}
From (\ref{proof_theorem6_part_b_line4_another}) and knowing that
$\gamma_1 \leq \frac{2}{\beta} \gamma_N - a q^\ast$, it is required
that
\begin{equation}
\gamma_1 = \frac{2}{\beta} \gamma_N - a q^\ast \ \ \ \
\overset{\text{By (\ref{proof_theorem6_part_b_line3})}}{\Rightarrow}
\ \ \ \ x_1^\ast = \frac{\gamma_N}{\beta a} - q^\ast.
\end{equation}

Thus, in both Case I and Case II, the best responses are as in
(\ref{theorem6_part_b}).

Part (c): We consider two cases:

Case I) Assume that $x_1^\ast > x_N$. In that case,
(\ref{proof_theorem6_part_a_line1}) and
(\ref{proof_theorem6_part_a_line2}) hold.
%
%
%
First, assume that $\gamma_N = \beta a q^\ast + \beta a x_1^\ast$.
Thus, $x_1^\ast = \frac{\gamma_N}{\beta a} - q^\ast$. Replacing this
in (\ref{proof_theorem6_part_a_line1}), the data rate for user $N$
is obtained as
\begin{equation}
x_N^\ast = \frac{\frac{2}{\beta} \gamma_N - \gamma_1 - a q^\ast}{a
(1 - \beta)} \leq 0,
\end{equation}
where the inequality results from the fact that $\gamma_1 \geq
\frac{2}{\beta} \gamma_N - a q^\ast$. Clearly, since the data rates
are non-negative, the above implies that $x_N^\ast = 0$. Replacing
this in (\ref{proof_theorem6_part_a_line1}), we have
\begin{equation}
 x_1^\ast = \frac{\gamma_1 - a q^\ast}{2 a}.
\label{proof_theorem6_part_c_line3}
\end{equation}
Next, assume that (\ref{proof_theorem6_part_a_line2}) holds as a
\textit{strict} inequality. That is,
\begin{equation}
 \gamma_N < \beta a q^\ast + \beta a x_1^\ast \ \ \ \ \Rightarrow \ \ \ \ x_N^\ast = 0.
\end{equation}
Replacing this in (\ref{proof_theorem6_part_a_line1}), the data rate
model in (\ref{theorem6_part_c}) is obtained.

Case II) Assume that $x^\ast_1 = x^\ast_N$. In that case, from
Proposition \ref{proposition2}, we have
\begin{equation}
\gamma_1 \leq (1 + \beta) a x_1^\ast + a q^\ast,
\label{proof_theorem6_part_c_line4}
\end{equation}
and
\begin{equation}
\gamma_N \geq \beta a q^\ast + \beta a x^\ast_1 \ \ \ \ \Rightarrow
\ \ \ \ \ \frac{2}{\beta} \gamma_N - a q^\ast \geq a q^\ast + 2 a
x_1^\ast. \label{proof_theorem6_part_c_line5}
\end{equation}
From (\ref{proof_theorem6_part_c_line4}) and
(\ref{proof_theorem6_part_c_line5}) and since by assumption
$\gamma_1 \geq \frac{2}{\beta} \gamma_N - a q^\ast$, it is required
that
\begin{equation}
 (1 + \beta) a x_1^\ast + a q^\ast \geq a q^\ast + 2 a x_1^\ast \ \ \ \ \ \Rightarrow \ \ \ \ \ a (1-\beta) x_1^\ast \leq 0.
\end{equation}
Thus, either $x_1^\ast = x_N^\ast = 0$ and $\gamma_1 = a q^\ast$ or
$\beta = 1$ and $\gamma_1 = 2 a x_1^\ast + a q^\ast$. In the latter
case,
\begin{equation}
x_1^\ast = \frac{\gamma_1 - a q^\ast}{2 a}.
\label{proof_theorem6_part_c_line6}
\end{equation}
In addition, from Proposition \ref{proposition2}, we have
\begin{equation}
x^\ast_1 = \frac{\gamma_1 - a q^\ast + a (1 - \beta) x_N^\ast}{2 a}.
\label{proof_theorem6_part_c_line7}
\end{equation}
From (\ref{proof_theorem6_part_c_line6}) and
(\ref{proof_theorem6_part_c_line7}), $x_1^\ast = x_N^\ast = 0$ and
$\gamma_1 = a q^\ast$. Clearly, these results satisfy
(\ref{theorem6_part_c}).

Part (d): For each node $n \! \in \! \mathcal{N} \backslash \{1,
N\}$, at each Nash equilibrium $\boldsymbol{x}^\ast \! \in \!
\mathcal{X}^\ast$ of Game \ref{game2}, it is required that
$x_n^B(\boldsymbol{x}^\ast_{-n}) = x^\ast_n$. Thus, in case of
having linear utility functions, the derivative of the objective
function in problem
(\ref{best_response_proposition_routing_equation}) with respect to
variable $x_n$ is obtained as
%
$ \gamma_n - a \left( q^\ast + x_1^\ast \right) - x_n a$.
%
If $\gamma_n \leq a \left( q^\ast + x_1^\ast \right)$, the
derivative is always \emph{non-positive} and the objective function
is \emph{decreasing} in data rate $x_n$. In that case, $x_n^\ast =
0$. Otherwise, i.e., if $\gamma_n \geq a \left( q^\ast + x_1^\ast
\right)$, then since the objective function is \emph{convex}, we
have $x_n^\ast = \frac{\gamma_n}{a} - q^\ast - x_1^\ast$. Together,
these two cases result in (\ref{theorem6_part_d}).  $\hfill
\blacksquare$

\subsection{Proof of Theorem \ref{theorem8}} \label{app:proof_theorem8}

We first obtain the optimal solution of Problem \ref{problem2} when
$N = 2$.
Since Problem \ref{problem2} is convex, we can use the KKT
 optimality conditions and confirm that the \emph{optimal} data rates are
\begin{equation}
x^S_1 = x_2^S = \frac{\gamma_1 + \gamma_2}{a}.
\end{equation}
Thus, at optimality, the network aggregate surplus becomes
\begin{equation}
\gamma_1 x_1^S + \gamma_2 x_2^S - \frac{a}{2} \left( \max\{x_1^S,
x_2^S\} \right)^2 = \frac{\left( \gamma_1 + \gamma_2 \right)^2}{a} -
\frac{a}{2} \left( \frac{\gamma_1 + \gamma_2}{a} \right)^2 =
\frac{\left( \gamma_1 + \gamma_2 \right)^2}{2 a}.
\end{equation}
Next, we examine the efficiency for all the scenarios in Theorem
\ref{theorem6}(a), (b), (c), where $q^\ast  = 0$ as there is
\emph{no} routing user in the network. First, we assume that
\begin{equation}
\gamma_2 \leq \gamma_1 \leq \left(1 + \frac{1}{\beta} \right)
\gamma_2. \label{proof_t7_line0}
\end{equation}
From Theorem \ref{theorem6}(a), the Nash equilibria are as in
(\ref{theorem6_part_a}). Since there are \emph{multiple} Nash
equilibria,

\noindent the \emph{worst-case} efficiency for Game \ref{game2} is
obtained by solving the following optimization problem
\begin{equation}
\begin{aligned}
& \underset{x_1^\ast}{\text{minimize}} & \! \frac{\left( \gamma_1 + \gamma_N \right) x_1^\ast - \frac{a}{2} {x_1^\ast}^2 }{ \frac{\left( \gamma_1 + \gamma_2 \right)^2 }{2 a} } \\
& \text{subject to} & \frac{\gamma_1}{(1+\beta) a } \leq x_1^\ast
\leq \frac{\gamma_2}{\beta a}. \label{proof_t7_line1}
\end{aligned}
\end{equation}
Problem (\ref{proof_t7_line1}) is a concave \emph{minimization} (not
maximization) problem. Thus, the optimality occurs at one of the
\emph{boundary} points. That is, either at $x_1^\ast =
\frac{\gamma_2}{\beta a}$ or at $x_1^\ast =
\frac{\gamma_1}{(1+\beta) a }$. The derivative of the objective
function in Problem (\ref{proof_t7_line1}) with respect to
$x_1^\ast$ can be written as
\begin{equation}
\frac{2 a}{\left( \gamma_1 + \gamma_2 \right)^2} \left( (\gamma_1 +
\gamma_2 ) - a x^\ast_1 \right). \label{proof_t7_line2}
\end{equation}
The sign of the derivative in (\ref{proof_t7_line2}) depends on the
choice of parameter $\beta$. We assume that
%
$\frac{1}{2} \leq \beta \leq 1$. 
%
This includes the two cases of $\beta = \frac{1}{2}$ and $\beta =
1$. From (\ref{proof_t7_line0}), we have
\begin{equation}
\gamma_1 \beta \geq \gamma_2 (1 - \beta) \ \ \ \ \ \ \Rightarrow \ \
\ \ \ \ \gamma_1 + \gamma_2 \geq \frac{\gamma_2}{\beta} \geq a
x_1^\ast, \label{proof_t7_line5}
\end{equation}
where the last inequality is due to constraint $x_1^\ast \leq
\frac{\gamma_2}{\beta a}$ in (\ref{proof_t7_line1}). From
(\ref{proof_t7_line5}), the derivative of the
objective function in Problem  (\ref{proof_t7_line1}) is
\emph{positive}. Thus, the objective function is \emph{increasing}
in $\: x^\ast_1$

\noindent and the minimum occurs at \emph{lower-bound} $x_1^\ast \!
= \! \frac{\gamma_1}{a (1+\beta)}$. In this case, the efficiency
ratio becomes
\begin{equation}
\nonumber \frac{2 \gamma_1}{(1\!+\!\beta) \left( \gamma_1 \! + \!
\gamma_2 \right)^2} \left( \gamma_1 \! + \! \gamma_2 \! - \!
\frac{\gamma_1}{2 ( 1 + \beta) } \right) = \frac{2}{1 + \beta}
\left( \frac{1}{1 + \gamma_2 / \gamma_1} \right) - \frac{1}{(1 +
\beta)^2} \left( \frac{1}{1 + \gamma_2 / \gamma_1} \right)^2.
\end{equation}
%
%
%
%
For the ease of exposition, we define
%
$\Gamma = \frac{1}{1 + \gamma_2 / \gamma_1}$.
%
Since $\gamma_1 \geq \gamma_2$, we have
\begin{equation}
\frac{\gamma_2}{\gamma_1} \leq 1 \ \Rightarrow \ 1 +
\frac{\gamma_2}{\gamma_1} \leq 2 \ \Rightarrow \ \frac{1}{1 +
\gamma_2 / \gamma_1} \geq \frac{1}{2} \ \Rightarrow \Gamma \geq
\frac{1}{2}.
\end{equation}
On the other hand, from the lower and upper bounds in problem
(\ref{proof_t7_line1}), we also have
\begin{equation}
\begin{split}
1 + \frac{\gamma_2}{\gamma_1} \geq \frac{1}{1 + 1/\beta} + 1 \
\Rightarrow \ \Gamma \leq \frac{1 + \beta}{1 + 2 \beta}.
\end{split}
\end{equation}
Thus, the worst-case efficiency is obtained by solving the following
 problem over variable $\Gamma$:
\begin{equation}
\begin{aligned}
& \underset{\Gamma}{\text{minimize}} & \frac{2}{1 + \beta} \Gamma - \frac{1}{(1+\beta)^2} \Gamma^2 \\
& \text{subject to} & \! \frac{1}{2} \leq \Gamma \leq \frac{1 +
\beta}{1 + 2 \beta}. \ \ \ \ \ \: \label{proof_t7_line6}
\end{aligned}
\end{equation}
The derivative of the objective function in problem
(\ref{proof_t7_line6}) can be obtained as
\begin{equation}
\frac{2}{1 + \beta} - \frac{2}{(1 + \beta)^2} \Gamma = \frac{2}{1 +
\beta} \left( 1 - \frac{\Gamma}{1 + \beta} \right).
\label{proof_t7_line7}
\end{equation}
On the other hand, we can show that
\begin{equation}
\Gamma \leq \frac{1 + \beta}{1 + 2 \beta} \ \ \ \Rightarrow \ \ \
\frac{\Gamma}{1 + \beta} \leq \frac{1}{1 + 2 \beta} \ \ \
\Rightarrow \ \ \ 1 - \frac{\Gamma}{1 + \beta} \geq 1 - \frac{1}{1 +
2 \beta} = \frac{2 \beta}{1 + 2 \beta} \geq 0.
\end{equation}
%
%
%
Thus, the derivative in (\ref{proof_t7_line7}) is always
\emph{positive} and minimum efficiency occurs at the \emph{lower
bound} $\Gamma = \frac{1}{2}$. Therefore, the worst-case efficiency
when (\ref{proof_t7_line0}) holds is obtained as
\begin{equation}
\frac{2}{1 + \beta} \times \frac{1}{2} - \frac{1}{(1+\beta)^2}
\times \frac{1}{4} = \frac{1}{1+\beta} - \frac{1}{4 (1 + \beta)^2}.
\label{proof_t7_line8}
\end{equation}
If $\beta = 1$, then (\ref{proof_t7_line8}) becomes
\begin{equation}
\abovedisplayskip 5pt \belowdisplayskip 5pt 1/2 - 1/16 = 7/16
\approx 0.438. \label{proof_t7_line9}
\end{equation}
On the other hand, if $\beta = \frac{1}{2}$, then
(\ref{proof_t7_line8}) becomes
\begin{equation}
\abovedisplayskip 5pt \belowdisplayskip 5pt 6 / 9 - 1 / 9 = 5 / 9
\approx 0.556. \label{proof_t7_line10}
\end{equation}
Next, we assume that $\beta < 1$ and
\begin{equation}
\abovedisplayskip 5pt \belowdisplayskip 5pt \left( 1 + 1 / \beta
\right) \gamma_2 \leq \gamma_1 \leq \frac{2}{\beta} \gamma_2.
\label{proof_t7_line10_remained}
\end{equation}
%
From Theorem \ref{theorem6}, the data rates of users 1 and 2 at Nash
equilibrium are as in (\ref{theorem6_part_b}) where $q^\ast \! = \!
0$. Assuming that $\gamma_2$ is \emph{fixed}, the worst-case
efficiency is obtained by solving the following problem

\begin{equation}
\begin{aligned}
& \underset{\gamma_1}{\text{minimize}} & \frac{\gamma_2}{\left(
\gamma_1 + \gamma_2 \right)^2} \left( 2 \left( \frac{1}{\beta} \! -
\frac{1}{1-\beta} \right) \gamma_1 +
\frac{1}{\beta} \left( \frac{4}{1 - \beta} - \frac{1}{\beta} \right) \gamma_2 \right) \\
& \text{subject to} & \left( 1 + 1 / \beta \right) \gamma_2 \leq
\gamma_1 \leq \frac{2}{\beta} \gamma_2. \ \ \ \ \ \ \ \ \ \ \ \ \ \
\ \ \ \ \ \ \: \ \ \ \ \ \ \ \ \ \ \ \ \ \ \label{proof_t7_line11}
\end{aligned}
\end{equation}
The derivative of the objective function in (\ref{proof_t7_line11})
can be obtained as
%
%
\begin{equation}
- \frac{1}{\left( \gamma_1 + \gamma_2 \right)^3} \left( \Phi \left(
\gamma_1 - \gamma_2 \right) + 2 \Psi \right),
\label{proof_t7_line12}
\end{equation}
where
\begin{equation}
\Phi = 2 \gamma_2 \left( \frac{1}{\beta} - \frac{1}{1- \beta}
\right) \ \ \ \ \ \ \text{ and } \ \ \ \ \ \ \Psi = \frac{1}{\beta}
\left( \frac{4}{1-\beta} - \frac{1}{\beta} \right) {\gamma_2}^2.
\end{equation}
If $\beta = \frac{1}{2}$, then $\Phi = 0$ and $\Psi = 12
{\gamma_2}^2$. Thus, the derivative in (\ref{proof_t7_line12}) is
\emph{negative} and the minimum in (\ref{proof_t7_line11}) occurs at
\emph{upper bound}
%
$\gamma_1 = \frac{2}{\beta} \gamma_2 = 4 \gamma_2$.
%
Replacing this in the objective function in (\ref{proof_t7_line11}),
the worst-case efficiency when $\beta = \frac{1}{2}$ and
(\ref{proof_t7_line10_remained}) holds is obtained as
\begin{equation}
\frac{12 {\gamma_2}^2}{\left( 4 \gamma_2 + \gamma_2 \right)^2} =
\frac{12 {\gamma_2}^2}{\left( 5 \gamma_2 \right)^2} = \frac{12}{25}
= 0.48. \label{proof_t7_line13}
\end{equation}
%
Finally, we assume that
\begin{equation}
\frac{2}{\beta} \gamma_2 \leq \gamma_1. \label{proof_t7_line14}
\end{equation}
From Theorem \ref{theorem6}(c), the Nash equilibrium is as in
(\ref{theorem6_part_c}) and the worst-case efficiency is obtained by
solving
 the following
optimization problem
\begin{equation}
\begin{aligned}
& \underset{\gamma_1, \gamma_2}{\text{minimize}} & \frac{\gamma_1 \frac{\gamma_1}{2 a} - \frac{a}{2} \left( \frac{\gamma_1}{2 a} \right)^2}{\frac{\left(\gamma_1 + \gamma_2 \right)^2}{2 a}} \\
& \text{subject to} & 0 \leq \frac{2}{\beta} \gamma_2 \leq \gamma_1.
\ \ \label{proof_t7_line15}
\end{aligned}
\end{equation}
The objective function in problem (\ref{proof_t7_line15}) is
\emph{decreasing} in $\gamma_2$. Thus, the minimum occurs at
\emph{upper-bound}
%
$\gamma_2 = \frac{\beta}{2} \gamma_1$. 
%
The worst-case efficiency when (\ref{proof_t7_line14}) holds is
obtained as
\begin{equation}
\frac{\frac{1}{2a} - \frac{a}{2} \left( \frac{1}{2 a} \right)^2 }{
\frac{\left(1 + \frac{\beta}{2}\right)^2 }{2 a} } = \frac{3}{4}
\left( \frac{1}{1 + \beta/2}\right)^2 = \frac{3}{\left( 2 + \beta
\right)^2}. \label{proof_t7_line17}
\end{equation}
If $\beta = 1$, then (\ref{proof_t7_line17}) becomes
\begin{equation}
\frac{3}{\left( 2 + 1 \right)^2} = \frac{1}{3} \approx 0.33.
\label{proof_t7_line18}
\end{equation}
On the other hand, if $\beta = \frac{1}{2}$, then
(\ref{proof_t7_line17}) becomes
\begin{equation}
\frac{3}{\left( 2 + \frac{1}{2} \right)^2} = \frac{12}{25} = 0.48.
\label{proof_t7_line19}
\end{equation}
Considering all the possible choices of system parameters, if
pricing parameter $\beta = 1$, then
\begin{equation}
\text{PoA}\left(\text{Game \ref{game2}}, \text{Problem
\ref{problem2}}\right) \ \ \underset{\text{and }
(\ref{proof_t7_line18})}{\overset{\text{By
(\ref{proof_t7_line9})}}{=}} \ \
\min\left\{ \frac{7}{16} , \frac{1}{3} \right\} = \frac{1}{3}.
\end{equation}
On the other hand, if $\beta = \frac{1}{2}$, then
\begin{equation}
\text{PoA}\left(\text{Game \ref{game2}}, \text{Problem
\ref{problem2}}\right) \ \ \underset{\text{and }
(\ref{proof_t7_line19})}{\overset{\text{By (\ref{proof_t7_line10}),
(\ref{proof_t7_line13}),}}{=}} \ \
\min\left\{ \frac{5}{9} , \frac{12}{25}, \frac{12}{25} \right\} =
\frac{12}{25}.
\end{equation}
This concludes the proof.  $\hfill \blacksquare$

\vspace{-0.1cm}

\subsection{Proof of Theorem \ref{theorem9}} \label{app:proof_theorem9}

Using the KKT optimality conditions, the optimal network aggregate
surplus, i.e., the optimal solution of Problem \ref{problem2}, for
linear utilities is obtained as
%
$\sigma^2 / ({2 a})$.
%
%
Next, we study two cases:

Case I) We assume that $\gamma_1 + \gamma_N = \sigma$. Similar to
the proof of Theorem \ref{theorem8}, here we obtain the PoA by
examining all the scenarios in Theorem \ref{theorem6}(a), (b), (c).
First, assume that
\begin{equation}
\abovedisplayskip 5pt \belowdisplayskip 5pt \gamma_N \leq \gamma_1
\leq \left( 1 + {1}/{\beta}\right) \gamma_N - \beta a q^\ast,
\label{condition_theorem8_caseI_part_a}
\end{equation}
and
\begin{equation}
\abovedisplayskip 5pt \belowdisplayskip 0pt
 \gamma_1 \geq a q^\ast. \label{condition_theorem8_caseI_part_a_other_one}
\end{equation}
In that case, Nash equilibrium is obtained as in
(\ref{theorem6_part_a}). We notice that
%
%
\begin{equation}
0 \overset{\text{By
(\ref{condition_theorem8_caseI_part_a_other_one})}}{\leq}
\frac{\gamma_1 - a q^\ast}{a (1+\beta)} \overset{\text{By
(\ref{condition_theorem8_caseI_part_a})}}{\leq} \frac{\gamma_N -
\beta a q^\ast}{\beta a}.
\end{equation}
To obtain the worst-case efficiency for this scenario, we need to
solve the following problem
\begin{eqnarray}
& \underset{\boldsymbol{x}^\ast, \boldsymbol{\gamma}, a, N, q^\ast}{\text{minimize}} \ & \frac{\sigma x^\ast_1 + \sum_{n=2}^{N-1} \gamma_n x^\ast_n - \frac{a}{2} (q^\ast + x^\ast_1)^2 }{{\sigma^2}/({2 a})} \label{problem1_line1} \\
& \text{subject to} \ & \gamma_n = a \left( q^\ast + x^\ast_n + x^\ast_1 \right), \ \ \ \ \text{if } x^\ast_n > 0, \ \ n = 2, \ldots, N-1, \label{problem1_line2}  \\
& \ & \gamma_n \leq a (q^\ast + x^\ast_1), \ \ \ \ \ \ \ \ \ \ \ \:\! \text{if } x^\ast_n = 0, \ \ n = 2, \ldots, N-1, \label{problem1_line2_added}  \\
& \ & \textstyle \sum_{n=2}^{N-1} x^\ast_n = q^\ast, \label{problem1_line3} \\
& \ & \gamma_1 + \gamma_N = \sigma, \label{problem1_line4}  \\
& \ & \gamma_1 \geq a q^\ast, \label{problem1_line5}  \\
& \ & 0 < \gamma_n \leq \sigma, \ \ \ \ n = 2, \ldots, N-1, \label{problem1_line6}  \\
& \ & \gamma_N \leq \gamma_1 \leq \left( 1 + 1 / \beta \right) \gamma_N - \beta a q^\ast, \label{problem1_line7}  \\
& \ & \frac{\gamma_1 - a q^\ast}{a (1 + \beta)} \leq x^\ast_1 = x^\ast_N \leq \frac{\gamma_N - \beta a q^\ast}{\beta a}, \label{problem1_line8}  \\
& \ & x^\ast_n \geq 0, \ \ \ \ n = 1, \ldots, N.
\label{problem1_line9}
\end{eqnarray}

\noindent Here, $\boldsymbol{\gamma} = (\gamma_1, \ldots, \gamma_N)$
denotes the vector of utility parameters for all users. We assume,
without loss of generality, that $\gamma_n = a (q^\ast + x^\ast_n +
x^\ast_1)$ for all users $n \in \mathcal{N} \backslash \{1, N\}$. In
fact, if for a feasible efficiency, we have $x^\ast_n > 0$ for some
$n \in \mathcal{N} \backslash \{1, N\}$, then this assumption simply
implies constraint (\ref{problem1_line2}). On the other hand, if for
a feasible efficiency, we have $x^\ast_n = 0$ for some $n \in
\mathcal{N} \backslash \{1, N\}$, then assuming $\gamma_n = a
(q^\ast + x^\ast_n + x^\ast_1)$ does \emph{not} have any impact on
the objective function as the term $\gamma_n x^\ast_n = 0$,
regardless of the value of $\gamma_n$. Therefore, we can restrict
our attention only to those feasible solutions for which we have
%
$\gamma_n = a (q^\ast + x^\ast_n + x^\ast_1)$ for all $n = 2,
\ldots, N-1$.
%
Having done so, to solve problem
(\ref{problem1_line1})-(\ref{problem1_line9}), we first assume that
all variables, except for $x_1 = x_N$, are \emph{fixed}.
The derivative of the objective function in (\ref{problem1_line1})
with respect to variable $x_1$ is obtained as
\begin{equation}
\frac{2 a}{\sigma^2} \left( \sigma - a (q^\ast + x^\ast_1) \right).
\label{derivative_x_1}
\end{equation}
For any $\frac{1}{2} \leq \beta \leq 1$ and due to the fact that
$\gamma_1 \geq \gamma_N$, we have
%
$\sigma = \gamma_1 + \gamma_N \geq \frac{\gamma_N}{\beta}$.
%
By adding $- a (q^\ast + x^\ast_1)$ to both sides of this
inequality, we have
\begin{equation}
\sigma - a (q^\ast + x^\ast_1) \geq \frac{\gamma_N}{\beta} - a
(q^\ast + x^\ast_1) \overset{\text{By
(\ref{condition_theorem8_caseI_part_a})}}{\geq} 0.
\label{derivative_sign_step2}
\end{equation}
Therefore, the derivative in (\ref{derivative_x_1}) is always
\emph{non-negative} and the worst-case efficiency occurs at
\textit{lower bound}
%
$x^\ast_1 = x^\ast_N = \frac{\gamma_1 - a q^\ast}{a (1 + \beta)}$.
%
From this, together with (\ref{problem1_line2}), for each $n = 2,
\ldots, N-1$,
\begin{equation}
 \gamma_n = a x^\ast_n + a( q^\ast + x^\ast_1 ) = a x^\ast_n + a \left( \frac{\gamma_1}{a (1 \! + \! \beta)} -
 \frac{q^\ast}{1 \! + \! \beta} + q^\ast \right) = a x^\ast_n + a \left( \frac{\gamma_1}{a (1 \! + \! \beta)} + q^\ast
 \left( \frac{\beta}{1 \! + \! \beta} \right) \right).
\label{gamma_n_model}
\end{equation}
Thus, the objective function in (\ref{problem1_line1}) becomes
\begin{equation}
\!\!\!\! \frac{2 a}{\sigma^2} \left[ \sigma \! \left( \frac{\gamma_1
\!-\! a q^\ast}{a(1\!+\!\beta)}\right) \! + \! \sum_{n=2}^{N-1} a \!
\left( \frac{\gamma_1}{a (1 \! + \! \beta)} \! + \! q^\ast \! \left(
\frac{\beta}{1 \! + \! \beta}\right) \! + \! x^\ast_n \right)
x^\ast_n \! - \! \frac{a}{2} \! \left( \frac{\gamma_1}{a (1 \! + \!
\beta)} \! + \! q^\ast \! \left( \frac{\beta}{1 \! + \! \beta}
\right) \right)^{\!2} \right]. \label{objective2}
\end{equation}
%
On the other hand, after reordering the terms, constraint
(\ref{problem1_line6}) becomes
\begin{equation}
x^\ast_n \leq \frac{\sigma}{a} - \frac{\gamma_1}{a(1+\beta)} -
q^\ast \left( \frac{\beta}{1 + \beta} \right), \ \ \ \ \ \ n = 2,
\ldots, N-1. \label{constraint_x_n}
\end{equation}
The right hand side in (\ref{constraint_x_n}) is
\emph{non-negative}. In fact, since $\sigma \geq \gamma_1$ and due
to (\ref{problem1_line5}), we have
\begin{equation}
\left( \sigma - \gamma_1 \right) + \beta \left( \sigma - a q^\ast
\right) \geq 0 \ \ \Rightarrow \ \ \sigma (1 + \beta) \geq \gamma_1
+ a \beta q^\ast \ \ \Rightarrow \ \ \frac{\sigma}{a} \geq
\frac{\gamma_1}{a (1\!+\!\beta)} + q^\ast \! \left( \frac{\beta}{1
\!+ \!\beta} \right).
\end{equation}
Replacing (\ref{gamma_n_model}), (\ref{objective2}), and
(\ref{constraint_x_n}) in problem
(\ref{problem1_line1})-(\ref{problem1_line9}), it reduces to the
following problem
\begin{eqnarray}
%
& \underset{\boldsymbol{\gamma}, a, N, q^\ast}{\text{minimize}} &
\frac{2 a}{\sigma^2} \left[  \frac{\sigma(\gamma_1 \! - \! a
q^\ast)}{a(1+\beta)} + \sum_{n=2}^{N-1} a \left( \frac{\gamma_1}{a
(1 \! + \! \beta)} \! + \! \frac{q^\ast \beta}{1 \! + \!
\beta} \! + \! x^\ast_n \right) x_n^\ast \! - \! \frac{a}{2} \left( \frac{\gamma_1}{a (1 \! + \! \beta)} \! + \! \frac{q^\ast \beta}{1 \! + \! \beta} \right)^{\!2} \right]  \label{problem1_2_line1} \ \ \ \ \ \ \ \\
& \text{subject to} & \sum_{n=2}^{N-1} x^\ast_n = q^\ast, \label{problem1_2_line2}  \\
& & \gamma_1 + \gamma_N = \sigma, \label{problem1_2_line3}  \\
& & \gamma_1 \geq a q^\ast, \label{problem1_2_line4}  \\
& & 0 < \gamma_n \leq \sigma, \ \ \ \ n = 2, \ldots, N-1, \label{problem1_2_line5}  \\
& & \gamma_N \leq \gamma_1 \leq \left( 1 + \frac{1}{\beta} \right) \gamma_N - \beta a q^\ast \label{problem1_2_line6}  \\
& & x^\ast_n \leq \frac{\sigma}{a} - \frac{\gamma_1}{a(1+\beta)} - q^\ast \left( \frac{\beta}{1 + \beta} \right), \ \ \ \ n = 2, \ldots, N-1, \label{problem1_2_line7}  \\
& & x^\ast_n \geq 0, \ \ \ \ \ \ \ \ \ \ \ \ \ \ \ \ \ \ \ \ \ \ \ \
\ \ \ \ \ \ \ \ \ \ \ \ \ \;\! n = 2, \ldots, N-1.
\label{problem1_2_line8}
\end{eqnarray}
Problem (\ref{problem1_2_line1})-(\ref{problem1_2_line8}) is
\emph{symmetric} in $x^\ast_2, \ldots, x^\ast_N$. Therefore, the
worst-case efficiency occurs when at Nash equilibrium we have
\begin{equation}
x^\ast_2 = x^\ast_3 = \ldots = x^\ast_{N-2} = x^\ast_{N-1} =
\frac{q^\ast}{N-2}. \label{all_equal}
\end{equation}
From (\ref{problem1_2_line7}) and (\ref{constraint_x_n}), it is
required that
\begin{equation}
\frac{q^\ast}{N-2} \leq \frac{\sigma}{a} -
\frac{\gamma_1}{a(1+\beta)} - q^\ast \left( \frac{\beta}{1 + \beta}
\right). \label{constraint_x_n_modified}
\end{equation}
By replacing (\ref{all_equal}) in (\ref{problem1_2_line1}), the
objective function in (\ref{problem1_2_line1}) reduces to
\begin{equation}
\frac{2 a}{\sigma^2} \left[ \sigma \left( \frac{\gamma_1 \! - \! a
q^\ast}{a(1\!+\!\beta)}\right) \! + \! \sum_{n=2}^{N-1} a \left(
\frac{\gamma_1}{a (1 \! + \! \beta)} + \! \frac{q^\ast \beta}{1 \! +
\! \beta} \! + \! \frac{q^\ast}{N\!-\!2} \right)
\frac{q^\ast}{N\!-\!2} \! - \! \frac{a}{2} \left( \frac{\gamma_1}{a
(1 \! + \! \beta)} \! + \! \frac{q^\ast \beta}{1 \! + \! \beta}
\right)^{\!2} \right]. \label{objective3}
\end{equation}
%
%
%
%
The objective function in (\ref{objective3}) is \emph{increasing} in
$N$ and constraint (\ref{constraint_x_n_modified}) becomes
\emph{less restrictive} as $N$ \emph{increases}. Thus, the
worst-case efficiency occurs as $N \rightarrow \infty$. We also
notice that
\begin{equation}
\lim_{N \rightarrow \infty} \ \sum_{n = 2}^{N-1} a \left(
\frac{\gamma_1}{a (1 + \beta)} + \frac{q^\ast \beta}{1 +  \beta} +
\frac{q^\ast}{N-2} \right) \frac{q^\ast}{N-2} = a q^\ast \left(
\frac{\gamma_1}{a (1 + \beta)} + \frac{q^\ast \beta}{1 + \beta}
\right). \label{objective_limit}
\end{equation}
On the other hand, as $N \rightarrow \infty$, constraint
(\ref{constraint_x_n_modified}) becomes
%
$0 \leq \frac{\sigma}{a} - \frac{\gamma_1}{a(1+\beta)} -
\frac{q^\ast \beta}{1 + \beta} $.
%
However, we already know that this constraint always holds as shown
in (\ref{objective2}). Thus, we can simply \emph{eliminate} this
constraint. By replacing (\ref{objective_limit}) in
(\ref{objective3}), the objective function becomes
\begin{equation}
\frac{2 a}{\sigma^2} \left[\sigma \left( \frac{\gamma_1 - a
q^\ast}{a(1+\beta)}\right)  + a q^\ast \left( \frac{\gamma_1}{a (1 +
\beta)} + \frac{q^\ast \beta}{1 + \beta} \right) - \frac{a}{2}
\left( \frac{\gamma_1}{a (1 + \beta)} + \frac{q^\ast \beta}{1 +
\beta} \right)^{\!2} \right]. \label{objective4}
\end{equation}
For notational simplicity, we define $\bar{q}^\ast = a q^\ast$.
After reordering the terms, (\ref{objective4}) becomes
\begin{equation}
\frac{2}{\sigma^2 (1+\beta)} \left( \gamma_1 \sigma -
\frac{\gamma_1^2}{2(1+\beta)} + \bar{q}^{\ast 2} \beta \left( 1 -
\frac{\beta}{2(1+\beta)} \right) + \bar{q}^\ast \left( - \sigma +
\gamma_1 \left( \frac{1}{1+\beta} \right) \right) \right).
\label{objective7}
\end{equation}
%
%
By looking at the remaining constraints, we can see that constraint
(\ref{problem1_line5}) can be written as
\begin{equation}
 \gamma_1 \geq \bar{q}^\ast \geq 0.
\label{condition_q}
\end{equation}
Given the definition of $\bar{q}^\ast$ and by combining constraints
(\ref{problem1_line4}) and (\ref{problem1_line7}), we also have
\begin{equation}
\frac{\sigma}{2} \leq \gamma_1 \leq \frac{1 + \beta}{1 + 2 \beta} \:
\sigma - \frac{\beta^2 \bar{q}^\ast}{1 + 2
\beta}.\label{condition_q_extra}
\end{equation}
%
%
Next, we take the derivative of the objective function in
(\ref{objective7}) with respect to $\gamma_1$ which yeilds
\begin{equation}
\frac{2}{\sigma^2 (1+\beta)} \left( \sigma - \frac{\gamma_1}{1 +
\beta} + \frac{\bar{q}^\ast}{1+\beta} \right) \geq \frac{2}{\sigma^2
(1+\beta)} \left( \sigma - \gamma_1 + \frac{\bar{q}^\ast}{1+\beta}
\right) \geq 0,
\end{equation}
where the last inequality is due to $\sigma \geq \gamma_1$ and
$\bar{q}^\ast \geq 0$. Since the derivative is \emph{non-negative},
the

\noindent worst-case efficiency occurs at the \emph{lower bound} of
$\gamma_1$. Comparing constraints (\ref{condition_q}) and
(\ref{condition_q_extra}),
\begin{equation}
\frac{\sigma}{2} \leq \bar{q}^\ast \ \ \ \Rightarrow \ \ \ \
\gamma_1 = \bar{q}^\ast, \label{condition_q_gamma_max_1}
\end{equation}
and
\begin{equation}
\frac{\sigma}{2} \geq \bar{q}^\ast \ \ \ \Rightarrow \ \ \ \
\gamma_1 = \frac{\sigma}{2}. \label{condition_q_gamma_max_2}
\end{equation}
If (\ref{condition_q_gamma_max_1}) holds, problem
(\ref{problem1_2_line1})-(\ref{problem1_2_line8}) becomes
\begin{eqnarray}
& \underset{\bar{q}^\ast}{\text{minimize}} \ & \frac{\bar{q}^{\ast 2}}{\sigma^2} \label{problem1_3_line1} \\
& \text{subject to} \ & \frac{\sigma}{2} \leq \bar{q}^\ast \leq
\frac{\sigma}{1 + \beta}. \label{problem1_3_line2}
\end{eqnarray}
On the other hand, if (\ref{condition_q_gamma_max_2}) holds, problem
(\ref{problem1_2_line1})-(\ref{problem1_2_line8}) becomes
\begin{eqnarray}
& \underset{\bar{q}^\ast}{\text{minimize}} \ & \frac{2}{\sigma^2
(1+\beta)} \left( \frac{3}{8 (1+\beta)} \sigma^2
+ \bar{q}^{\ast 2} \frac{\beta(2+\beta)}{2 (1+\beta)} - \sigma \frac{1+2\beta}{2(1+\beta)} \right) \label{problem1_4_line1} \\
& \text{subject to} \ & 0 \leq \bar{q}^\ast \leq \frac{\sigma}{2}.
\label{problem1_4_line2}
\end{eqnarray}
The objective function in (\ref{problem1_3_line1}) is
\emph{increasing} in $\bar{q}^\ast$, while the objective function in
(\ref{problem1_4_line1}) is \emph{decreasing} in $\bar{q}^\ast$.
Thus, for both optimization problems
(\ref{problem1_3_line1})-(\ref{problem1_3_line2}) and
(\ref{problem1_4_line1})-(\ref{problem1_4_line2}), the worst-case
efficiency occurs at
%
$\bar{q}^\ast = \frac{\sigma}{2}$. 
%
Exploiting this in (\ref{problem1_3_line1}) and
(\ref{problem1_4_line1}), the worst-case efficiency when $\gamma_1 +
\gamma_N = \sigma$ and inequalities
(\ref{condition_theorem8_caseI_part_a}) and
(\ref{condition_theorem8_caseI_part_a_other_one}) hold is obtained
as
\begin{equation}
\frac{1}{\sigma^2} \left( \frac{\sigma}{2} \right)^2 = \frac{1}{4}.
\label{efficiency_bound_1_4}
\end{equation}
Interestingly, the worst-case efficiency in this scenario does
\emph{not} depend on the choice of pricing parameter $\beta$.
Next, assume that condition (\ref{condition_theorem8_caseI_part_a})
holds and we have
\begin{equation}
 \gamma_1 \leq a q^\ast.
 \label{condition_theorem8_caseI_part_a_extra}
\end{equation}
From Theorem \ref{theorem6}(a), the Nash equilibria are obtained as
%
$0 \leq x^\ast_1 = x^\ast_N \leq \frac{\gamma_N - \beta a
q^\ast}{\beta a}$.
%
We can show that, in this scenario, the worst-case efficiency occurs
if $N \to \infty$ and we have $x^\ast_1=x_N^\ast = 0$ and $a q^\ast
= \frac{1 + 2 \beta \sigma}{2 \beta^2 + 4 \beta + 3}$. Thus, the
worst-case efficiency when (\ref{condition_theorem8_caseI_part_a})
and (\ref{condition_theorem8_caseI_part_a_extra}) hold is obtained
as
\begin{equation}
\frac{2}{2 \beta^2 + 4 \beta + 3}. \label{somelines_4}
\end{equation}
Details are omitted for brevity. Notice that if $\beta =
\frac{1}{2}$, then (\ref{somelines_4}) becomes $\frac{4}{11} \approx
0.36$.

Next, we assume that
\begin{equation}
\left( 1 + \frac{1}{\beta} \right) \gamma_N - \beta a q^\ast \leq
\gamma_1 \leq \frac{2}{\beta} \gamma_N - a q^\ast.
\label{condition_theorem8_caseI_part_b}
\end{equation}
From Theorem \ref{theorem6}(b), at Nash equilibrium we have
%
$x_1^\ast = \frac{\gamma_N}{\beta a} - q$ and $x^\ast_N =
\frac{\frac{2}{\beta} \gamma_N - \gamma_1}{a (1 - \beta)} -
\frac{q^\ast}{1 - \beta}$.
%
We can show that the worst-case efficiency when
(\ref{condition_theorem8_caseI_part_b}) holds is again as in
(\ref{somelines_4}). Finally, we assume
\begin{equation}
\gamma_1 \geq \frac{2}{\beta} \gamma_N - a q^\ast.
\label{condition_part_c_case_1_1}
\end{equation}
In that case, from Theorem \ref{theorem6}, if $\gamma \geq a
q^\ast$, then $x_1^\ast = \frac{\gamma_1}{2 a} - \frac{q^\ast}{2}$
and $x_N^\ast = 0$. On the other hand, if $\gamma \leq a q^\ast$,
then $x_1^\ast = x_N^\ast = 0$. In either case, the worst-case
efficiency is still obtained as in (\ref{somelines_4}).

Case II) We assume that
%
$\gamma_1 + \gamma_N < \sigma$. Without loss of generality, we also
assume that $\gamma_2 = \sigma$. In that case,
%
we can show that the worst-case efficiency in this scenario becomes
\begin{equation}
\frac{2}{3} \approx 0.67. \label{2_over_3_results}
\end{equation}
In fact, the results in this case are similar to the results in
Theorem \ref{theorem1} (see \cite[Theorem 3]{johari_jsac2006}).

Combining the results from Case I and Case II, if pricing parameter
$\beta = 1$, then
\begin{equation}
\text{PoA}\left(\text{Game \ref{game2}}, \text{Problem
\ref{problem2}}\right) \ \ \underset{\text{and }
(\ref{2_over_3_results})}{\overset{\text{By
(\ref{efficiency_bound_1_4}), (\ref{somelines_4}),}}{=}} \ \
\min\left\{
\frac{1}{4}, \frac{2}{9}, \frac{2}{3} \right\} = \frac{2}{9}.
\end{equation}
On the other hand, if $\beta = \frac{1}{2}$, then
\begin{equation}
\text{PoA}\left(\text{Game \ref{game2}}, \text{Problem
\ref{problem2}}\right) \ \ \underset{\text{and }
(\ref{2_over_3_results})}{\overset{\text{By
(\ref{efficiency_bound_1_4}), (\ref{somelines_4}),}}{=}} \ \
\min\left\{ \frac{1}{4}, \frac{4}{11}, \frac{2}{3} \right\} =
\frac{1}{4}.
\end{equation}
This concludes the proof.  $\hfill \blacksquare$

\subsection{Proof of Theorem \ref{theorem11}} \label{app:proof_theorem11}

We first notice that at optimality, we always have $z_1^S = z_N^S =
v_1^S = v_N^S$. This can be easily proved by contradiction.
Therefore, we can rewrite the objective function of Problem
\ref{problem3} as
\begin{equation}
\sum_{n=1}^{N} \gamma_n y_n + z_1 \left( \gamma_1 + \gamma_N \right)
- \frac{a_1 + a_N}{2} {z_1}^2 - \frac{a}{2} \left( \sum_{n=1}^{N}
y_n + z_1 \right)^2. \label{objective_function_theorem11}
\end{equation}
The above objective function is \emph{concave} in $y_1, \ldots,
y_N$. Using KKT optimality conditions, we can show that for each $n
\in \mathcal{N} \backslash \mathcal{M}$ we have $x^S_n = 0$ and for
each $n \in \mathcal{M}$, we have
\begin{equation}
y^S_n = \left\{ \begin{array}{ll} \frac{1}{M} \left(
\frac{\gamma_{\max}}{ a} - z^S_1 \right), & \ \  \text{if }
\frac{\gamma_{\max}}{a} \geq z^S_1, \\ 0, & \ \ \text{otherwise}.
\end{array} \right. \label{x_n_model_theorem11}
\end{equation}

\noindent Part (a): Examining both cases $\frac{\gamma_{\max}}{a}
\geq z_1$ and $\frac{\gamma_{\max}}{a} \leq z_1$, we can show that
optimality occurs if $\frac{\gamma_{\max}}{a} \leq z_1$. In that
case, $y^S_n = 0$ for all $n \in \mathcal{N}$ and the objective
function in (\ref{objective_function_theorem11}) becomes
\begin{equation}
z_1 \left( \gamma_1 + \gamma_N \right) - \frac{a + a_1 + a_N}{2}
{z_1}^2, \label{theorem10_objective_1}
\end{equation}
which is \emph{concave} in $z_1$. Thus, at optimality, we have
$z_1^S = \frac{\gamma_1 + \gamma_N}{a + a_1 + a_N}$.

\noindent Part (b): We can show that in this case, optimality occurs
when $\frac{\gamma_{\max}}{a} \geq z_1$. Replacing
(\ref{x_n_model_theorem11}) in (\ref{objective_function_theorem11}),
the objective function becomes
\begin{equation}
z_1 \left( \gamma_1 + \gamma_N - \gamma_{\max} \right) - \frac{a_1 +
a_N}{2} {z_1}^2 + \frac{\gamma_{\max}^2}{2 a},
\label{objective_function_theorem11_parts_b_c}
\end{equation}
which is \emph{concave} in $z_1$. Thus,
\begin{equation}
z_1^S \! = \! \frac{\gamma_1 + \gamma_N - \gamma_{\max}}{a_1 + a_N}.
\label{proof_theorem11_partb}
\end{equation}
By replacing (\ref{proof_theorem11_partb}) in
(\ref{x_n_model_theorem11}), the data rates in (\ref{theorem11_b})
are resulted.

\noindent Part (c): Again, we can show that in this case, optimality
occurs when $\frac{\gamma_{\max}}{a} \geq z_1$. The objective
function is the same as that in
(\ref{objective_function_theorem11_parts_b_c}). However, since
$\gamma_1 + \gamma_N \leq \gamma_{\max}$, at optimality we have
$z_1^S = 0$. Replacing this in (\ref{x_n_model_theorem11}), the data
rates in (\ref{theorem11_c}) are obtained. $\hfill \blacksquare$

\subsection{Proof of Theorem \ref{theorem12}} \label{app:proof_theorem12}

From (\ref{NE_non_zero}), and given the payoff functions in Game
\ref{game3}, for each user $n \in \mathcal{N}$, we have
\begin{equation}
y_n^\ast = \left\{ \begin{array}{ll} \left(\gamma_n - a \sum_{r=1, r
\neq n}^N y_r^\ast \right)/(2 a), & \text{if } \gamma_n
> a \sum_{r=1, r\neq n}^N y_r^\ast, \\ 0, & \text{if } \gamma_n \leq
a \sum_{r=1, r\neq n}^N y_r^\ast. \end{array} \right.
\end{equation}
First assume that
\begin{equation}
\gamma_1 + \gamma_N \geq \left( 1 + \frac{a_1 + a_N}{a} \right)
\gamma_{\max}. \label{proof_t11_condition_a}
\end{equation}
In that case, from (\ref{theorem11_a}), the maximum network
aggregate surplus becomes $(\gamma_1 + \gamma_N)^2 / (2 (a + a_1 +
a_N) )$. The worst-case efficiency is obtained by solving the
following optimization problem
\begin{eqnarray}
& \underset{\boldsymbol{y}^\ast, \boldsymbol{\gamma}, a, a_1, a_N, N, q^\ast}{\text{minimize}} \ & \frac{\sum_{n=1}^N \gamma_n y^\ast_n - \frac{a}{2} q^{\ast 2}}{\frac{(\gamma_1 + \gamma_N)^2}{2 (a+a_1+a_N)}} \label{problem3_line1} \\
& \text{subject to} \ &
\gamma_n = a q^\ast + a y^\ast_n, \ \ \ \ \text{if } y_n > 0, \ \ n = 1, \ldots, N, \label{problem3_line2a}  \\
& \ & \gamma_n \leq a q^\ast, \ \ \ \ \ \ \ \ \ \ \ \; \text{if } y_n = 0, \ \ n = 1, \ldots, N, \label{problem3_line2b} \\
& \ & \textstyle \sum_{n=1}^{N} y^\ast_n = q^\ast \geq 0, \label{problem3_line3} \\
& \ & 0 \leq \gamma_n \leq \gamma_{\max}, \ \ \ \ n = 1, \ldots, N, \label{problem3_line6}  \\
& \ & \gamma_1 + \gamma_N \geq \left( 1 + \frac{a_1 + a_N}{a} \right) \gamma_{\max}, \label{problem3_line7}  \\
& \ & y^\ast_n \geq 0, \ \ \ \ n = 1, \ldots, N. \\
& \ & a, a_1, a_N > 0. \label{problem3_line9}
\end{eqnarray}
%
Since the objective function in (\ref{problem3_line1}) is
\emph{increasing} in $a_1>0$ and $a_N>0$, the minimum occurs if $a_1
\! \rightarrow \! 0$ and $a_N \! \rightarrow \! 0$. We can also
assume, without loss of generality, that $\gamma_n = a q^\ast + a
y^\ast_n$ for each $n \! \in \! \mathcal{N}$.
%
%
By replacing (\ref{problem3_line2a}) in (\ref{problem3_line1}),
(\ref{problem3_line6}), and (\ref{problem3_line7}), problem
(\ref{problem3_line1})-(\ref{problem3_line9}) becomes
\begin{eqnarray}
& \underset{\boldsymbol{y}^\ast, a, N, q^\ast}{\text{minimize}} &
\frac{2 a}{(2 a q^\ast \! + \! a y^\ast_1 \! + \! a y^\ast_N)^2} \!
\left( a (q^\ast \! + \! y^\ast_1) y^\ast_1 \! + \! a (q^\ast \! +
\! y^\ast_N) y^\ast_N \! + \! \sum_{n=2}^{N-1} a
(q^\ast\!+\!y^\ast_n) y^\ast_n \!
- \! \frac{a}{2} q^{\ast 2} \! \right) \label{problem3_2_line1} \ \ \ \ \ \ \ \ \\
& \text{subject to} & \textstyle \sum_{n=2}^{N-1} y^\ast_n = q^\ast - y^\ast_1 - y^\ast_N \geq 0, \label{problem3_2_line2} \\
& & 0 \leq y^\ast_n \leq \frac{\gamma_{\max}}{a} - q^\ast, \ \ \ \ n = 1, \ldots, N, \label{problem3_2_line3}  \\
& & a q^\ast \leq \gamma_{\max}, \label{problem3_2_line4}  \\
& & q^\ast \geq y^\ast_1 + y^\ast_N, \label{problem3_2_line5}  \\
& & 2 q^\ast + y^\ast_1 + y^\ast_N \geq \frac{\gamma_{\max}}{a}, \\
& & a > 0.  \label{problem3_2_line6}
\end{eqnarray}
Problem (\ref{problem3_2_line1})-(\ref{problem3_2_line6}) is
\emph{symmetric} in $y^\ast_2, \ldots, y^\ast_{N-1}$.
Thus, the worst-case efficiency occurs if
\begin{equation}
y^\ast_2 = y^\ast_3 = \ldots = y^\ast_{N-2} = y^\ast_{N-1} =
\frac{q^\ast - y_1^\ast - y_N^\ast}{N-2}. \label{all_equal_y}
\end{equation}
Furthermore, problem
(\ref{problem3_2_line1})-(\ref{problem3_2_line6}) is
\emph{decreasing} in $N$. Thus, the worst-case efficiency occurs
when $N \to \infty$. From this, together with (\ref{all_equal_y}),
optimization problem
(\ref{problem3_2_line1})-(\ref{problem3_2_line6}) reduces to
%
%
\begin{eqnarray}
& \underset{y_1^\ast, y_N^\ast, a, q^\ast}{\text{minimize}} \ & \frac{2 (y_1^{\ast 2} + y_N^{\ast 2} +
\frac{q^{\ast 2}}{2})}{(2 q^\ast + y^\ast_1 + y^\ast_N)^2} \label{problem3_3_line1} \ \ \ \ \\
& \text{subject to} \ & y^\ast_1 + y^\ast_N \leq q^\ast \leq  \frac{\gamma_{\max}}{a}, \label{problem3_3_line2} \\
& \ & 2 q^\ast + y^\ast_1 + y^\ast_N \geq \frac{\gamma_{\max}}{a}
\label{problem3_3_line3}
\end{eqnarray}
Problem (\ref{problem3_3_line1})-(\ref{problem3_3_line3}) is
\emph{symmetric} in $y_1$ and $y_N$. In fact, we can show that the
worst-case efficiency occurs at $y_1^\ast = y_N^\ast =
\frac{q^\ast}{4}$. Replacing this in the objective function in
(\ref{problem3_3_line1}), the worst-case efficiency when
(\ref{proof_t11_condition_a}) holds is obtained as
%
%
\begin{equation}
2 \left( \frac{2 \times \frac{1}{16} + \frac{1}{2}}{(\frac{5}{2})^2}
\right) = \frac{1}{5}. \label{1_over_5_results}
\end{equation}
We can also show that if
%
$\gamma_{\max} \leq \gamma_1 + \gamma_N \leq \left( 1 + \frac{a_1 +
a_N}{a} \right) \gamma_{\max}$ 
%
or
%
$\gamma_{\max} \geq \gamma_1 + \gamma_N$,
%
then the worst-case efficiency is \emph{equal} or \emph{higher}
(i.e., better) than $\frac{1}{5}$. In particular, if $\gamma_{\max}
\geq \gamma_1 + \gamma_N$, then the worst-case efficiency is
$\frac{2}{3}$ which resembles the results in \cite{johari_jsac2006}.
Details are omitted here for brevity. In summary, we have
\begin{equation}
\text{PoA}\left(\text{Game \ref{game3}}, \text{Problem
\ref{problem3}}\right) \ \ \overset{\text{By
(\ref{1_over_5_results})}}{=} \ \
\min \left\{ \frac{1}{5}, \frac{2}{3} \right\} = \frac{1}{5}.
\end{equation}
This concludes the proof. $\hfill \blacksquare$

\bibliographystyle{IEEEtran}
\bibliography{IEEEabrv,ISNCG}

\newpage

\begin{table} \label{tab:summary_of_results}  \centering{ \caption{
$ \ \ \ \ \ \ \ \ \ \ \ \ \ \ \ \ \ \: $ Summary of the key results
in this paper and comparison with \newline related state-of-the-art
results without considering network coding in
\cite{johari_jsac2006}.} } \vspace{-0.25cm}
\begin{tabular}{|c|c|c|c|c|}
  \hline
  Networking & Routing & \multicolumn{2}{c}{Network Coding and
  Routing} \vline
  & Network Coding and Routing \\
  Setting & Only & \multicolumn{2}{c}{with \emph{Zero} Side Link
  Costs} \vline
  & with \emph{Non-zero} Side Link Costs \\ \hline \hline
 & & \multicolumn{2}{c}{} \vline & \\
 Optimization & Problem \ref{problem1} & \multicolumn{2}{c}{Problem \ref{problem2}} \vline & Problem
 \ref{problem3} \\ & & \multicolumn{2}{c}{} \vline & \\ \hline
 & & \multicolumn{2}{c}{} \vline & \\
 Game & Game \ref{game1} & \multicolumn{2}{c}{Game \ref{game2}} \vline &
 Game \ref{game3} \\ & & \multicolumn{2}{c}{} \vline &  \\ \hline
Number of & & \multicolumn{2}{c}{} \vline & \\
Nash & One (Unique) & \multicolumn{2}{c}{Can be \emph{infinite}} \vline & One (Unique) \\
Equilibria & & \multicolumn{2}{c}{} \vline & \\ \hline
%
%
%
%
%
%
%
%
%
%
%
Price-of & & \multicolumn{2}{c}{} \vline & \\
Anarchy$^\dag$ & $\displaystyle \frac{2}{3}$ & \multicolumn{2}{c}{$\displaystyle \frac{1}{4}$} \vline & $\displaystyle \frac{1}{5}$ \\
(PoA) & & \multicolumn{2}{c}{} \vline & \\ \hline
%
%
%
%
%
 & & \multicolumn{2}{c}{} \vline & \\
Theorem & Theorem \ref{theorem1} & \multicolumn{2}{c}{Theorem
\ref{theorem9}} \vline & Theorem
\ref{theorem12} \\
 & & \multicolumn{2}{c}{} \vline & \\ \hline \hline
Reference & \cite{johari_jsac2006} & \multicolumn{3}{c}{This Paper}
 \vline \\ \hline
\end{tabular}

\vspace{0.5cm}

\begin{tabular}{l}
\ $\dag \ \:\! $ Here, the PoA for the network coding scenario with
\emph{zero} side link costs is calculated based on
\\
\ the assumption that we use \emph{price discrimination} with
parameter $\beta = 0.5$. If single pricing is used,
\\
\ i.e., if $\beta = 1$, then the PoA can be less (i.e., worse) than
$\frac{1}{4}$, as discussed in Theorem \ref{theorem8}.
\end{tabular}
\end{table}

\begin{table} \centering \caption{List of Key Notations}
\vspace{-0.25cm}
\begin{tabular}{|l|l|}
\hline $\text{PoA}\left(\text{Game $\Omega$}, \text{Problem
$\Omega$} \right) \!\! $ & Price-of-anarchy for Game $\Omega$ with
respect to Problem $\Omega$, where $\Omega = 1, 2, 3$. \\ \hline
$\mathcal{N}$ & Set of all users in the network. \\ \hline
$N$ & Number of all users in the network. \\
\hline
$s_n, t_n$ & Transmitter and receiver nodes of user $n \in
\mathcal{N}$, respectively. \\ \hline
$(i, j)$ & Shared bottleneck link between intermediate nodes $i$ and
$j$. \\ \hline
$x_n$ & Data rate of user $n \in \mathcal{N}$ in the networks in Figs. \ref{fig:f1} and \ref{fig:f2}. \\
\hline
$\boldsymbol{x}_{-n}$ & Vector of data rates of all users other than
user $n$ in the networks in Figs. \ref{fig:f1} and \ref{fig:f2}. \\
\hline
$\boldsymbol{x}$ & Vector of data rates of all users in the networks
in Figs. \ref{fig:f1} and \ref{fig:f2}.
\\ \hline
$U_n(\cdot)$ & Utility function of user $n \in \mathcal{N}$. \\
\hline
$\gamma_n$ & Slope of linear utility function of user $n \in
\mathcal{N}$. \\ \hline
$C(\cdot)$ & Cost function of shared bottleneck link $(i, j)$. \\
\hline
$p(\cdot)$ & Price function of shared bottleneck link $(i, j)$. \\
\hline
$a$ & Price parameter, $p(q) = a q$. \\ \hline
$\mu$ & Price value for routed packets. \\ \hline
$\delta$ & Price value for network coded packets. \\ \hline
$\beta$ & Price discrimination parameter. \\ \hline
$P_n$ & Payoff function of user $n \in \mathcal{N}$ in Game
\ref{game1}. \\ \hline
$Q_n$ & Payoff function of user $n \in \mathcal{N}$ in Game
\ref{game2}. \\ \hline
$\boldsymbol{x}^S$ & Optimal solution for Problems \ref{problem1}
and \ref{problem2}. \\ \hline
$\boldsymbol{x}^\ast$ & Nash equilibrium for Games \ref{problem1}
and \ref{problem2}. \\ \hline
$x^B_n(\cdot)$ & Best response data rate for user $n \in
\mathcal{N}$ in Games \ref{game1} and \ref{game2}. \\ \hline
$X_1, X_N$ & Packets/symbols sent from source nodes $s_1$ and $s_N$,
respectively. \\ \hline
$X_1 \oplus X_N$ & Packet/symbol obtained by joint encoding of
packets/symbols $X_1$ and $X_N$. \\ \hline
$C_1(\cdot)$, $C_N(\cdot)$ & Cost functions of side links $(s_1, t_N)$ and $(s_N, t_1)$ in the network in Fig. \ref{fig:f7}. \\
\hline
$p_1(\cdot)$, $p_N(\cdot)$ & Price functions of side links $(s_1, t_N)$ and $(s_N, t_1)$ in the network in Fig. \ref{fig:f7}. \\
\hline
$a_1$, $a_N$ & Price parameters, $p_1(q) = a_1 q$ and $p_N(q) = a_N
q$. \\ \hline
$y_n$ & Data rate for routed packets of user $n \in \mathcal{N}$ in the network in Fig. \ref{fig:f7}. \\
\hline
$z_1, z_N$ & Data rate for encoded packets of users 1 and $N$ on link $(i, j)$ in the network in Fig. \ref{fig:f7}. \\
\hline
$v_1, v_N$ & Data rate for remedy packets of users 1 and $N$ on the side links in the network in Fig. \ref{fig:f7}. \\
\hline
$\boldsymbol{y}_{-n}$ & Vector of data rates for routed packets of all users other than user $n$ in the network in Fig. \ref{fig:f7}. $ \!\!\!\!\! $ \\
\hline
$W_n$ & Payoff function of user $n \in \mathcal{N}$ in Game
\ref{game3}. \\ \hline
$\boldsymbol{y}^S, \boldsymbol{z}^S, \boldsymbol{v}^S$ & Optimal
solution for Problem \ref{problem3}. \\ \hline
$\boldsymbol{y}^\ast, \boldsymbol{z}^\ast, \boldsymbol{v}^\ast$ &
Nash equilibrium for Game \ref{problem3}. \\
\hline
$y^B_n(\cdot)$ & Best response data rate for routed packets of user
$n \in \mathcal{N}$ in Game \ref{game3}. \\ \hline
$z_1^B(\cdot), z_N^B(\cdot)$ & Best response data rate for
encoded packets of users 1 and $N$ in Game \ref{game3}. \\
\hline
$v_1^B(\cdot), v_N^B(\cdot)$ & Best response data rate for
remedy packets of users 1 and $N$ in Game \ref{game3}. \\
\hline
\end{tabular}
\label{tab:notations}
\end{table}

\newpage

\end{document}